%% file: main.tex
\documentclass[lettersize,journal]{IEEEtran}
\usepackage{amsmath,amsfonts}
\usepackage{booktabs} 
\usepackage{multirow} 
\usepackage{float}
\usepackage{prettyref}
\usepackage[skins]{tcolorbox}
\usepackage{xcolor}
\newcommand{\TWORCell}[2]{\begin{tabular}{@{}c@{}}#1 \\ #2\end{tabular}}

\usepackage{booktabs} 
\usepackage{multirow} 
\usepackage{cuted}     
\usepackage{capt-of}  
\usepackage[percent]{overpic}   
\usepackage{placeins}           
\usepackage{stfloats}           
\usepackage{prettyref}
\newrefformat{fig}{Figure~\ref{#1}}
\newrefformat{par}{Section~\ref{#1}}
\newrefformat{appen}{Appendix~\ref{#1}}
\newrefformat{sec}{Section~\ref{#1}}
\newrefformat{sub}{Section~\ref{#1}}
\newrefformat{table}{Table~\ref{#1}}
\newrefformat{ass}{Assumption~\ref{#1}}
\newrefformat{alg}{Algorithm~\ref{#1}}
\newrefformat{def}{Definition~\ref{#1}}
\newrefformat{thm}{Theorem~\ref{#1}}
\newrefformat{cor}{Corollary~\ref{#1}}
\newrefformat{lem}{Lemma~\ref{#1}}
\newrefformat{step}{Step~\ref{#1}}
\newrefformat{ln}{Line~\ref{#1}}
\newrefformat{rem}{Remark~\ref{#1}}
\newrefformat{eq}{Equation~\ref{#1}}
\newrefformat{pb}{Problem~\ref{#1}}
\newrefformat{it}{Item~\ref{#1}}
\newrefformat{te}{Term~\ref{#1}}
\def\Eqref Eq:#1:{\eqref{eq:#1}}
\newrefformat{Eq}{Equation~\Eqref#1:}

\usepackage{amsmath,amsfonts,amsthm}
\usepackage{graphicx}
\usepackage{mathrsfs}
\usepackage{wrapfig}
\usepackage{bbold}
\usepackage{tabu}
\usepackage{multirow}
\usepackage{enumitem}
\usepackage{booktabs}
\usepackage{xcolor}
\usepackage{adjustbox}
\usepackage{algpseudocode}
\usepackage{svg}
\usepackage{graphpap}
\DeclareMathOperator*{\argmin}{arg\,min}

\usepackage[ruled]{algorithm2e} 

\SetAlFnt{\small}
\SetAlCapFnt{\small}
\SetAlCapNameFnt{\small}
\SetAlCapHSkip{0pt}

\usepackage{array}
\usepackage[caption=false,font=normalsize,labelfont=sf,textfont=sf]{subfig}
\usepackage{textcomp}
\usepackage{stfloats}
\usepackage{url}
\usepackage{verbatim}
\usepackage{graphicx}
\def\BibTeX{{\rm B\kern-.05em{\sc i\kern-.025em b}\kern-.08em
    T\kern-.1667em\lower.7ex\hbox{E}\kern-.125emX}}
\usepackage{balance}
\begin{document}
\title{Handle-based Mesh Deformation\\Guided By Vision Language Model}

\author{Xingpeng Sun$^1$, Shiyang jia$^3$, Zherong Pan$^2$, Kui Wu$^2$, Aniket Bera$^1$ \thanks{$^1$ Purdue University, USA. $^2$ LightSpeed Studios, USA. $^3$ University of California San Diego, USA.}}

\markboth{Journal of \LaTeX\ Class Files,~Vol.~18, No.~9, September~2020}%
{How to Use the IEEEtran \LaTeX \ Templates}

\maketitle

\begin{strip}
\centering
\setlength\unitlength{1pt}
\begin{picture}(510,290)
  \put(0,0){\includegraphics[width=\textwidth]{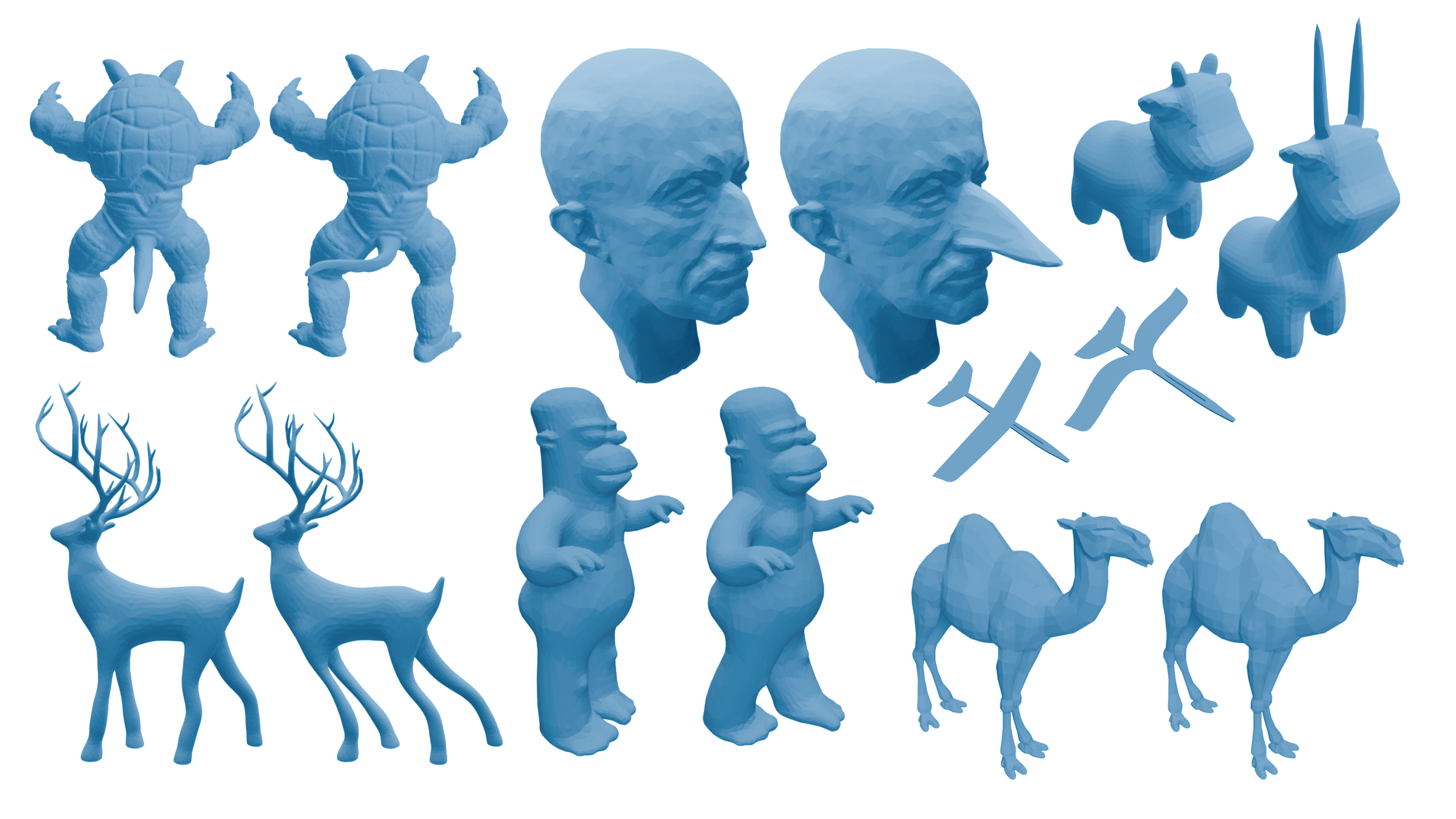}}
  \small
  \put(50,280){(a) Swing the tail leftward}
  \put(220,280){(b) Pinocchio telling lies}
  \put(370,287){(c) Raising its head with elongated horns}
  \put(50,5){(d) A deer leaning forward}
  \put(200,5){(e) Homer is walking}
  \put(325,5){(f) A camel with more fat stored in its hump}
  \put(400,118){\shortstack{(g) Airplane with backward-\\curving wings}}
\end{picture}
\captionof{figure}{A gallery of results generated using our text-driven handle-based mesh deformation, where the deformed meshes (right) are guided by the text prompt.}
\label{fig:teaser}
\end{strip}

\begin{abstract}
Mesh deformation is a fundamental tool in 3D content manipulation. Despite extensive prior research, existing approaches often suffer from low output quality, require significant manual tuning, or depend on data-intensive training. To address these limitations, we introduce a training-free, handle-based mesh deformation method.
Our core idea is to leverage a Vision-Language Model (VLM) to interpret and manipulate a handle-based interface through prompt engineering. We begin by applying cone singularity detection to identify a sparse set of potential handles. The VLM is then prompted to select both the deformable sub-parts of the mesh and the handles that best align with user instructions. Subsequently, we query the desired deformed positions of the selected handles in screen space. To reduce uncertainty inherent in VLM predictions, we aggregate the results from multiple camera views using a novel multi-view voting scheme.
Across a suite of benchmarks, our method produces deformations that align more closely with user intent, as measured by CLIP and GPTEval3D scores, while introducing low distortion---quantified via membrane energy. In summary, our approach is highly automated  and consistently delivers high-quality mesh deformations.
\end{abstract}

\begin{IEEEkeywords}
handle-based deformation, mesh deformation, vision language model
\end{IEEEkeywords}

\input{intro.tex}
\input{related.tex}
\input{problem.tex}
\input{method.tex}
\begin{figure*}
  \centering
  \begin{overpic}[width=\textwidth,trim=0 2cm 0 0,clip]{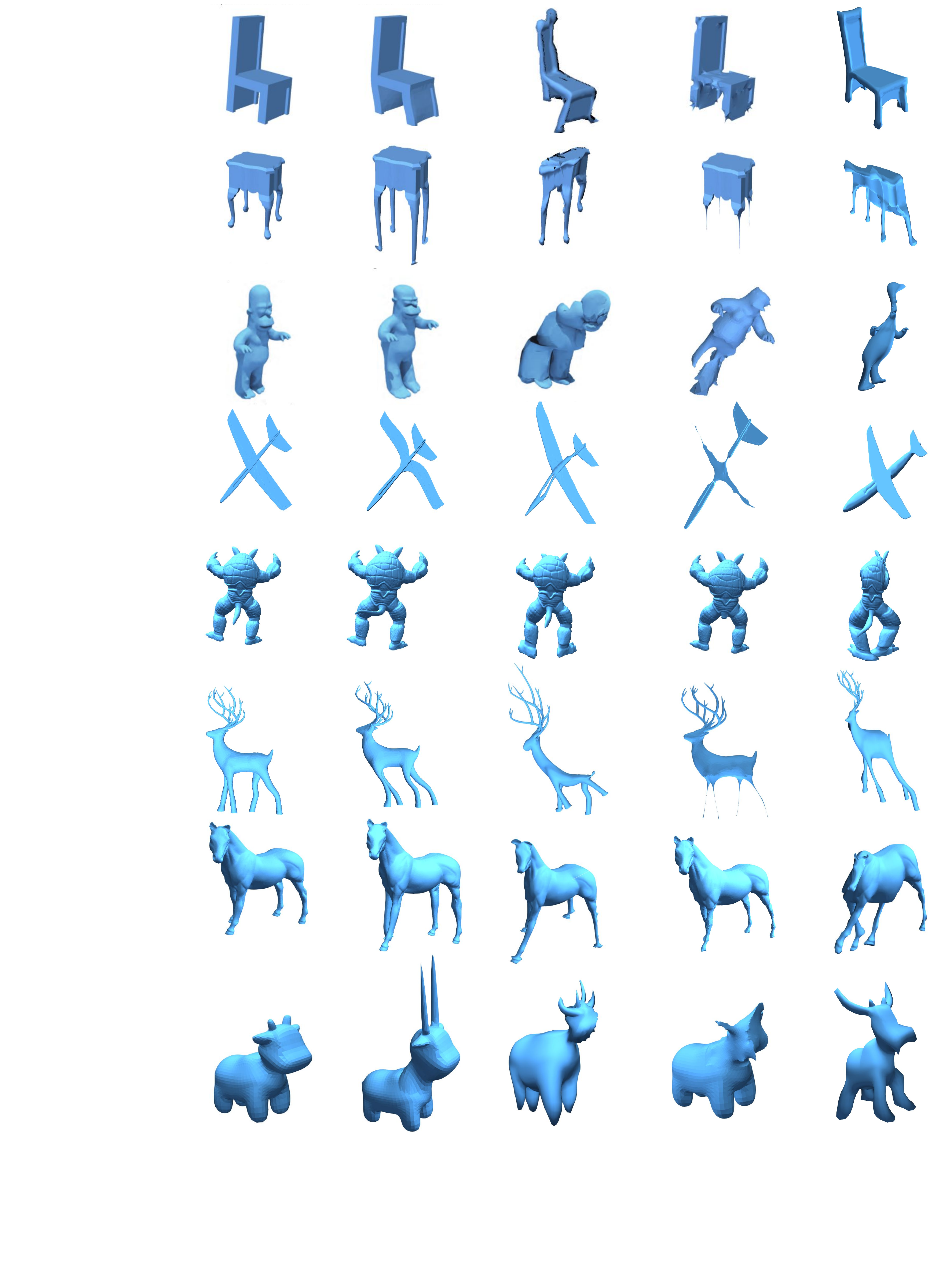}
    \put( 17,100){\textbf{Origin}}
    \put(30,100){\textbf{Ours}}
    \put(40,100){\textbf{TextDeformer}}
    \put(53,100){\textbf{MagicClay}}
    \put(65,100){\textbf{MeshUp}}

    \put(0,93){\shortstack[l]{a) suitable for a\\tall person to sit on}}
    \put(0,82){\shortstack[l]{b) stretching all\\chair legs to the\\front of the chair}}
    \put(0,70){\shortstack[l]{c) Homer with\\shorter head and\\legs extend forward}}
    \put(0,60){\shortstack[l]{d) Airplane with\\backward-curving wings}}
    \put(0,51){\shortstack[l]{e) Armadillo swifts\\tail to its left}}
    \put(0,38){\shortstack[l]{f) Deer leaning\\forward}}
    \put(0,28){\shortstack[l]{g) Horse with\\longer legs}}
    \put(0,12){\shortstack[l]{h) Spot raising\\its head with\\elongated horns}}
  \end{overpic}
  \caption{\label{fig:compareTextDeformer}
  From left to right, we show the original mesh, the deformed mesh using our method, TextDeformer, MagicClay, and MeshUp. Compared with our method, TextDeformer/MagicClay/MeshUp often introduce excessive deformations and unnecessary high-frequency details.}
\end{figure*}
\input{evaluation.tex}

\input{conclusion.tex}

\bibliographystyle{IEEEtran}

\bibliography{sample-bibliography}
\clearpage
\input{supp.tex}

\end{document}

%% file: intro.tex
\section{\label{sec:intro}Introduction}

Mesh deformation is a fundamental tool for many downstream applications in game and film production pipelines, including character rigging, terrain editing, and UV parameterization. Over the years, multiple research directions have enhanced the efficacy, accuracy, and usability of deformation techniques. For instance, various elastic energy formulations (e.g.,~\cite{Rabinovich:SLIM:2017}) have been introduced to model material behavior, enabling the automatic computation of mesh shapes from a sparse set of user-defined handles. In parallel, interpolation-based methods (e.g.,~\cite{floater2003mean, lipman2008green,jacobson2011bounded}) allow for rapid mesh shape computation without the need for numerical optimization. Over time, researchers have enhanced these methods by proposing flexible handle combinations~\cite{jacobson2011bounded} and joint-aware deformation techniques~\cite{xu2009joint}. Nonetheless, the process of manually authoring mesh shapes remains labor-intensive and costly, which typically involves the manual labels of deformable mesh sub-part, the identification of handles, and the dragging of handles to pose the mesh.

The recent success of machine learning has opened new opportunities for automating the mesh deformation pipeline through data-driven approaches. Machine learning techniques have been applied to learn end-to-end energy-based deformations~\cite{cao2023efficient}, as well as cage- or skeleton-driven deformations~\cite{yifan2020neural, li2021learning}. In parallel, several works~\cite{xu2020rignet, kyptd} have focused on learning to generate deformation rigs directly from mesh geometry.
While these methods can assist artists in setting up the deformation interface, they still often require extensive manual fine-tuning to manipulate handles and achieve desired poses. Moreover, many of these approaches are data-hungry, frequently necessitating retraining for each specific mesh category, which further limits their practicality and scalability.
Most recently, the rise of large generative models (e.g.,~\cite{rombach2022high}) has sparked growing interest in text-driven 3D model generation~\cite{zhang2024clay} and deformation~\cite{gao2023textdeformer}. Text-based interfaces offer a highly accessible and efficient means of interacting with 3D content, eliminating the need for expertise in modeling or rigging. However, the output quality of these methods often falls short.
For example, the text-driven deformation approach in~\cite{gao2023textdeformer} leverages the CLIP score~\cite{radford2021learning} to guide the optimization of mesh vertices from screen space. Yet, this optimization process is prone to challenges such as view occlusion and the generation of noisy surfaces. Although the method incorporates elastic energy regularization to improve mesh quality, the resulting deformations remain suboptimal. In summary, none of existing text-driven methods can generate high-quality mesh deformations.

We present a Vision-Language Model (VLM) approach for text-guided 3D content creation by leveraging a handle-based deformation interface to modify 3D objects. Our method combines the strengths of traditional handle-based deformation~\cite{jacobson2011bounded} with the accessibility of text-driven interfaces~\cite{gao2023textdeformer}. Rather than directly optimizing mesh vertices, we employ a VLM to interpret textual instructions and translate them into handle selections along with corresponding drag directions. By operating within the constraints of a handle-based framework, our approach preserves fine-grained details and produces high-quality, realistic deformations.
Our method takes as input an arbitrary 3D mesh and a text instruction specifying the desired deformation. First, we provide the VLM with multi-view renderings of the mesh and prompt it to break down the deformation task into a sequence of sub-instructions, each targeting a specific part of the mesh. For each sub-instruction, we implement a multi-view graph-cut algorithm that enables the VLM to segment the relevant deformable sub-part of the mesh.
Next, we apply a cone singularity detection algorithm~\cite{li2023efficient} to identify potential handle points within the segmented sub-part. The VLM is then tasked with selecting and dragging handle points under each camera view. These predicted handle manipulations are translated into mesh deformations using an optimization-based algorithm, with deformations averaged across views to mitigate the inherent uncertainty in VLM predictions and 2D-to-3D information loss.
Overall, our method enables text-guided 3D content generation. Unlike prior approaches~\cite{kyptd,gao2023textdeformer}, it eliminates the need for retraining on specific 3D object categories and consistently facilitates detail preservation. Technically, our contribution is to restrict the mesh modification to low-frequency spaces using handle-based interfaces, which mitigates the inherent ambiguity in text-based VLM output, thus consistently producing high-quality output meshes. Practically, our method lowers the barrier for non-experts and speeds up workflow for professionals; interprets high-level design intent and enables context-aware editing; and paves the way for procedural modeling with user prompts. Our contributions are as follows:
\begin{itemize}
\item A VLM-guided, few-shot, text-driven automatic 3D mesh deformation technique.
\item A VLM-guided 3D content generation technique with guaranteed output quality due to the handle-based interface.
\item Our experiments illustrated in~\prettyref{fig:teaser} show that our method generates deformed meshes that align better with text prompts in terms of CLIP and GPTEval3D scores and less distortion as measured by membrane energy.
\end{itemize}

%% file: related.tex
\section{Related Work}
We review related works on 3D deformation models and text-driven 3D content generation techniques.
\begin{figure*}[t]
\centering
\begin{picture}(510, 160)
    \put(0, 0){\includegraphics[width=\textwidth]{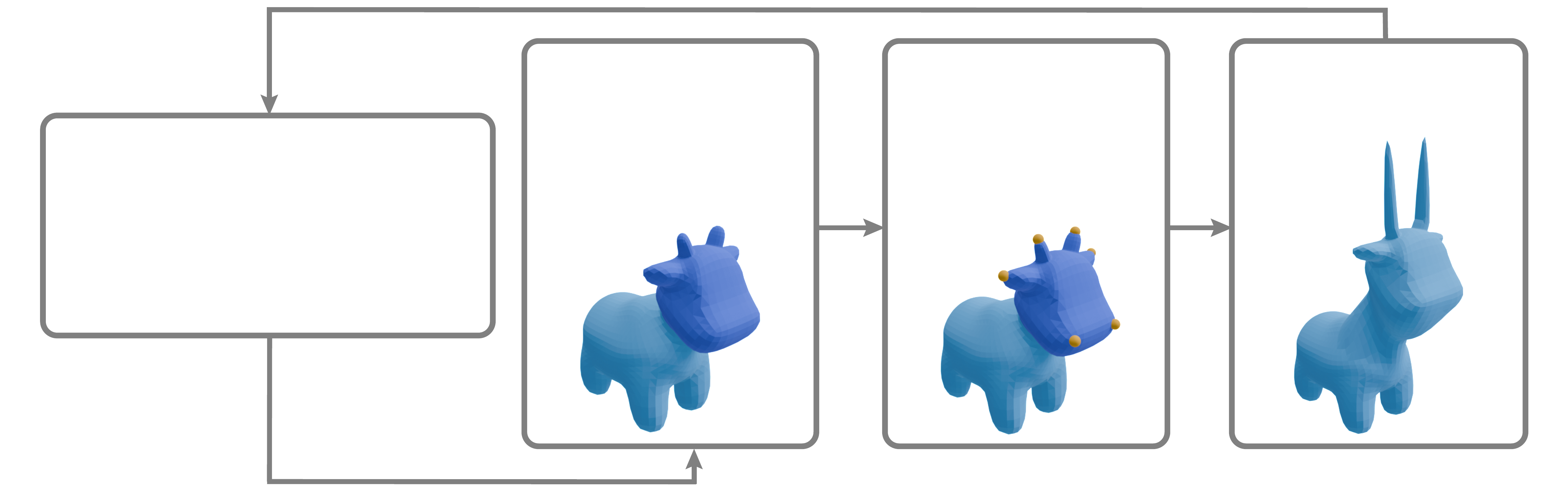}}
    \small
    
    \put(40, 110){\textbf{Sub-instruction Extraction}}
    \put(14, 90){\includegraphics[width=15px]{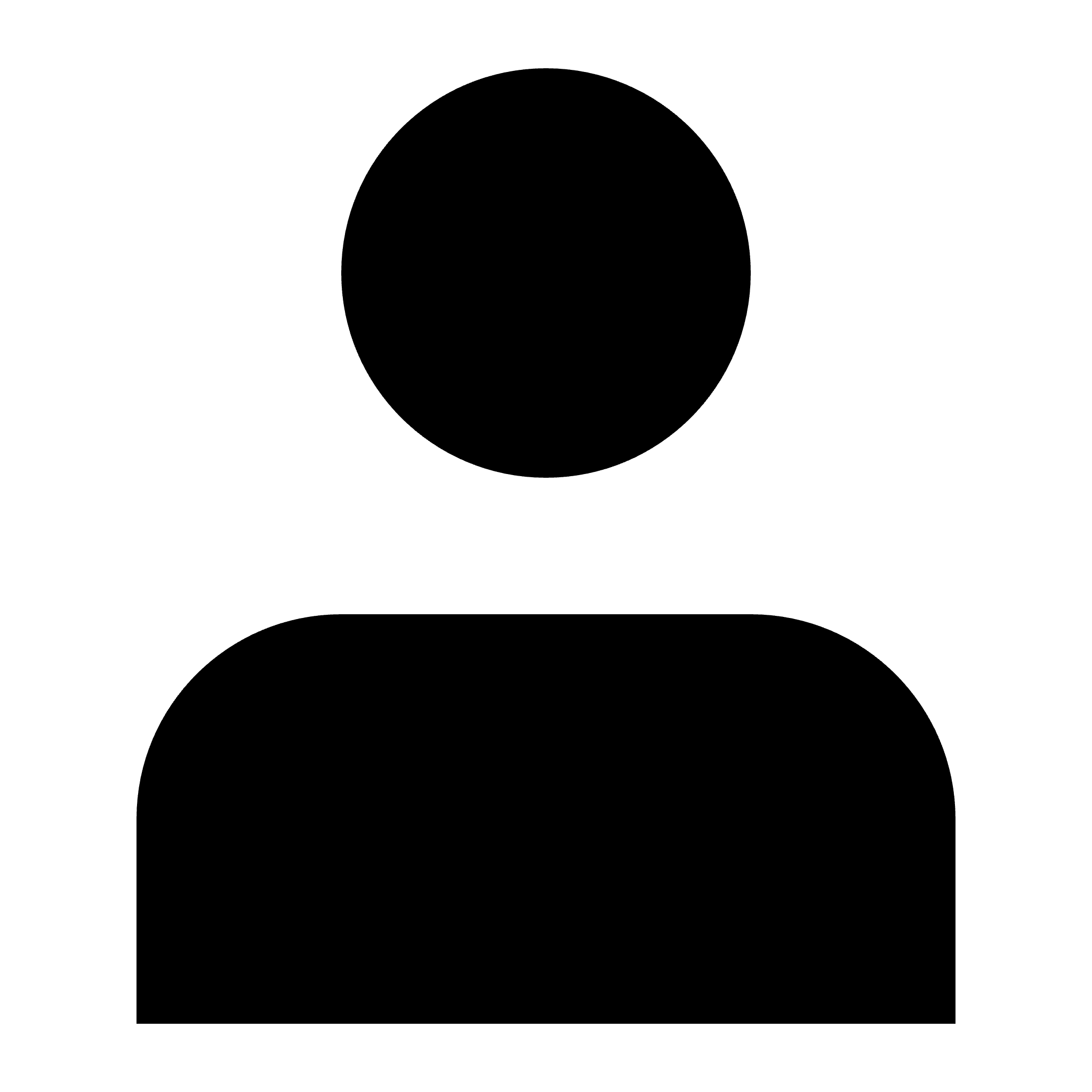}}
    \put(28, 93){Raise its head with elongated horns}
    \put(15, 66){\includegraphics[width=13px]{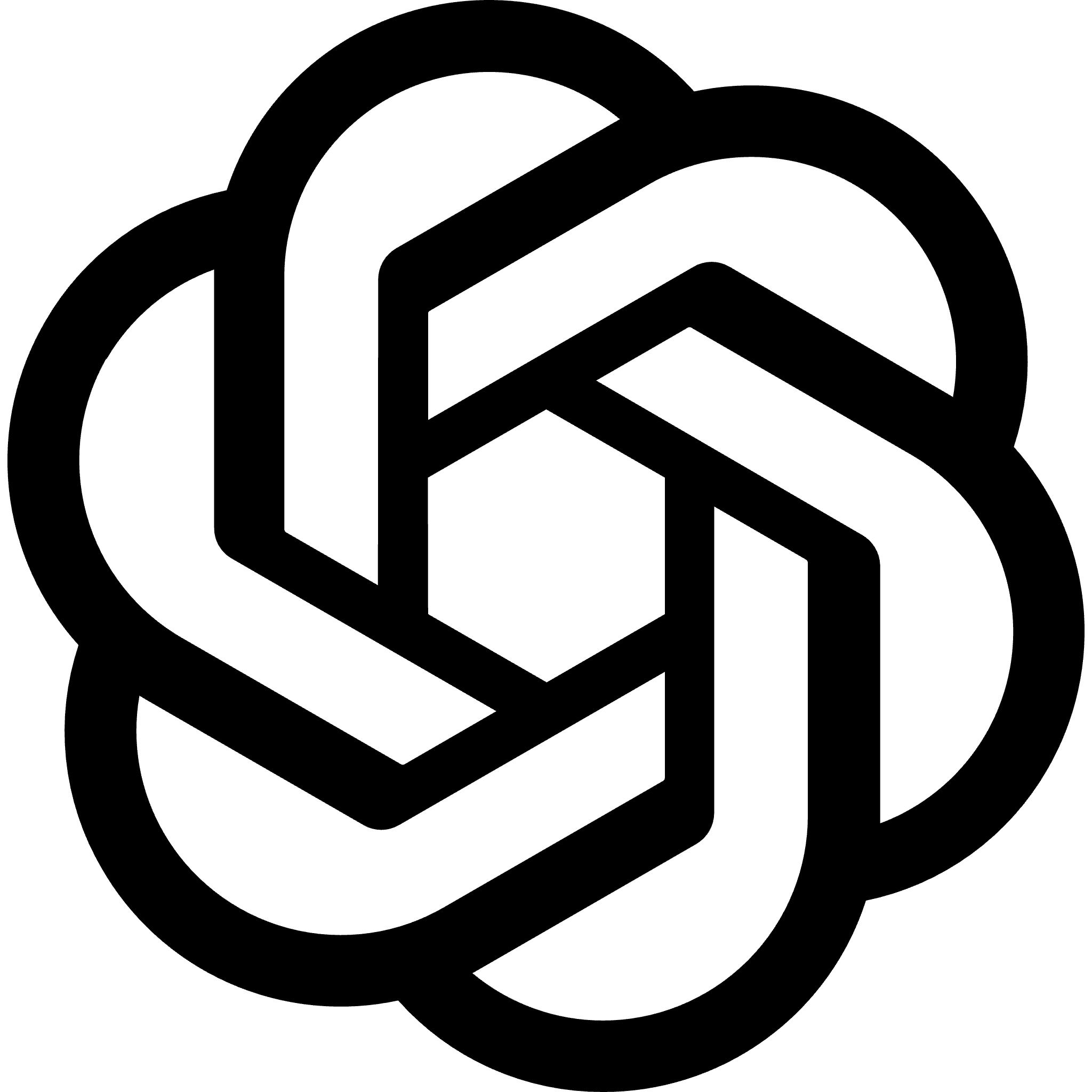}}
    \put(38, 76){Step 1: raise head}
    \put(38, 65){Step 2: elongate horns}
    
    \put(178, 137){\textbf{Sub-part Identification}}
    \put(175, 115){\includegraphics[width=15px]{figs/user.png}}
    \put(195, 120){Raise head}
    \put(176, 94){\includegraphics[width=13px]{figs/openai.png}}
    \put(195, 100){Related part: head}

    \put(305, 137){\textbf{Handle Selection}}
    \put(292, 119){\includegraphics[width=15px]{figs/user.png}}
    \put(310, 124){Raise head}
    \put(293, 98){\includegraphics[width=13px]{figs/openai.png}}
    \put(310, 110){Move the handle at}
    \put(310, 99){\footnotesize{(1012,500)} to}
    \put (310, 89){\footnotesize{(1012,650)}}

    \put(415, 128){\textbf{\shortstack{Mesh Deformation\\Multi-view Voting}}}

    \put(95,40){$\mathcal{M}_n'$}
    \put(95, 134){$\mathcal{I}_n$}
    \put(272,95){\small$l(f)$}
    \put(272,77){\small$l(v)$}
    \put(390,95){\small$\mathcal{H}$}
    \put(390,75){\small$\bar{\mathcal{H}}$}
\end{picture}
\caption{\label{fig:pipeline} An illustration of our pipeline working for the example in~\prettyref{fig:teaser} (a). From left to right: We first extract a series of sub-instructions $\mathcal{I}_1,\cdots,\mathcal{I}_n$ from $\mathcal{T}$. We then identify a deformable sub-part corresponding to each $\mathcal{I}_n$. Next, we detect a super-set of handles $\bar{\mathcal{H}}$ (green points) and have the VLM select a subset of handles $\mathcal{H}$ to be deformed, and predict their target positions in the image space. Finally, we optimize the deformed mesh for each view and merge the results via multi-view voting.} 
\end{figure*}

%
%

\paragraph{Deformation Models}
A deformation model defines how an shape undergoes global deformation when specific parts are modified. By employing a deformation model, the number of degrees of freedom exposed to users is significantly reduced, simplifying the manipulation process. Deformation models have widespread applications in fields such as computer vision~\cite{wen2019pixel2mesh++}, computer graphics~\cite{nealen2006physically}, and medical imaging~\cite{sotiras2013deformable}.
Over the years, several deformation models have been developed specifically for mesh-based representations. These models can be broadly classified into two categories: optimization-based and interpolation-based models. Optimization-based models typically assume that the mesh is governed by an internal potential energy, with the deformed state corresponding to a local energy minimum~\cite{Rabinovich:SLIM:2017, nealen2006physically}. As a result, when a sub-part of the mesh is altered, the entire structure deforms in a way that preserves detail. However, optimization-based models rely on numerical algorithms to compute the mesh shape, which can be computationally expensive. In contrast, interpolation-based models~\cite{floater2003mean, lipman2008green, jacobson2011bounded} define a closed-form interpolation scheme that can be efficiently evaluated across the entire mesh.
Most deformation models offer intuitive interfaces for selecting and modifying mesh sub-parts. The two most common interfaces are handle-based~\cite{jacobson2011bounded} and cage-based~\cite{stroter2024survey}. We opt for the handle-based interface, as it exposes an even smaller number of degrees of freedom, making it easier for VLMs to manipulate.

In addition to analytic deformation models, learning-based models have been developed, mostly for other shape representations than meshes, such as point clouds~\cite{yang2018foldingnet}, NeRF~\cite{park2021nerfies}, and Gaussian Splatting~\cite{lu20243d}. However, these models are still relatively immature compared to their mesh-based counterparts.  Interestingly, recent works~\cite{xu2022deforming,sun2024tuttenet} have leveraged mesh-based techniques to deform NeRFs, either by extracting a mesh~\cite{xu2022deforming} or by employing a background grid~\cite{sun2024tuttenet}. These findings suggest that our method can adapt to newer appearance models with minimal adjustments in the future.

\paragraph{Text-driven 3D Content Generation}
With the success of VLM-guided image generation, text-driven 3D content generation has gained significant attention. However, due to limited datasets and computational resources, text-to-3D generative models are still in their infancy~\cite{zhang2024clay}. As an alternative, researchers have explored extracting 3D information from text-to-image diffusion models~\cite{rombach2022high}. \cite{poole2022dreamfusion} derived NeRF representations from multi-view images using the SDS loss, ShapeWalk~\cite{slim2024shapewalk} and ShapeTalk~\cite{achlioptas2023shapetalk} perform global point cloud 3D shape editing, while~\cite{gao2023textdeformer} and \cite{michel2022text2mesh} deform 3D meshes by aligning them with textual descriptions via the CLIP model. Both approaches aim to recover fine details from multi-view images.
However, these methods often introduce artifacts due to discrepancies between 2D and 3D views, multi-view inconsistencies, and occlusion. To address this, our approach employs handle-based deformation, ensuring detail preservation of the input mesh. Another limitation of CLIP-based methods~\cite{gao2023textdeformer,michel2022text2mesh} is the difficulty in understanding complex text instructions~\cite{wu2024self}. In contrast, our method leverages the more advanced VLM model (GPT-4o) to handle intricate textual guidance effectively.
Recent efforts such as BlenderAlchemy~\cite{huang2024blenderalchemy} and BlenderGym~\cite{gu2025blendergym} also tried to introduce VLM-based agentic frameworks for graphic editing. However, these approaches target general graphic editing within only Blender, and rely heavily on human to first preprocessed inputs to tailor for the software, whereas our method operates directly on a watertight single mesh guided only by text input.

Complementary to our work, ~\cite{wang2024chat2layout} and ~\cite{yang2024holodeck} use VLMs to organize diverse 3D models from pre-existing datasets but do not allow for geometry modifications. Our method bridges the gap between these approaches. It balances the detail-oriented mesh deformation of CLIP-guided methods~\cite{gao2023textdeformer,michel2022text2mesh} with the structural organization of VLM-guided layouts~\cite{wang2024chat2layout,yang2024holodeck}. We enable modifications to low-frequency mesh geometries via handle-based manipulation, while preserving high-frequency details as much as possible.

%% file: problem.tex
\section{Problem Statement}
In this section, we formalize the problem of VLM-guided handle-based mesh deformation. The input to our method is a mesh $\mathcal{M}$ represented as a set of vertices $\mathcal{V}$ and triangles $\mathcal{F}$, denoted as $\mathcal{M}=\langle\mathcal{V},\mathcal{F}\rangle$. Additionally, we assume the user provides a text description $\mathcal{T}$ specifying the desired properties of the deformed mesh. The output of our method is a deformed mesh $\mathcal{M}'=\langle\mathcal{V}',\mathcal{F}\rangle$, featuring updated vertex positions while preserving the original topology.
To ensure the quality of the deformed mesh, we adopt a strategy similar to~\cite{kyptd}, where the deformation is guided by adjusting the positions of a small set of keypoints, known as handles $\mathcal{H}\subset\mathcal{V}$. The full shape $\mathcal{M}'$ is then determined from these handles using analytical deformation models, such as those proposed in~\cite{yifan2020neural, jacobson2011bounded}. However, unlike previous approaches, our method is designed for open-vocabulary mesh deformation, meaning it can handle meshes from any category without requiring retraining.
In contrast,~\cite{kyptd} necessitates training separate keypoint predictors for each object category. To overcome this limitation, we propose a novel framework that leverages a VLM to detect keypoints and perform mesh deformations in a few-shot manner, enabling generalization across diverse mesh categories.

%% file: method.tex
\section{Method}
To compute the desired deformation in $\mathcal{M}'$, we introduce a novel framework that leverages the strengths of non-learning-based geometric analysis algorithms~\cite{jacobson2011bounded,jacobson2013robust,li2023efficient} and the 3D understanding capabilities of VLM~\cite{hong20233d,yang2024llm,li20243dmit}. The key advantage of our approach lies in its open-vocabulary nature. Indeed, non-learning-based algorithms, which do not require training on specific 3D datasets, are immune to data scarcity and can handle meshes of arbitrary object categories. In contrast, the VLM, trained on vast text corpora, is capable of open-vocabulary comprehension.

Our framework consists of four stages as illustrated in~\prettyref{fig:pipeline}, and we refer readers to our supplementary materials for all the detailed VLM prompts. In the first stage, the text instruction $\mathcal{T}$ is decomposed into a set of $N$ actionable deformation sub-instructions ${\mathcal{I}_1, \cdots, \mathcal{I}_N}$, each of which can be achieved by manipulating a single handle to deform a sub-part of the mesh. These sub-instructions are then recursively realized and applied to the mesh through the remaining three stages.
During the second stage, we employ a novel VLM-guided inside-outside segmentation algorithm, inspired by~\cite{jacobson2013robust}, to identify the mesh sub-part associated with each sub-instruction. The third stage involves detecting potential handles using cone singularity detection~\cite{li2023efficient}, with the VLM tasked with deforming a single handle from each view. Finally, the fourth stage applies handle deformation to the entire mesh using an optimization-based deformation model~\cite{jacobson2011bounded,sifakis2012fem,liu2017quasi} and combines the multi-view deformations through multi-view voting. All three stages are performed for each sub-instruction.
We denote the mesh deformed using the first $n$ instructions as $\mathcal{M}_n'$, which serves as the reference mesh when applying the subsequent instruction $\mathcal{I}_{n+1}$. Ultimately, the fully deformed mesh is represented as $\mathcal{M}'=\mathcal{M}_N'$. We provide details of each stage in the following subsections.

\subsection{Sub-instruction Extraction}
\setlength{\columnsep}{10pt}
\begin{figure}
\centering
\includegraphics[width=\linewidth]{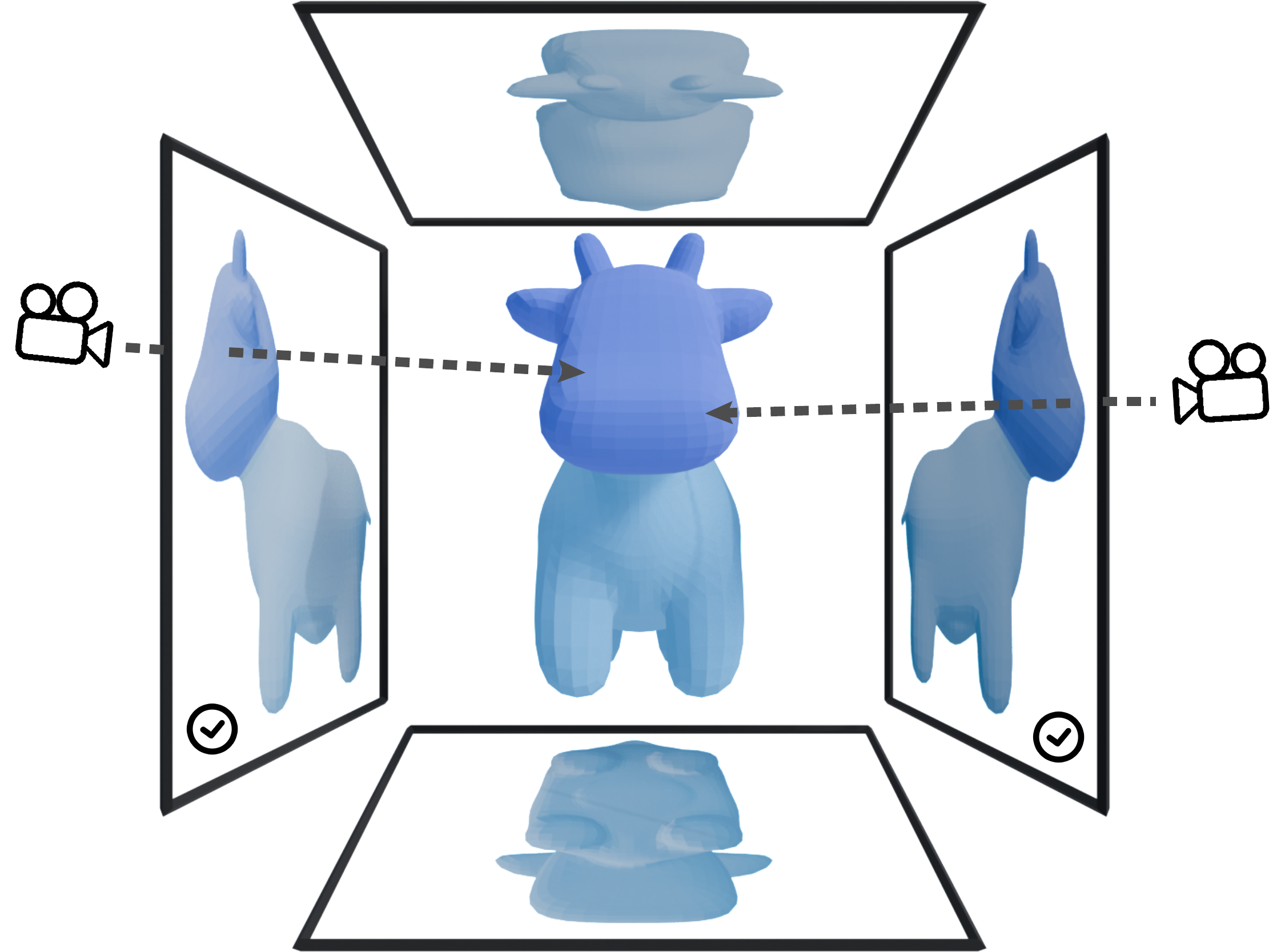}
\caption{\label{fig:graph-cut}In the example of~\prettyref{fig:teaser} (a), we use $\bar{M}=6$ camera views (4 are shown in this figure). The VLM selects two views (left and right ticks), and we use Grounding-DINO and SAM to derive the pixel-wise mask (red). The two views are then merged to yield a per-face label $l(f)$ on the mesh using graph cut.}
\end{figure}
During our first step, we need to extract the set of sub-instructions ${\mathcal{I}_1, \cdots, \mathcal{I}_N}$. Such decomposition is an essential step to reduce the workload of VLM. Indeed, as noticed by~\cite{qi2025shapellm}, VLM can be rather inaccurate in predicting precise 3D information, such as coordinate values. Therefore, we propose to reduce the workload of VLM by first segmenting the mesh into a fixed part and a deformable part, and then task it with dragging only a few handles (typically 1-2 handle points) on the deformable part. As such, VLM needs to determine mesh segmentation prior to each deformation. Our sub-instruction decomposition ensures that each sub-instruction is only related to a single deformable part, leading to much more accurate segmentation result and eased handle selection. Taking the model cow in~\prettyref{fig:teaser} (c) for example, our text prompt is ``raising its head with elongated horns'', which can be accomplished by first ($\mathcal{I}_1=$)``raise head'' and then ($\mathcal{I}_2=$)``elongate horns'', where $\mathcal{I}_1$ is only related to the heads and $\mathcal{I}_2$ is related to the horns. In practice, we extract these sub-instructions by providing the VLM with a couple of text-based examples. These sub-instructions can be inferred from the text prompt without even looking at the actual 3D model.

\subsection{Mesh Sub-part Identification}
For each sub-instruction $\mathcal{I}_n$, our next task is to segment the deformable sub-part of the mesh. As a major challenge to this end, existing open-domain segmentation models, such as Grounding DINO~\cite{liu2023grounding} and Segment Anything Model (SAM)~\cite{kirillov2023segment} only works in 2D image space, while our handle-based deformation method requires a mesh segmentation. To bridge the gap, we borrow the idea of graph-cut-based inside-outside segmentation~\cite{jacobson2011bounded}, which uses a graph-cut algorithm to revise a rough estimation of inside-outside labels on the mesh surface. As illustrated in~\prettyref{fig:graph-cut}, we propose to use multi-view segmentation label as our rough estimation. Specifically, we maintain a set of $M$ rendered images from different camera view angles, denoted as $\mathcal{C}=\{C_1,\cdots,C_M\}$. From each image $C_m$, we invoke Grounding-DINO to detect the bounding box of the deformable part, and then use SAM to extract a pixel-wise mask. Next, for all the mesh faces $f\in\mathcal{F}$, we solve a multi-view-guided graph-cut formulated as the following mixed-integer programming:
\begin{equation}
\begin{aligned}
\label{eq:graph-cut}
\underset{l(f)\in\{0,1\}}{\argmin}
&\sum_{f\in\mathcal{F}}\sum_{m=1}^M\mathbb{I}_\text{mask}(f,C_m)[1-l(f)] + \\
&\sum_{f\in\mathcal{F}}\sum_{m=1}^M\mathbb{I}_\text{no-mask}(f,C_m)l(f) + \\
&\sum_{f\cap f'\neq\emptyset}|l(f)-l(f')|w_{f,f'},
\end{aligned}
\end{equation}
which is solvable in polynomial time~\cite{Golovinskiy:2008:RCF}. Specifically, we optimize the binary label $l(f)$ for each face $f\in\mathcal{F}$ with $l(f)=1$, meaning that $f$ belongs to the deformable sub-part. If the face $f$ is visible from $C_m$ and SAM predicts that $f$ intersects the masked region, as indicated by $\mathbb{I}_\text{mask}(f,C_m)$, then our first term above encourages that $l(f)=1$ and introduces a cost otherwise. Similarly, if the face $f$ is visible from $C_m$ and SAM predicts that $f$ does not intersect the masked region, as indicated by $\mathbb{I}_\text{no-mask}(f,C_m)$, then our second term above encourages that $l(f)=0$. Note that the first two terms are only related to faces visible to the given camera view, and we leave all the other invisible face labels to be determined by the optimizer. Finally, our third term is a smoothness term that encourages neighboring faces to take on similar values. We weight these terms using a coefficient $w_{f,f'}$ that is proportional to the dihedral angle $\theta_{f,f'}$ between the two faces. In this way, the mesh tends to use the same labels for relatively flat areas, while label boundaries tend to appear in sharp features, as suggested by the implementation in CGAL~\cite{alliez2016cgal}. Our last step converts the face $l(f)$ to the vertex labels $l(v)$ for $v\in\mathcal{V}$ by labeling all vertices in a deformable face to be deformable as well.

The remaining task is to find a set of usable camera views, where label boundaries are clearly visible. Also, we need to provide a text prompt to Grounding-DINO and SAM, notifying the deformable part related to the sub-instruction $\mathcal{I}_n$. To this end, we propose to start from a super-set $\bar{\mathcal{C}}=\{C_1,\cdots,C_{\bar{M}}\}$ of $\bar{M}$ images. In practice, we find that only using the 6 axis-aligned view angles suffice, i.e., $\bar{M}=6$, $C_1$ rendered from $+X$, $C_2$ from $-X$, etc. More view angles could lead to better results at the cost of more images for VLM to digest. Following prior works in text-to-3D~\cite{chen2024sculpt3d} and texture synthesis~\cite{chen2023text2tex}, which find 4–12 views sufficient to cover an entire mesh, we adopt six views—enough for our simpler setting, where only a few handles need to be observed. Given the super-set, we engineer our prompt to query VLM for the mesh sub-part name related to $\mathcal{I}_n$. For example, for $\mathcal{I}_2=$``elongate horns'' in the example of~\prettyref{fig:teaser} (a), the related part is the ``horn''. We further ask the VLM to choose a subset $\mathcal{C}\subset\bar{\mathcal{C}}$ of $M\leq\bar{M}$ images from which the deformable sub-part is clearly visible. 

\subsection{Handle Detection and Selection}
With the deformable sub-part segmented in the previous section, the VLM could deform the mesh by moving any vertex belonging the sub-part. However,~\cite{kyptd} has demonstrated that the selection of keypoint is crucial to the quality of such deformation. Therefore, they choose to train a neural model to predict useful keypoints. Regretfully, the model used by~\cite{kyptd} is re-trained for each 3D model category and cannot scale across multiple domains, which significantly compromise the benefits of VLM. During the early stage of this research, we have also tried to have the VLM mark the key points' locations on the image. However, we found that VLM can struggle to select points on meshes without textures, especially for meshes with large, nearly flat areas.

\begin{figure}[ht]
\centering
\includegraphics[width=\linewidth]{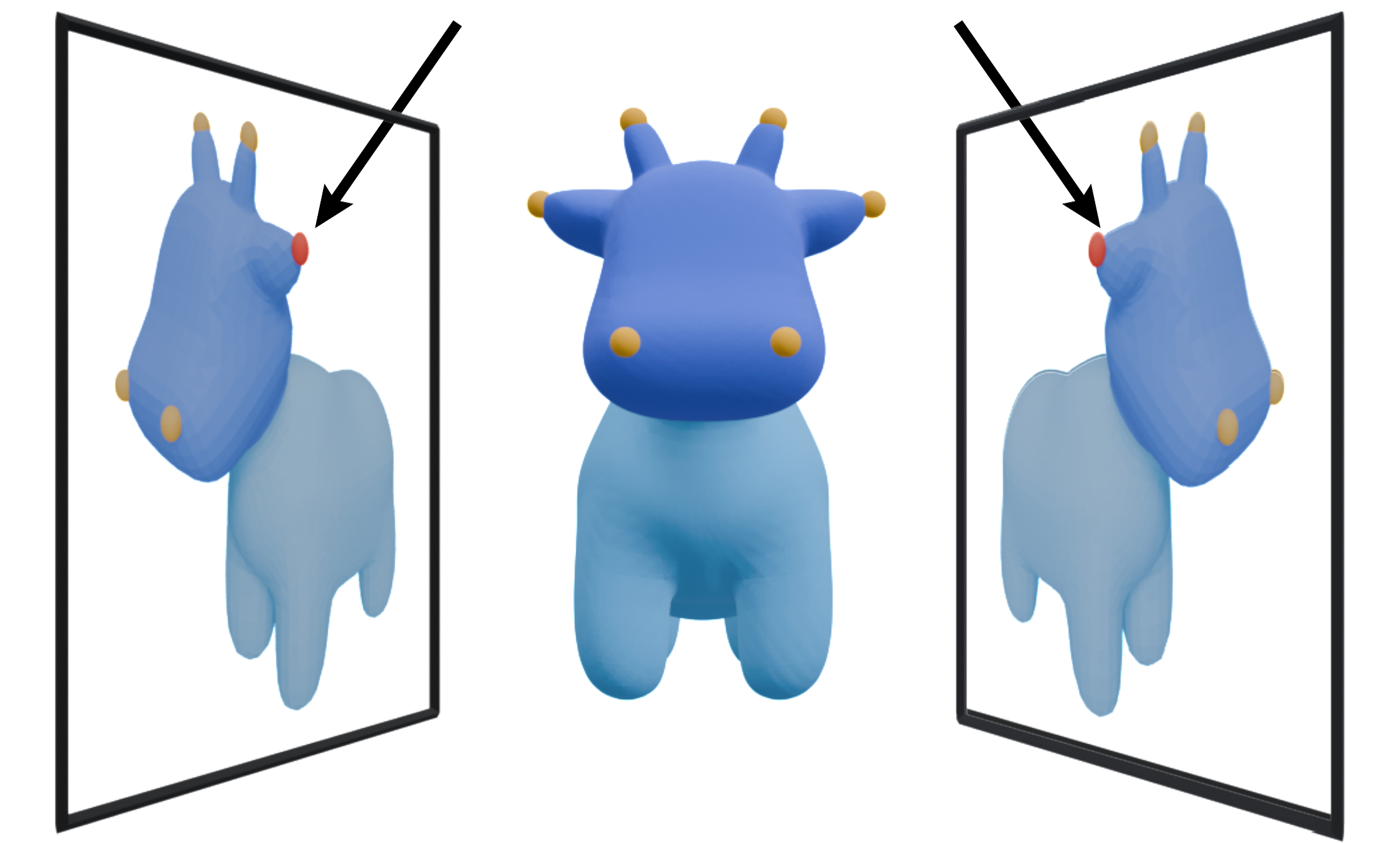}
\caption{\label{fig:handle-selection} For the two camera views selected during the sub-part identification stage, we show VLM a super-set of handles $\bar{\mathcal{H}}$ (drawn in yellow) and ask it to select a desired set of at least one handles $\mathcal{H}$ (arrow indicated red points).}
\end{figure}
To mitigate this problem, we observe that keypoints are typically placed at certain feature points, e.g., the tip of fingers, the foot of a human, and the tip of wings of an airplane. Fortunately, these feature points have distinguishing geometric properties of being the concentration points of Gauss curvature. In view of this, we propose to borrow the fast cone singularity detection algorithm~\cite{li2023efficient} to identify a small super-set of potential handles $\bar{\mathcal{H}}\subset\mathcal{V}$. This formulation minimizes the distortion of conformal mapping as predicted by the Yamabe equation, and we refer readers to~\cite{li2023efficient} for more details. The number of cone singularity keypoints is regulated by the distortion bound, with a lower bound generating more singularities. If the distortion bound is set to zero, every mesh vertex theoretically becomes a cone singularity, as the condition for smooth parameterization becomes maximally restrictive~\cite{li2023efficient, fang2021computing}. To ensure viable keypoints for downstream deformation, our algorithm adaptively reduces the distortion bound if no singularities initially appear on segmented mesh regions. With the handle super-set detected, we then engineer the VLM to select a subset $\mathcal{H}\subset\bar{\mathcal{H}}\subset\mathcal{V}$ of one or more handles to drag for each select camera view in $\mathcal{C}$. Specifically, we draw all the detected handles $\bar{\mathcal{H}}$ as small yellow dots in each image $C_m\in\mathcal{C}$. We then present these images to VLM and task it with selecting a set of handle coordinates that are useful to realize the sub-instruction $\mathcal{I}_n$. The final handles $\mathcal{H}$ are selected to be the closest vertices to the selected coordinates in $\bar{\mathcal{H}}$ in the image space, as illustrated in~\prettyref{fig:handle-selection}. 

\input{table.tex}

\subsection{Mesh Deformation via Multi-view Voting}
With the desired handles $\mathcal{H}$ identified, we then engineer VLM to predict the desirable target handle position in the image space. Our goal in this last stage is to produce a mesh deformation that realizes sub-instruction $\mathcal{I}_n$ to deform $\mathcal{M}_{n-1}'=\langle \mathcal{V}_{n-1}',\mathcal{F} \rangle$ into $\mathcal{M}_n'=\langle \mathcal{V}_n',\mathcal{F} \rangle$. We face two major challenges during this stage. First, since the target handle position is predicted in the 2D image space, the information is under-determined and we have to recover the 3D coordinates. For example, the perspective camera distortion can lead to undesirable distortion in the deformed mesh $\mathcal{M}_n'$. Second, the predicted 2D position can be biased towards the selected camera view since the VLM is unaware of the full 3D geometry. 

To deal with the first issue, we propose a distortion regularized optimization. Specifically, let us denote $\mathcal{H}(C_m)$ as the set of handles selected under camera view $C_m$ and the target position for some $v\in\mathcal{H}(C_m)$ be denoted as $v^\star$ in 2D image space. We optimize for the target positions for all the handle points in the super-set $\bar{\mathcal{H}}(C_m)$, regularized by the finite-element-discretized membrane energy~\cite{sifakis2012fem}. Specifically, our optimization formulation is written as follows:
\begin{equation}
\begin{aligned}
\label{eq:deform-opt}
\underset{v\in\bar{\mathcal{H}}}{\argmin}&
\sum_{v\in\mathcal{H}\subset\mathcal{V}_n'}\|P_m v-v^\star\|^2+\\
&\sum_{\underset{v_1,v_2,v_3\in\mathcal{V}_n'}{\langle v_1,v_2,v_3 \rangle\in\mathcal{F}}} \lambda E_\text{membrane}(v_1,v_2,v_3,\mathcal{M}_{n-1}').
\end{aligned}
\end{equation}
Here the first objective term penalizes the distance between the deformed handle position $v$ and the VLM-predicted target handle position $v^\star$ in the image space, with $P_m\in\mathbb{R}^{2\times3}$ being the 3D-to-2D camera projection matrix associated with $C_m$. The second objective term is the membrane energy using $\mathcal{M}_{n-1}'$ as the reference mesh with $\lambda=0.01$ being the regularization coefficient. Note that under the biharmonic deformation framework~\cite{jacobson2011bounded}, any vertex $v\in\mathcal{V}_n'$ is a linear function of all $v\in\bar{\mathcal{H}}$. Therefore, the only nonlinear term in~\prettyref{eq:deform-opt} is the membrane energy $E_\text{membrane}$. Fortunately, the number of handles is very small and~\prettyref{eq:deform-opt} can be solved via a couple of iterations of Newton's method.

\begin{figure}[ht]
\centering
\includegraphics[width=\linewidth]{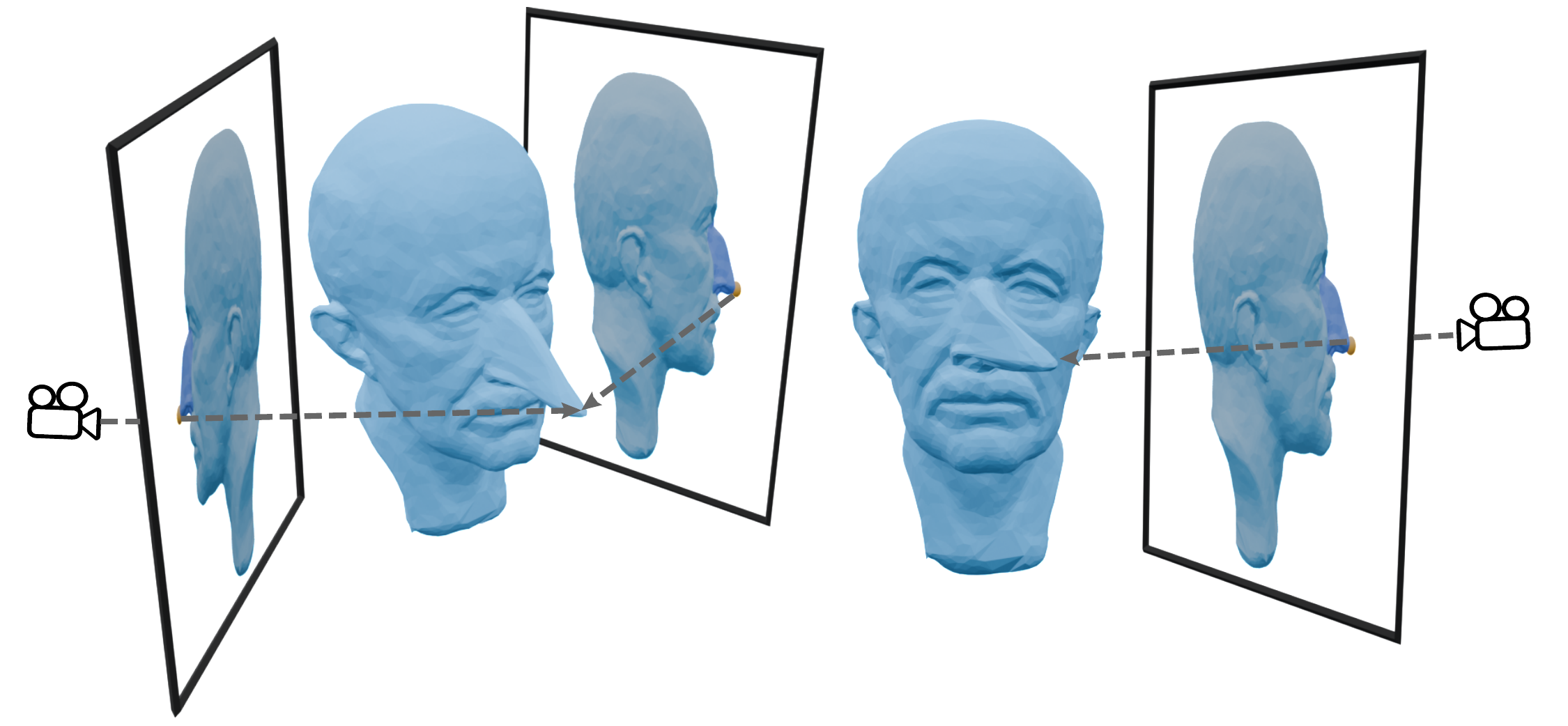}
\caption{\label{fig:voting}When guided by two views, the noise is straightly elongated (left). But if only one view is used, the noise bends to the right (right).}
\end{figure}
To deal with the second issue, we propose to blend the results over multiple views. Specifically, we add additional subscript and denote the result of~\prettyref{eq:deform-opt} under camera view $C_m$ as $v_m\in\bar{\mathcal{H}}_m$. The final handle positions are then derived by $v=\sum_{m=1}^M v_m/M$, and the other vertex positions are derived via biharmonic deformation. BBW inherently preserve the sharp features of meshes. This operator is essentially doing voting over the VLM-predicted deformations over multiple views. As illustrated in~\prettyref{fig:voting}, we find this technique very effective in dealing with the bias caused by 2D-to-3D transformation. Again, one could use more camera views to achieve better results at the higher VLM inference cost, while we find that $\bar{M}=6$ is sufficient for all results in our evaluation. To mitigate the VLM hallucination in predicting the handle position, we self-collect a dataset to fine-tune the VLM. Particularly, for each data, we have an image with handles drawn in red dots, a deformation prompt, a list of handles as input, and reasoning, selected handles, handle directions as output. As noted in prior work~\cite{sun2024layoutvlm}, VLMs may struggle with zero-shot accurate 2D image coordinates prediction. To improve spatial reasoning, we additionally prompt the VLM to output the handle movement direction (e.g., “Up,” “Right”) alongside the final position. More details can be found in supplementary material. 

%% file: table.tex
\begin{table*}[t]
\centering
\footnotesize
\caption{\label{table:ablation}We conduct ablation study with two variants of our method, one without GPT handle selection (i.e. we randomly pick a handle from $\bar{\mathcal{H}}$), and the other one without multi-view voting (i.e. using a single view). We show that both strategies are necessary in improving the quality of results in terms of CLIP and GPTEval3D scores.}
\vskip 0in
\begin{tabular}{lcccccc}
\toprule
\multirow{2}{*}{\textbf{Method}}&\multicolumn{2}{c}{\textbf{CLIP Score} $\uparrow$} &\multicolumn{4}{c}{\textbf{GPTEval3D Score} $\uparrow$} \\

\textbf{} & \textbf{\textit{ViT-B/32}} & \textbf{\textit{ViT-L/14}}& \textbf{\textit{Prompt Asset Alignment}}& \textbf{\textit{3D Plausibility}}& \textbf{\textit{Geometry Details}}& \textbf{\textit{Overall}}  \\
\midrule
Without GPT Handle Selection &0.2299& 0.1884& 1022.9122&913.3979& 924.0328& 928.8978
\\
Single View & 0.2314&0.1942& 940.8532&969.1528&985.2911& 976.4946
\\
\textbf{Ours}& \textbf{0.2524}& \textbf{0.2033}&\textbf{1205.1430}& \textbf{1029.1463}& \textbf{1091.5224} &\textbf{1147.5173}
\\
\bottomrule
\end{tabular}
\vskip -0in
\end{table*}
\begin{table}[t]
\centering
\footnotesize
\caption{\label{table:finetune} Performance between original VLM and fine-tune VLM}
\vskip 0in
\begin{tabular}{lccc}
\toprule
\textbf{Method} & Winning Rate$\uparrow$&CLIP$\uparrow$&Overall GPTEval3D$\uparrow$\\
\midrule
GPT-4o & 77.14\%&0.2408&1107.861\\
Fine-tune GPT-4o & \textbf{88.57\%} & \textbf{0.2524}&\textbf{1147.5173} \\
\bottomrule
\end{tabular}
\vskip -0.1in
\end{table}
\begin{table*}[t]
\centering
\footnotesize
\caption{\label{table:comparison}When compared with the results of TextDeformer, MagicClay, and MeshUp, our method generates slightly lower CLIP ViT-B/32 score, which is unsurprising because TextDeformer uses CLIP ViT-B/32 score as a loss function. However, our method generates consistent higher GPTEval3D score and enjoys a lower inference cost.}
\vskip -0.1in
\begin{tabular}{lccccccc}
\toprule
\multirow{2}{*}{\textbf{Method}}&\multirow{2}{*}{\textbf{Run Time} $\downarrow$} &\multicolumn{2}{c}{\textbf{CLIP Score} $\uparrow$} &\multicolumn{4}{c}{\textbf{GPTEval3D Score} $\uparrow$} \\

\textbf{} & \textbf{} & \textbf{\textit{ViT-B/32}}& \textbf{\textit{ViT-L/14}}& \textbf{\textit{Prompt Asset Alignment}}& \textbf{\textit{3D Plausibility}}& \textbf{\textit{Geometry Details}}& \textbf{\textit{Overall}}  \\
\midrule
Origin Mesh & -&0.2277&0.1949& 1000.0000&1000.0000& 1000.0000& 1000.0000
\\
TextDeformer~\cite{gao2023textdeformer} &15 mins& \textbf{0.2543}&0.1874& 981.0867&960.3860&933.6462& 952.6972
\\
MeshUp~\cite{kim2024meshup} & 40 mins &0.2495&0.1981&1037.0134&990.4378&963.2348&988.4987
\\
MagicClay~\cite{Barda24} &74 mins&0.2373 &0.1962 &1013.9473 &984.9727 &1029.0764 & 1016.4348
\\
\textbf{Ours}&\textbf{8 mins} & 0.2524& \textbf{0.2033}&\textbf{1205.1430}& \textbf{1029.1463}& \textbf{1091.5224} &\textbf{1147.5173}
\\
\bottomrule
\end{tabular}
\vskip -0.1in
\end{table*}
\begin{table}[t]
\centering
\footnotesize
\caption{\label{table:membrane} We compare the average/variance of mesh distortion induced by our method and two baselines. We don't compare with MagicClay because it generates shape parts rather than deformations, and $E_\text{membrane}$ is undefined for different discretizations.}
\vskip -0.1in
\begin{tabular}{lc}
\toprule
\textbf{Method} & {$\mathbf{E}_\mathbf{membrane}$} (avg./var.)  $\downarrow$\\
\midrule
TextDeformer~\cite{gao2023textdeformer} & 349.28/108.54\\
MeshUp~\cite{kim2024meshup} & 227.60/74.66\\
\textbf{Ours} & \textbf{99.92/44.78}  \\
\bottomrule
\end{tabular}
\vskip -0in
\end{table}
\begin{table}[t]
\centering
\footnotesize
\caption{\label{table:time} We report the approximate run time for each step of our pipeline: VLM-guided camera view selection, graph-cut optimization, handle detection \& selection, and mesh deformation.}
\vskip 0in
\begin{tabular}{lcccccc}
\toprule
\textbf{} & \TWORCell{View}{Selection} & \TWORCell{Graph}{Cut} & \TWORCell{Select}{$\bar{\mathcal{H}}$ \& $\mathcal{H}$} & \TWORCell{Mesh}{Deform}\\
\midrule
Time (min) & 1.5 & 3.5 & 3 & 0.2\\
\bottomrule
\end{tabular}
\vskip -0.2in
\end{table}
\begin{table*}[t]
\centering
\footnotesize
\renewcommand{\arraystretch}{1.1}
\setlength{\tabcolsep}{5pt}
\caption{Evaluation results for semantic similar prompts for the same mesh.}
\label{table:prompt_results}
\begin{tabular}{lcc|cccc}
\toprule
\multirow{2}{*}{\textbf{Prompt}} & \multicolumn{2}{c}{\textbf{CLIP Score} $\uparrow$} & \multicolumn{4}{c}{\textbf{GPTEval3D Score} $\uparrow$} \\
\cmidrule(lr){2-3} \cmidrule(lr){4-7}
& \textbf{ViT-B/32} & \textbf{ViT-L/14} & \textbf{Alignment} & \textbf{Plausibility} & \textbf{Details} & \textbf{Overall} \\
\midrule
\multicolumn{7}{l}{\textbf{Human}} \\
\textbf{A}: ``Bring both arms down to a lower position'' & 0.2533 & 0.2469 & 1168 & 1015 & 1045 & 1103 \\
\textbf{B}: ``Shifted to a calmer pose, with both sides eased downward'' & 0.2565 & 0.2498 & 1174 & 1010 & 1049 & 1100 \\
\textbf{C}: ``Relaxed posture with sides drawn lower'' & 0.2546 & 0.2485 & 1176 & 1008 & 1052 & 1095 \\
\midrule
\multicolumn{7}{l}{\textbf{Chair}} \\
\textbf{A}: ``A tall barchair'' & 0.2671 & 0.2394 & 1342 & 1063 & 1040 & 1257 \\
\textbf{B}: ``A chair raised for extra height, perfect for high counters'' & 0.2685 & 0.2279 & 1338 & 1067 & 1038 & 1233 \\
\textbf{C}: ``A chair designed to bring you closer to the sky'' & 0.2552 & 0.2401 & 1295 & 1060 & 1042 & 1184 \\
\midrule
\multicolumn{7}{l}{\textbf{Airplane}} \\
\textbf{A}: ``Pull the airplane wings backwards'' & 0.2163 & 0.1808 & 1237 & 1008 & 1038 & 1216 \\
\textbf{B}: ``Shape it into a plane that can fly faster'' & 0.2145 & 0.1821 & 1242 & 1012 & 1038 & 1211 \\
\textbf{C}: ``Deform to a jet with a swept-back design'' & 0.2130 & 0.1835 & 1245 & 1015 & 1026 & 1209 \\
\bottomrule
\end{tabular}
\end{table*}

%% file: evaluation.tex
\section{Evaluation}
We utilze a dataset from~\cite{dataset}, originally from Thingi10K~\cite{Thingi10K}, and focus on the 1,171 organic models due to their semantically meaningful parts aligned with natural language prompts. To reduce VLM cost, we filter out geometrically similar meshes, resulting in 150 diverse, semantic-rich shapes. About 33\% of our original models are asymmetric. For user instructions requiring symmetric deformation,~70\% of our results are correctly symmetric thanks to our axis-aligned cameras. We use GPT-4o as our VLM and will release the curated dataset upon acceptance.
%
%
All experiments were run on Apple M1 CPU and Nvidia A100 GPU. To profile computational times, we measured the cost of both CPU and GPU computations, including networking overhead. Under this configuration, generating a single result takes an average runtime of 8 minutes. We only make~10 VLM-API calls on average per example. We also clarify the limitation that our method relies on stable networks. Additional results can be found in supplementary material.

\paragraph{Metrics}
We evaluate our method using two metrics: the CLIP score~\cite{radford2021learning} and the GPTEval3D score~\cite{wu2023gpteval3d}. CLIP is a joint embedding model for text and images. To leverage this model, we embed the user input text $\mathcal{T}$ and rendered images from $30$ evenly sampled camera views, generated using the Fibonacci sphere algorithm. We compute the cosine similarity between the text and image embeddings, averaging the scores across all views to obtain the final CLIP score.
We evaluate performance using two different CLIP models: ViT-B/32 and ViT-L/14.
GPTEval3D, on the other hand, leverages GPT-4V to evaluate text-to-3D models across several human-aligned criteria and can handle complex text prompts. Specifically, we assess performance based on four criteria: Prompt Asset Alignment, 3D Plausibility, Geometry Details, and Overall.

\paragraph{Ablation Study}
 In the initial study, we replace the VLM-guided handle selection mechanism with a random selection scheme. Specifically, instead of having the VLM choose the handle set $\mathcal{H}$, we randomly select a handle $v_\text{random} \in \bar{\mathcal{H}}$ to form $\mathcal{H}$. All other factors are kept constant.
In our standard algorithm, a handle position $v\in\mathcal{H}$ is moved from $P_m v$ to $v^\star$ in image space, where $P_m$ is the 3D-to-2D camera projection matrix associated with $C_m$. To ensure consistency, we apply the same transformation to the randomly selected handle by setting $v_\text{random}^\star=P_mv_\text{random}+v^\star-P_mv$.
In the second study, we bypass the multi-view voting process and instead use a single, randomly chosen view for deformation. 
The performance of our method, compared to these two variants, is summarized in~\prettyref{table:ablation}. The results show that our approach consistently achieves higher CLIP and GPTEval3D scores, highlighting the necessity of both the VLM-guided handle selection and multi-view voting.

We also study the benefits of VLM fine-tuning in~\prettyref{table:finetune}. We hire artists to create groundtruth handle deformations for a small dataset of shapes, on which we fine-tune the VLM. We make use of the default supervised fine-tuning parameters and finetune the GPT-4o through SFT API function provided by OpenAI online. We additionally calculate the winning rate in displacing parts that need to be deformed in the correct direction. We use annotated directions in the test set of our fine-tuning dataset as the ground-truth direction. The results show that the fine-tuned model outperforms in all metrics, demonstrating greater robustness against LLM hallucinations. However, the standard VLM without fine-tuning already generates satisfactory results on all examples.

\begin{figure}[ht]
\centering
\includegraphics[width=\linewidth]{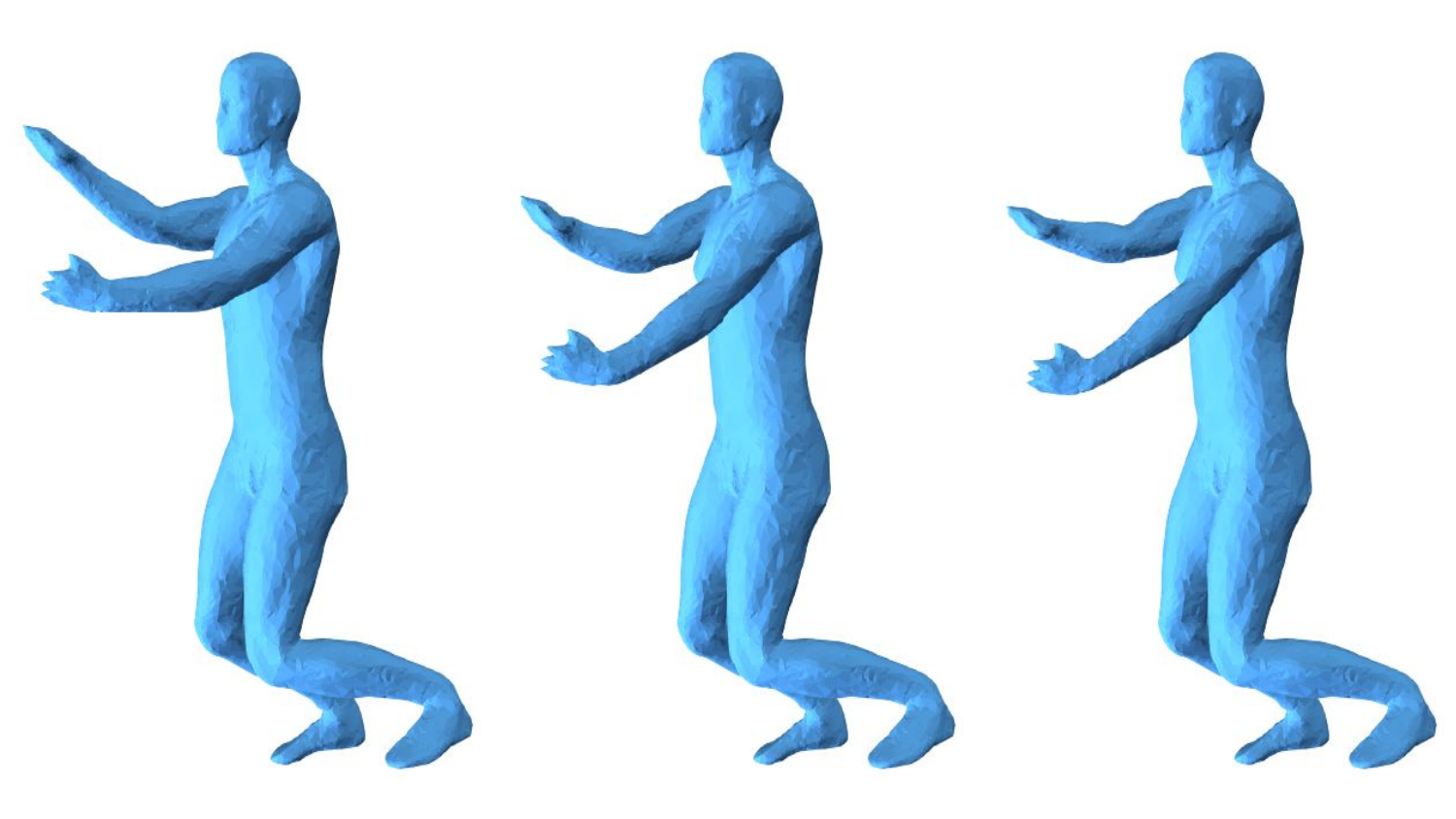}
\caption{\label{fig:arap} Results of using ARAP deformation and biharmonic deformation. Text: Bring both arms down to a lower position. Left: Origin, Middle: ARAP, Right: Biharmonic}
\vskip -0.1in
\end{figure}

We further perform the qualitative experiment in~\prettyref{fig:arap} where we use both ARAP and biharmonic deformation models. We confirm that using ARAP leads to very similar results as our adopted biharmonic method, and thus, our method can be generalized to various handle-based deformations. This versatility also allows us to address known limitations of BBW, such as surface ripples~\cite{jacobson2011bounded}, by extending our method to other deformation models.

\paragraph{Comparison with Baselines}

TextDeformer~\cite{gao2023textdeformer}, another training-free framework for generating 3D content via mesh deformation, shares significant similarities with our method and serves as our baseline. We notice another relevant work, FusionDeformer~\cite{xu2024fusiondeformer}, but its closed-source code prevents direct comparison. However, its results are likely similar to TextDeformer. We also note BlenderAlchemy~\cite{huang2024blenderalchemy}, which employs a VLM-based agentic framework for Blender-based graphic editing. However, it targets a different task and input setting, focusing on general graphic editing rather than deformation. Their setup requires user-specified shape keys for each mesh part, whereas our method supports a more intuitive interface—taking only a single text description and a watertight mesh as input. Thus, our approach is complementary rather than directly comparable. 

TextDeformer directly optimizes mesh vertices by minimizing the CLIP score as a loss function across multiple views. However, a major limitation is that CLIP lacks an understanding of the plausible 3D shape space, leading to noisy surface details during optimization. To address this, TextDeformer incorporates elastic energy regularization to constrain shape deformation. Unfortunately, as shown in~\prettyref{fig:compareTextDeformer}, this regularization still results in undesirable deformations, compromising 3D plausibility. In contrast, our method consistently produces more visually satisfactory results.
We present our quantitative results in~\prettyref{table:comparison}, comparing our deformed meshes to both the original mesh and the outputs of TextDeformer. Our method consistently outperforms the original mesh across all metrics, demonstrating our effectiveness. On the other hand, although TextDeformer achieves a higher CLIP ViT-B/32 score, which is expected, given its use of ViT-B/32 as a loss function, our method surpasses it when evaluated with ViT-L/14. Additionally, we outperform TextDeformer across all GPTEval3D metrics. 
These results underscore the sensitivity of the CLIP score to model size and highlight its limitations in reflecting 3D plausibility. As illustrated in~\prettyref{fig:compareTextDeformer}, the GPTEval3D score aligns more closely with human aesthetic judgments. 
Further, we note that our method enjoys a lower computational cost, with the performance breakdown summarized in~\prettyref{table:time}. This is because TextDeformer requires many evaluations of the loss function by running CLIP model inference, while we only requires a network-independent optimization (\prettyref{eq:deform-opt}) for mesh deformation that is very efficient to solve.
Finally, we compare the mesh distortion induced by both methods as measured by $E_\text{membrane}$. As summarized in~\prettyref{table:membrane}, we see that our method generates much smaller distortion on average. While the distortion variance of TextDeformer is also high, i.e. it can sometimes generate results corresponding to rather small distortion. This is because the optimizer in TextDeformer can sometimes get stuck in local minima, leading to very small overall deformation, which does not align well with $\mathcal{T}$.

We also compare with a part-based text-guided mesh editing method, MagicClay~\cite{Barda24}. MagicClay generates regions of a mesh according to textual prompts using a hybrid mesh-SDF representation. We use the segmented region as input to evaluate its ability to evolve the shape toward a text prompt. We further compare against an SDS-based baseline, MeshUp~\cite{kim2024meshup}, According to~\prettyref{fig:compareTextDeformer} \&~\prettyref{table:comparison}, our methods exhibit better results than those baselines both qualitatively and quantitatively.
We observe that our GPTEval3D scores for 3D plausibility and geometric detail are also higher than those of the original meshes. Since GPTEval3D evaluates all four metrics in a single batch using the user prompt, the scores are somewhat interdependent. As shown in~\prettyref{table:comparison}, the \textit{3D Plausibility} and \textit{Geometry Details} scores are relatively close between our method and the original, whereas the difference in Alignment is much larger, indicating that we produce meshes of comparable plausibility while achieving higher prompt alignment.

\paragraph{User Study}We conducted a user study with 21 participants representing diverse ages, genders, and educational backgrounds. Each participant was presented with 16 distinct mesh sets and asked to evaluate deformation results generated by three baseline methods and our approach. For each mesh set, participants answered two questions: (1) Which deformation best aligns with the given text prompt? and (2) Which deformation is less noisy and more visually salient? Results show that in 93.53\% of cases, participants rated our deformations as better aligned with the textual descriptions, and in 85.71\% of cases, they considered our outputs more visually salient. These findings demonstrate strong alignment between our method’s outputs and human perceptual judgments.

\paragraph{Prompt Sensitivity}
\begin{figure}[ht]
\centering
\includegraphics[width=\linewidth]{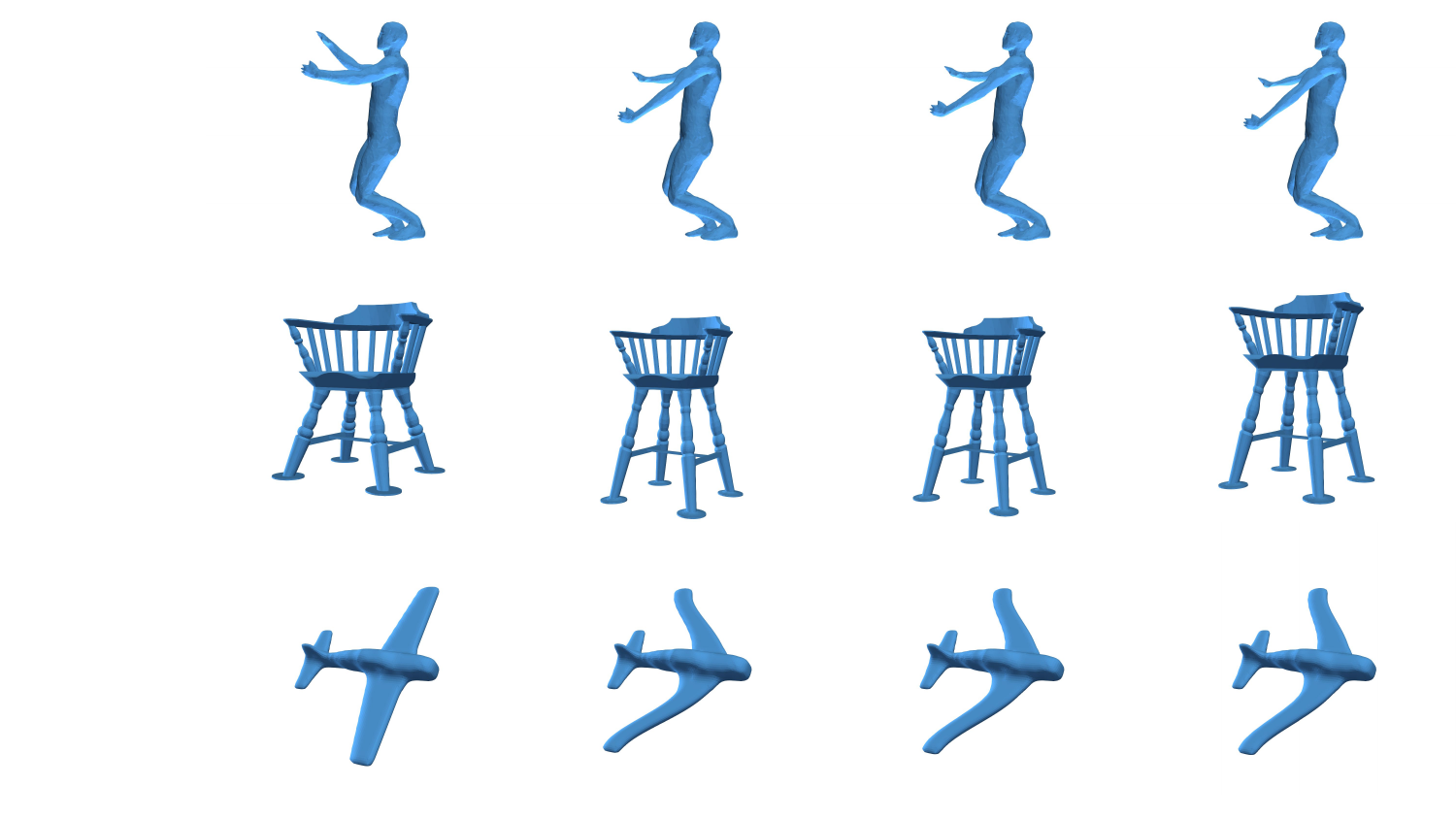}
\put(-250,116){Human}
\put(-250,76){Chair}
\put(-250,26){Airplane}
\put(-200,146){Origin}
\put(-135,146){A}
\put(-80,146){B}
\put(-27,146){C}
\caption{\label{fig:prompt_sens}Our method produces consistent mesh deformations for semantically similar prompts, demonstrating robustness to paraphrasing and wording variations.}
\vskip -0.1in
\end{figure}

We acknowledge the importance of prompt engineering for achieving robust results and provide the full prompt in the supplementary material. Nevertheless, we find that roughly 35\% of our deformation work without explicitly specifying target parts, such as the example ``Pinocchio telling lie'' in~\prettyref{fig:teaser}. We also show in~\prettyref{table:prompt_results} and~\prettyref{fig:prompt_sens} that our method is insensitive to prompts. The \textbf{Human}, \textbf{Chair}, and \textbf{Airplane} examples each include three semantically similar but differently worded prompts, and the resulting deformations remain highly consistent across both quantitative metrics and qualitative human evaluations. As VLM and prompt engineering are widely adopted to solve multiple graphic tasks~\cite{huang2024blenderalchemy, gu2025blendergym}, we further empirically demonstrated that it can also be robustly applied to handle-based mesh deformation tasks.

\paragraph{User Control}

\begin{figure}[ht]
\centering
\includegraphics[width=\linewidth]{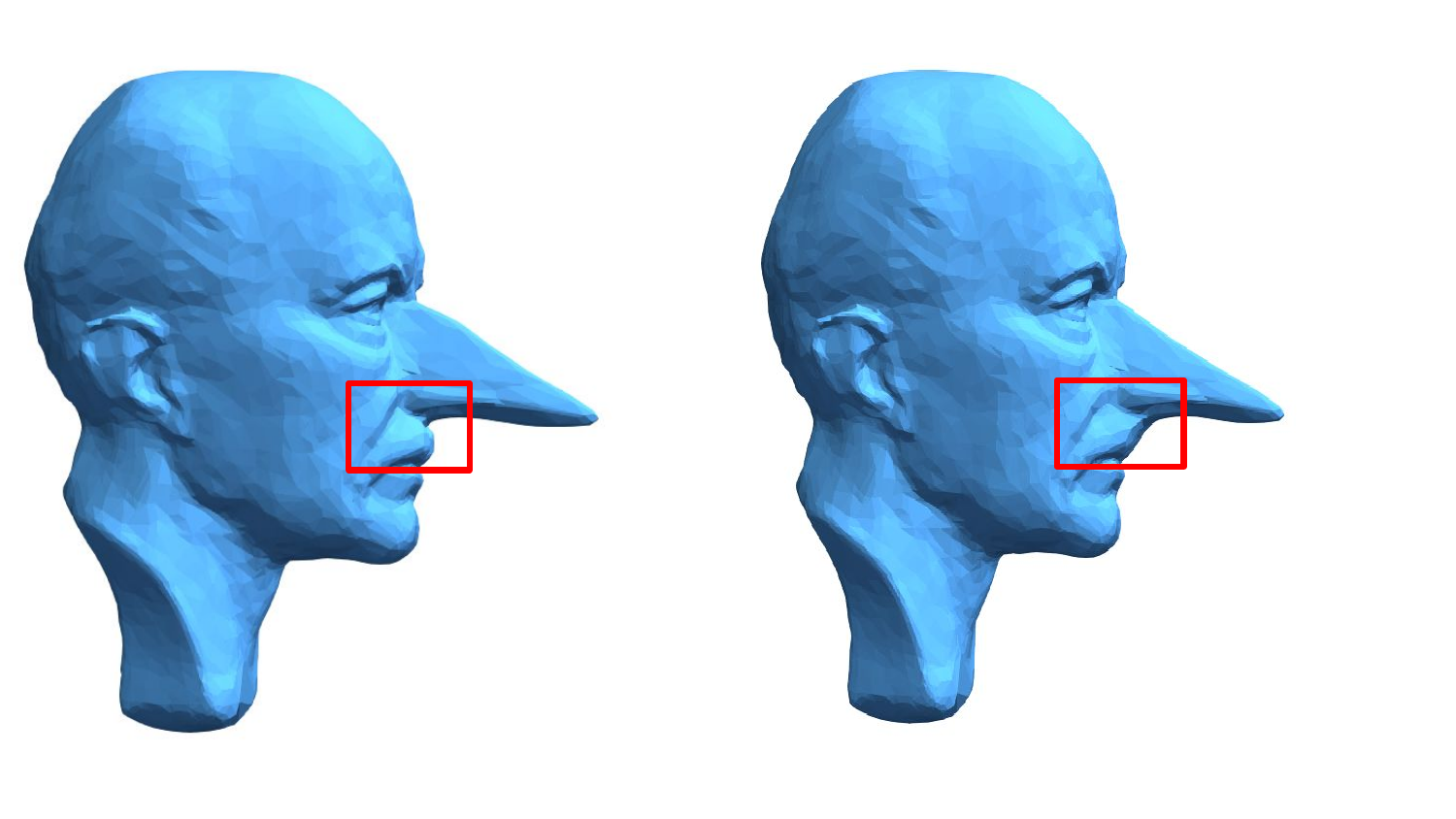}
\caption{\label{fig:user_area}We show user-guided sub-part identification. While certain regions of interest may be difficult to describe using natural language, artists may still wish to deform them in conjunction with other semantic parts (e.g., extending the red area along with the nose). To this end, we allow users to modify the 2D mask and we then solve~\prettyref{eq:graph-cut} to update our 3D segmentation. The right figure shows the result where the red-highlighted area is deformed coherently with the nose due to the user-updated segmentation.}
\vskip -0.1in
\end{figure}

\begin{figure}[ht]
\centering
\includegraphics[width=\linewidth]{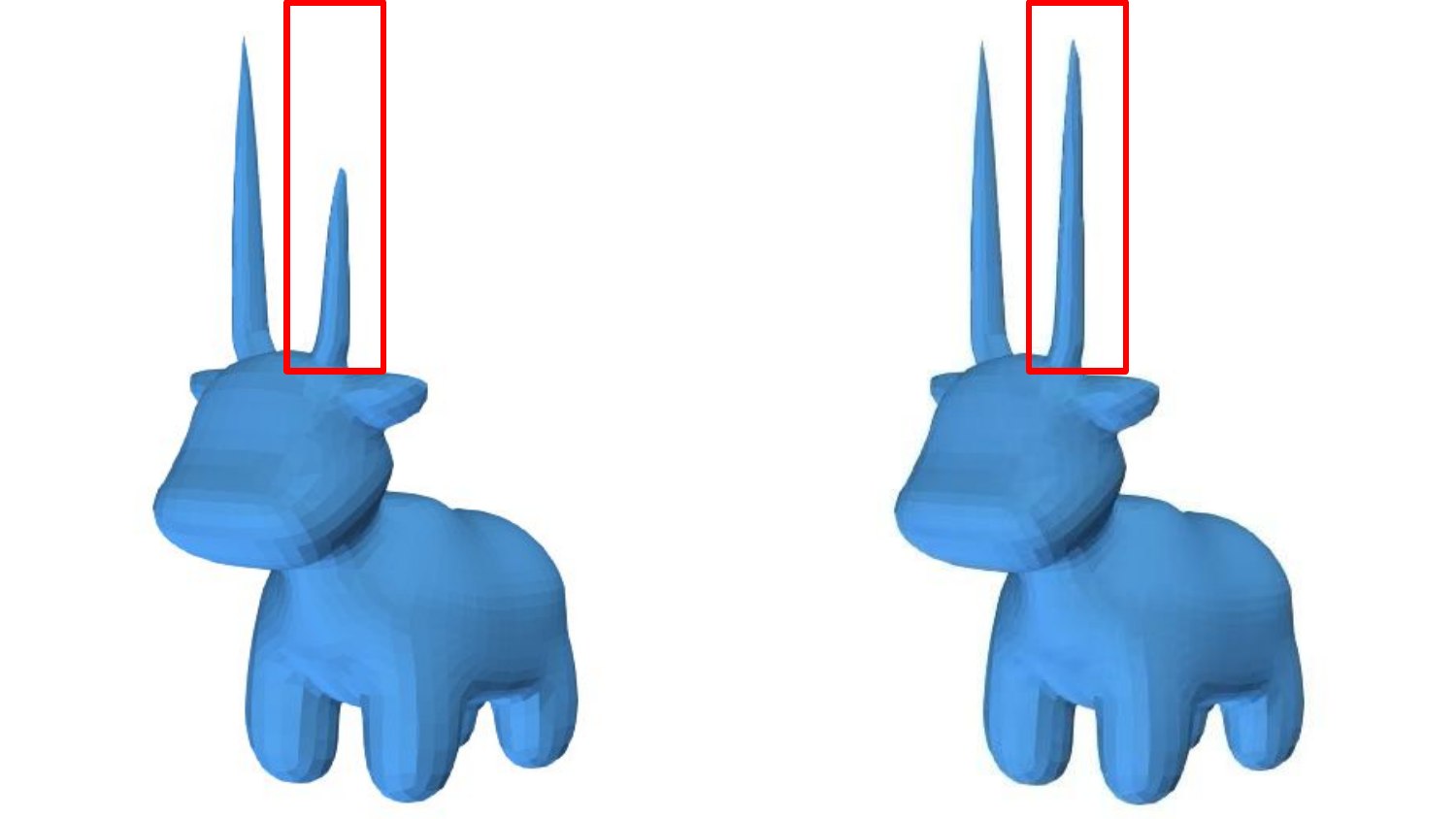}
\caption{\label{fig:user_deform}We show user-guided deformation magnitude control. Artists may require fine-grained control over deformation magnitude beyond vague descriptions like ``make it longer''. In this example, the initial prompt, ``elongate horns'', produces asymmetric results (left). To correct this, the user can manually drag the horns (right).}
\vskip -0.1in
\end{figure}

As a major benefit, our method generates a sequence of explainable intermediary outputs, including segmentations, handles, and handle translations. Therefore, users could intuitively modify the results by tuning any of these intermediary outputs and continue running the remaining steps of our pipline. In~\prettyref{fig:user_area} and~\prettyref{fig:user_deform}, we show two use cases, where user modify the segmentation mask and handle translations to fine-tune the results.

%% file: conclusion.tex
\section{Conclusion \& Limitation}
We propose a VLM-guided method for text-based 3D mesh deformation. Central to our approach is a handle-based deformation framework, which effectively regularizes the output mesh and ensures its quality. Our method carefully builds VLM into the pipeline of handle-based deformation. Specifically, VLM participates in the process of sub-part identification, handle identification, and mesh deformation. Through comprehensive evaluations, we demonstrate that our method consistently achieves high scores on CLIP and GPTEval3D metrics.

Despite its strengths, our method has some limitations. First, it is restricted to low-frequency modifications and cannot generate finer details, as achieved by methods like~\cite{chen2021decor}. Second, as our key design choice, we only focus on high-curvature, non-smooth regions. We clarify that our method may be flawed on all-smooth shapes, e.g. a sphere. Third, our approach does not utilize any shape priors. Deformations are driven solely by handle manipulations and energy-based regularizations, such as membrane energy. Lastly, our method may result in self-intersection or unsymmetric deformation due to the fact that VLM has no 3D shape prior knowledge and, therefore, cannot always perfectly give intersection-free and symmetric handle movement prediction. Such limitations can be fixed by using a barrier function and symmetric-aware geometric processing tools~\cite{golovinskiy2009symmetry}. 

%% file: supp.tex




{\fontfamily{phv}\selectfont
\flushleft\fontsize{9pt}{11pt} \selectfont \textbf {\scshape  \textbf{Supplementary Material}}
}

In this supplementary material, we provide the complete experimental details on segmentation, handle generation, deformation, and VLM prompts. We also provide a video of our 3D mesh results in \textit{Video.mp4} for reviewers to better visualize the qualitative results.

\subsection{Graph-Cut Segmentation Result}
In~\prettyref{table:segmentation_results}, we present the segmentation results obtained using our graph-cut-based 3D segmentation pipeline, demonstrating its effectiveness.
\begin{table}[!ht]
\centering
\vspace{-20px}
\resizebox{\linewidth}{!}{%
\begin{tabular}{p{0.3\linewidth}p{0.3\linewidth}p{0.3\linewidth}}
    \includegraphics[width=\linewidth]{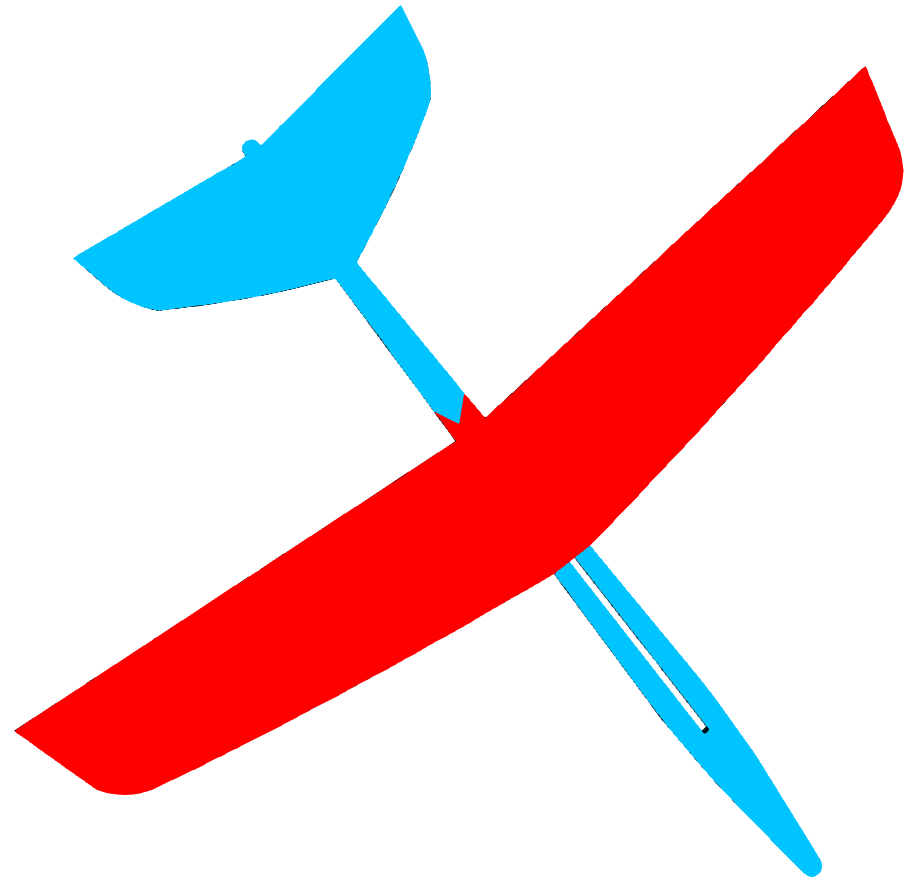} &
    \includegraphics[width=\linewidth]{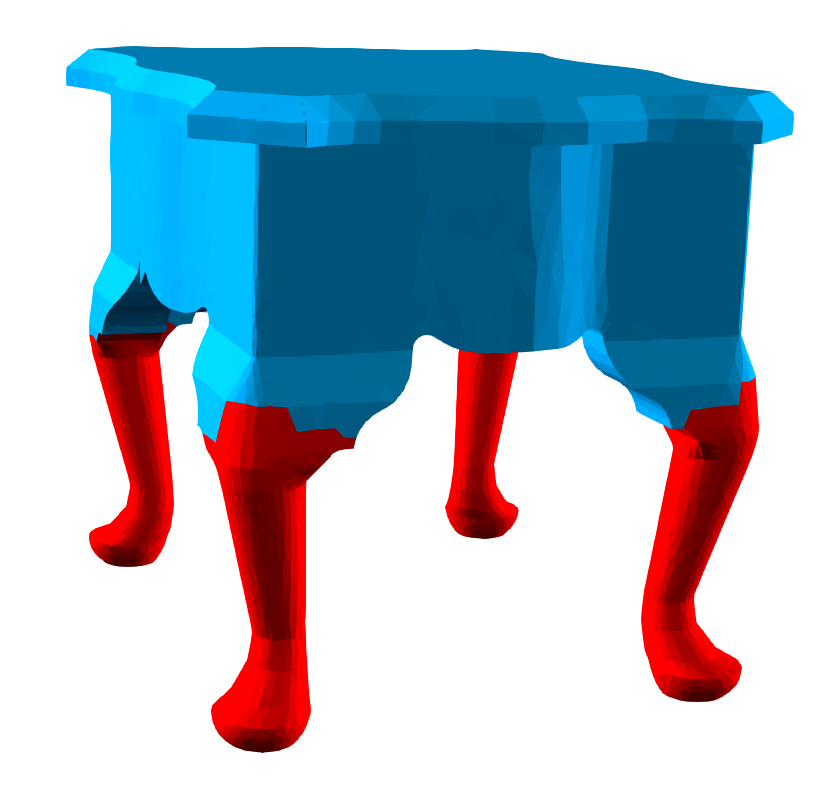} &
    \includegraphics[width=\linewidth]{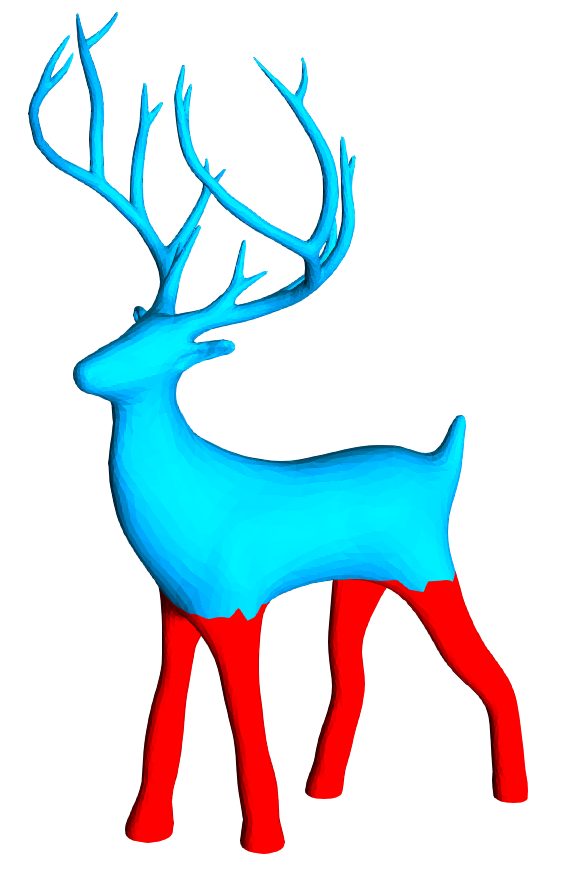}         \\
    \small{Airplane with wings backward closer to the end of the airplane, \textbf{wings}.} &
    \small{Suitable for a tall person to sit on, \textbf{chair legs}.} &
    \small{Deer leaning forward, \textbf{legs}.} \\
    \includegraphics[width=\linewidth]{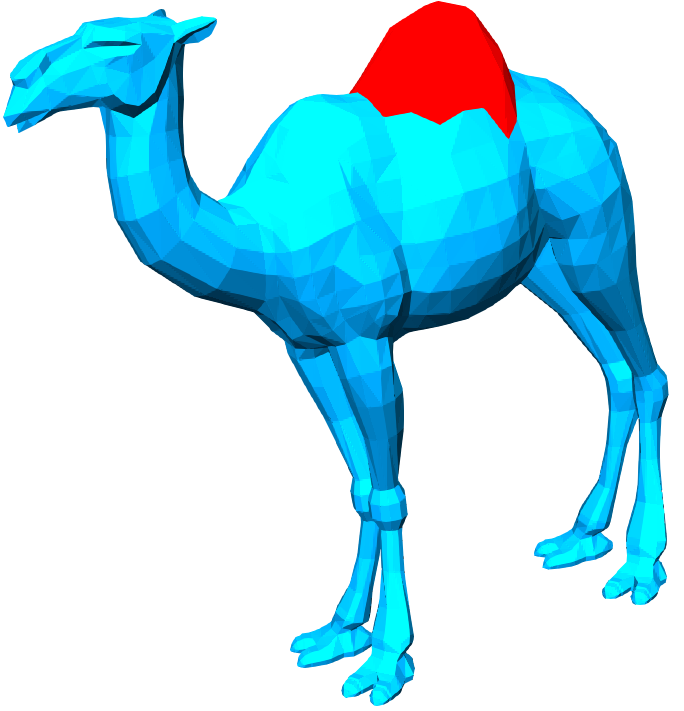} &
    \includegraphics[width=\linewidth]{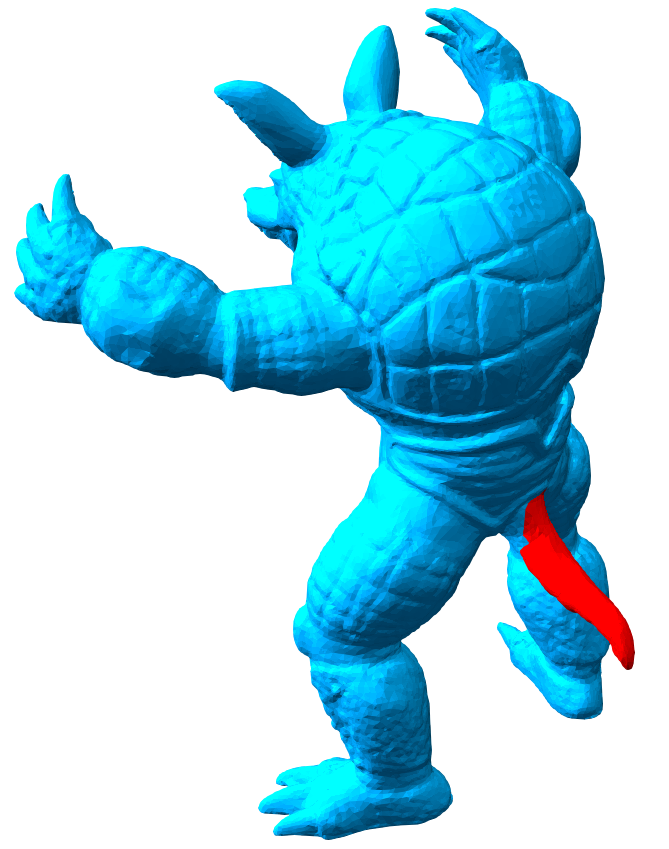} &
    \includegraphics[width=\linewidth]{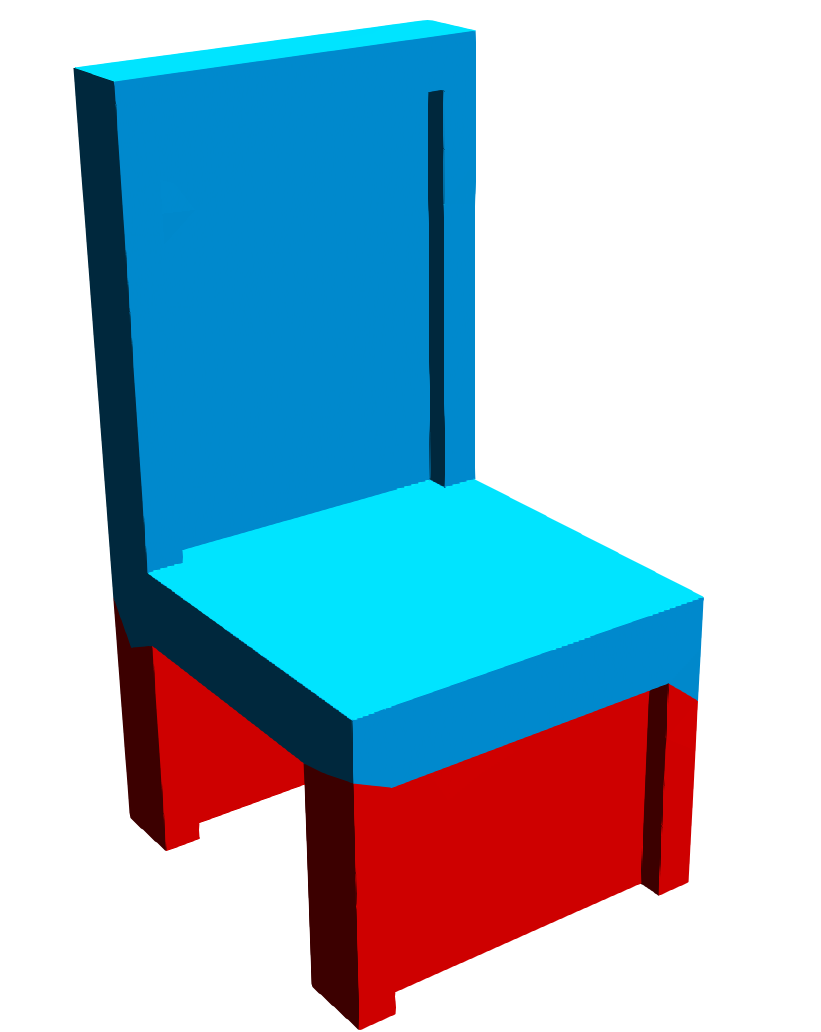}      \\
    \small{Camel with more fat stored in its hump, \textbf{hump}.} &
    \small{Swift the tail to its left, \textbf{tail}.} & 
    \small{Stretching all chair legs to the front of the chair, \textbf{chair legs}.}   \\
\end{tabular}%
}
\caption{Segmentation Results: Regions requiring deformation are highlighted in red. The captions include the corresponding user text input, with the deformable regions emphasized in bold.}
\label{table:segmentation_results}
\vskip -0.2in
\end{table}

\subsection{Handle Generation}
\begin{figure}[h]
\centering
\includegraphics[width=\linewidth]{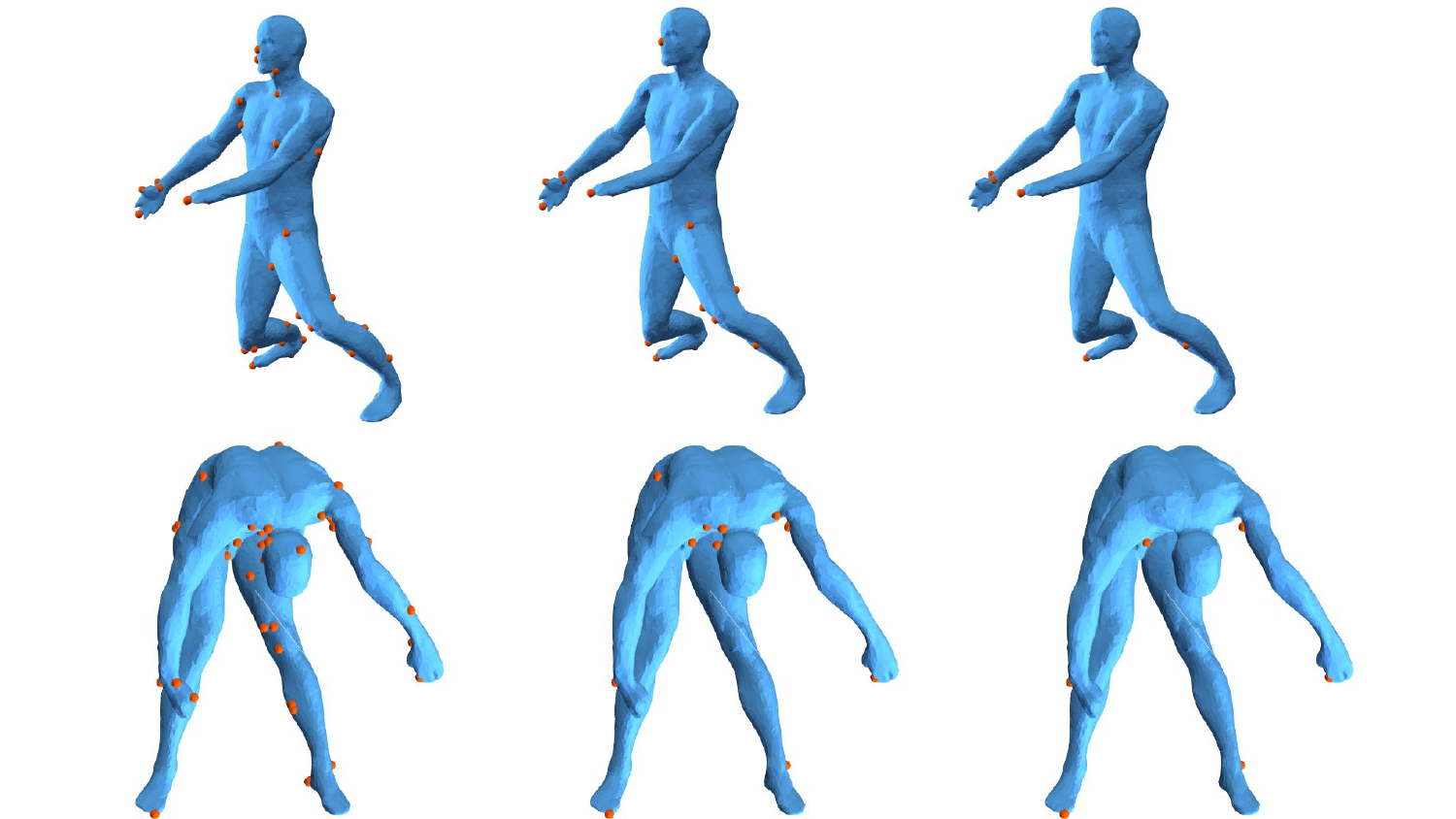}
\caption{\label{fig:cone} Detected cone singularity points are drawn in red dots. Distortion bounds are 0.1, 0.2, 0.4 from left to right respectively.}
\end{figure}

In~\prettyref{table:cone_avg}, we report the average input distortion bound, handle counts, and running time of cone singularity construction over all the models, which is used to construct the potential set of handles $\tilde{\mathcal{H}}$. We adjust the input distortion bound to generate a reasonable number of handles. Running time is measured on AMD Ryzen 9 7950X processor. We also showcase different handles under different distortion bounds in~\prettyref{fig:cone}.

\begin{table}[h]
\vspace{-10px}
\centering
\footnotesize
\caption{\label{table:cone_avg} Average cone generation statistics.}
\vskip 0in
\begin{tabular}{lccc}
\toprule
& \textbf{Distortion Bound} & \textbf{\#Cone} & \textbf{Run Time (s)} \\
\midrule
\textbf{Average} & 0.22 & 24.56 & 3.38 \\
\bottomrule
\end{tabular}
\vspace{-10px}
\end{table}

\subsection{More Results}
\begin{figure}[ht]
\centering
\includegraphics[width=\linewidth]{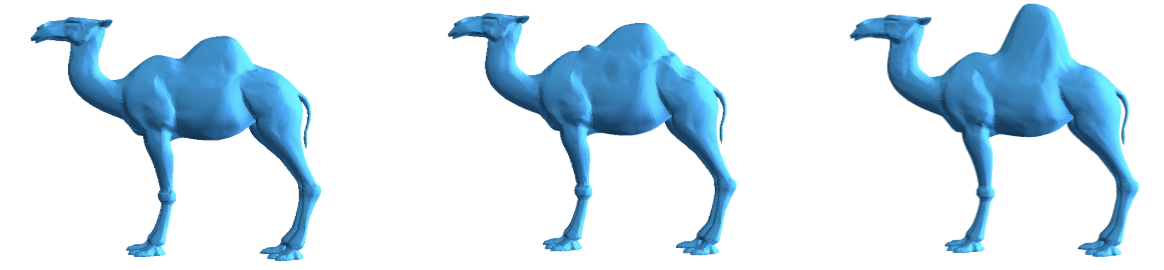}
\caption{\label{fig:control-saliency} Left: Origin; Middle: Camel with a slightly taller hump; Right: Camel with an extremely tall hump, as high as a mountain.}
\end{figure}

We visualize more results, including those appeared in main paper. First, We demonstrate that our pipeline controls deformation saliency via prompts in~\prettyref{fig:control-saliency}. This is achieved by fine-tuning the VLM with diverse data pairs, where different text inputs may correspond to varying movement magnitudes. However enhancing VLM fine-tuning with more geometric optimization constraints could improve controllable text-guided deformation.

\begin{figure}[H]
\centering
\scalebox{.9}{%
\begin{tabular}{p{0.45\linewidth}p{0.45\linewidth}}
    \includegraphics[width=\linewidth]{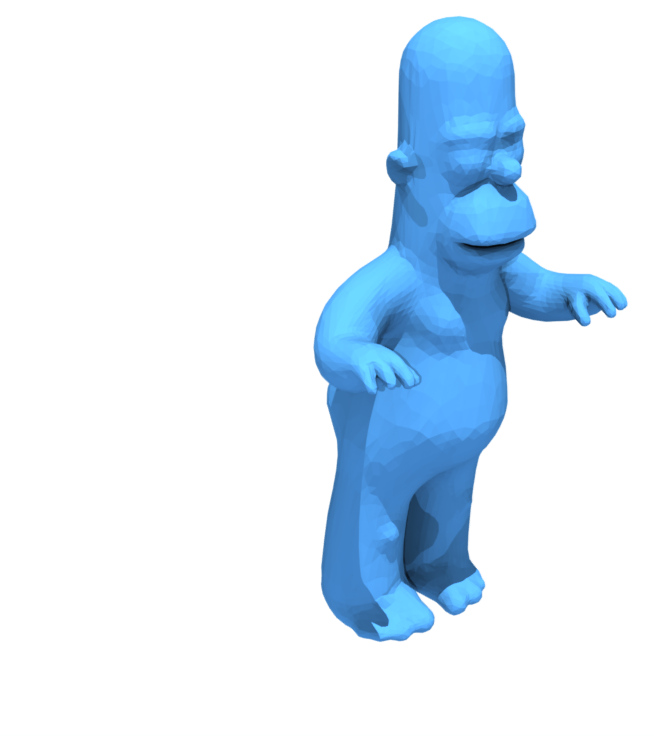} &
    \includegraphics[width=\linewidth]{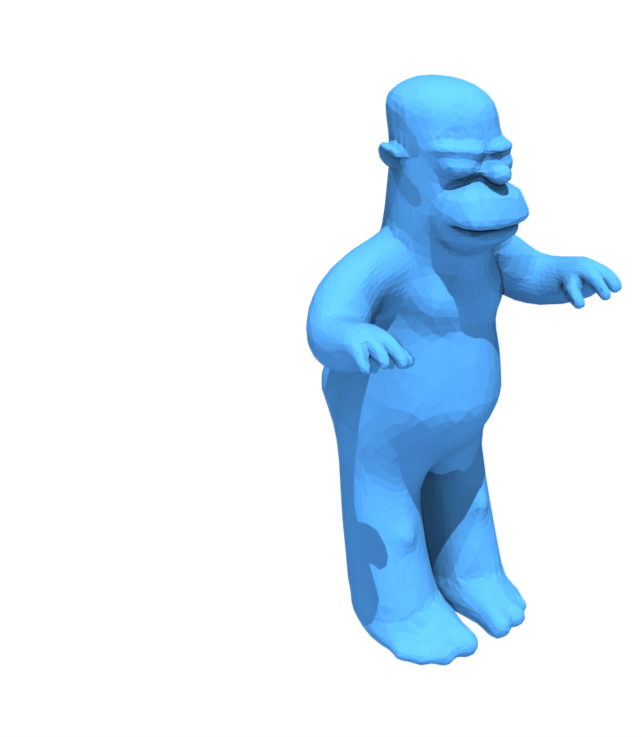} \\
    \multicolumn{2}{c}{Homer with shorter head and legs extended forward.} \\
    \includegraphics[width=\linewidth]{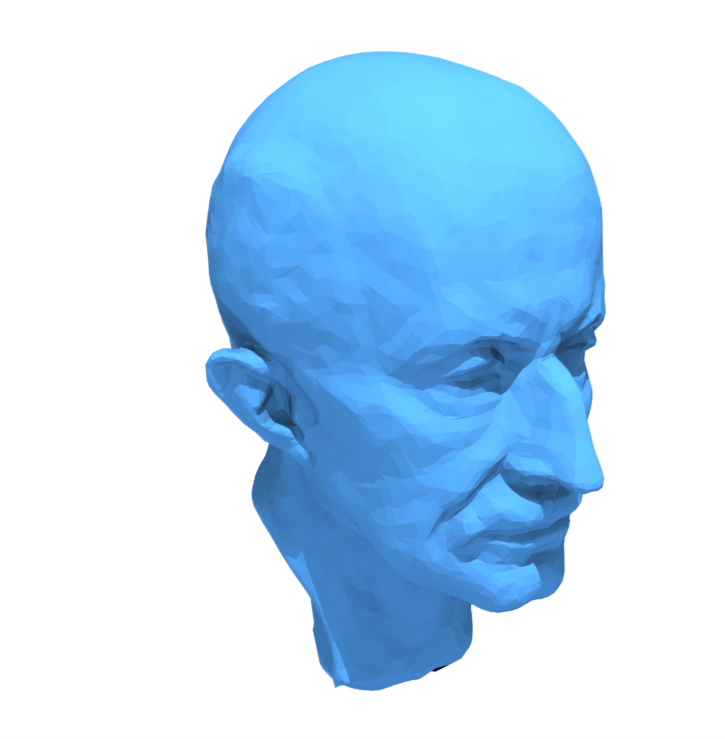} &
    \includegraphics[width=\linewidth]{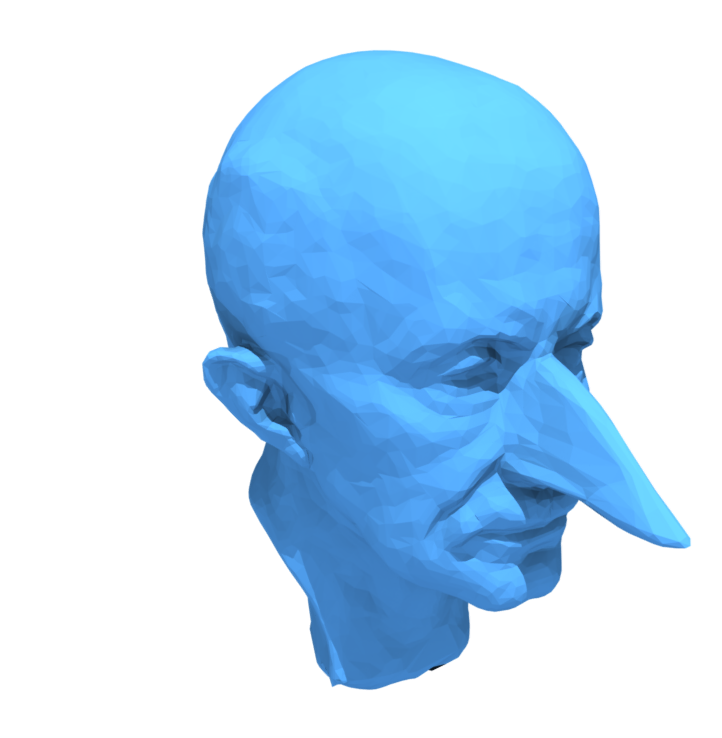} \\
    \multicolumn{2}{c}{Pinocchio telling lie.} \\
\end{tabular}%
}
\caption{\label{fig:deformation_results_4}Example visualization.}
\end{figure}

\begin{figure}[H]
\centering
\scalebox{.9}{%
\begin{tabular}{p{0.45\linewidth}p{0.45\linewidth}}
    \includegraphics[height=1.5\linewidth,trim=4cm 3cm 1cm 4.5cm,clip]{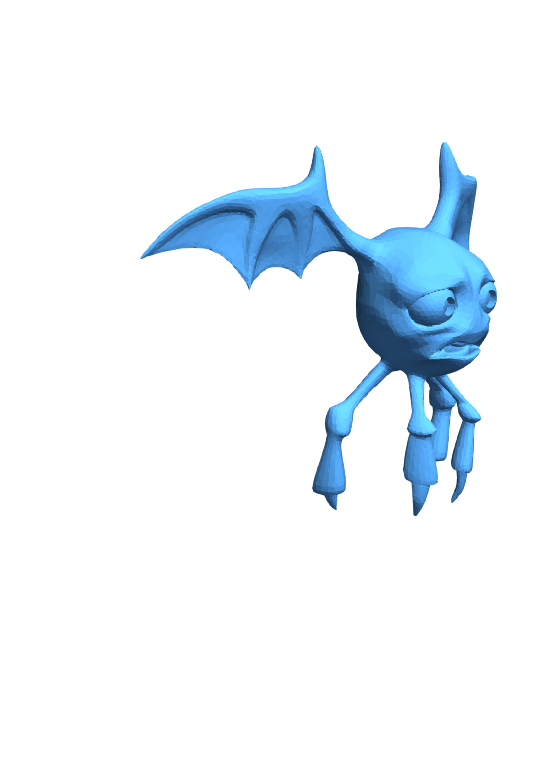} &
    \includegraphics[height=1.5\linewidth,trim=4cm 3cm 1cm 4.5cm,clip]{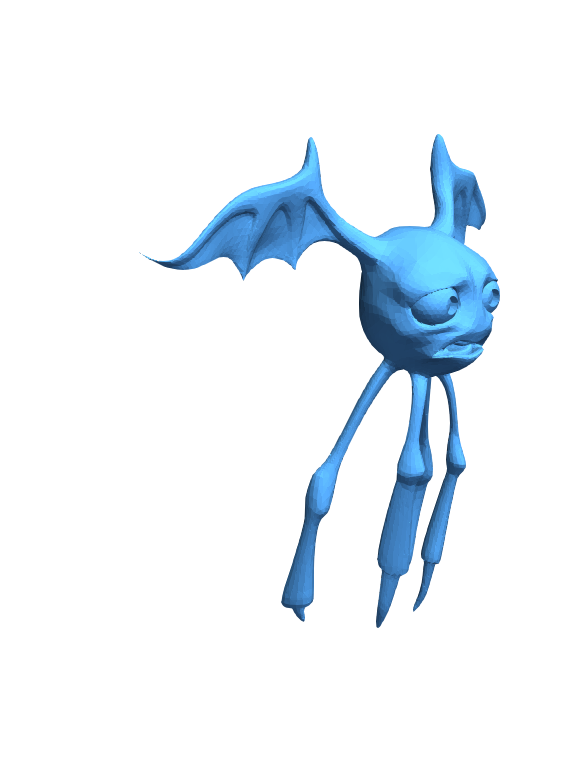} \\
    \multicolumn{2}{c}{\TWORCell{Adjust the demon's wings by pulling them backward}{while straightening and elongating its legs.}} \\
    \includegraphics[width=\linewidth]{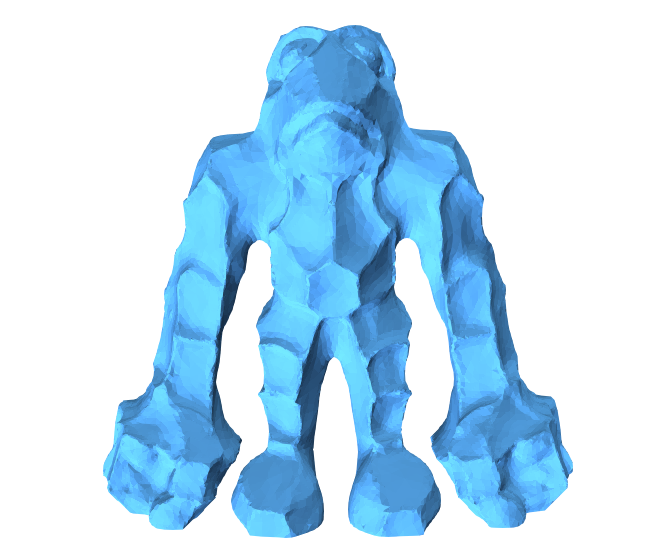} &
    \includegraphics[width=\linewidth]{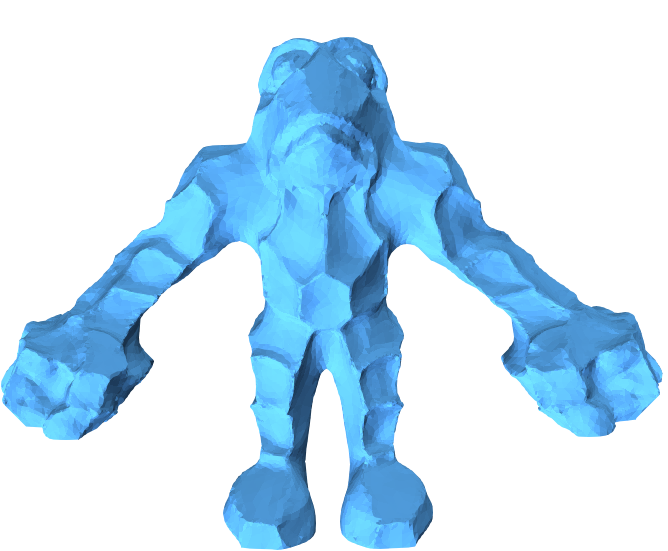} \\
    \multicolumn{2}{c}{Shorten the reach of arms and stretch arms outward.} \\
    \includegraphics[width=\linewidth]{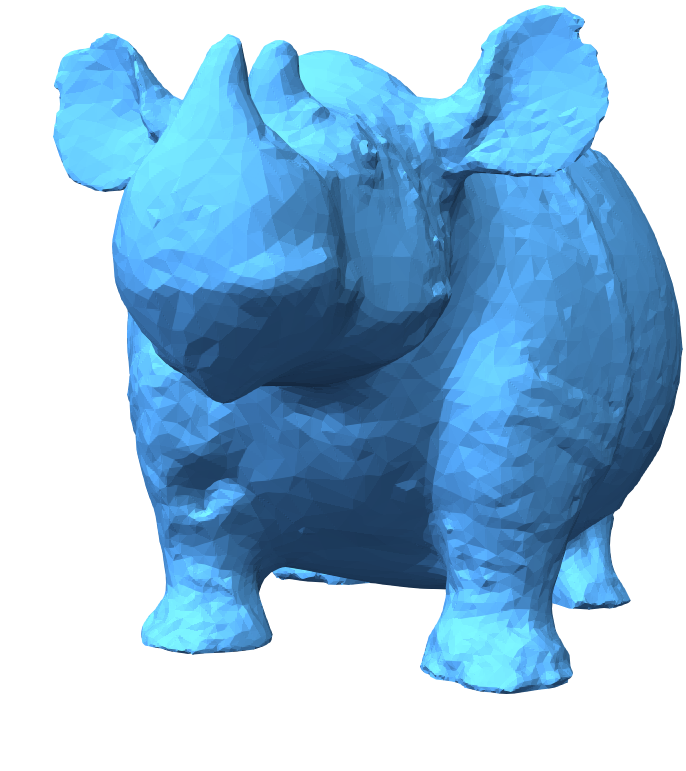} &
    \includegraphics[width=\linewidth]{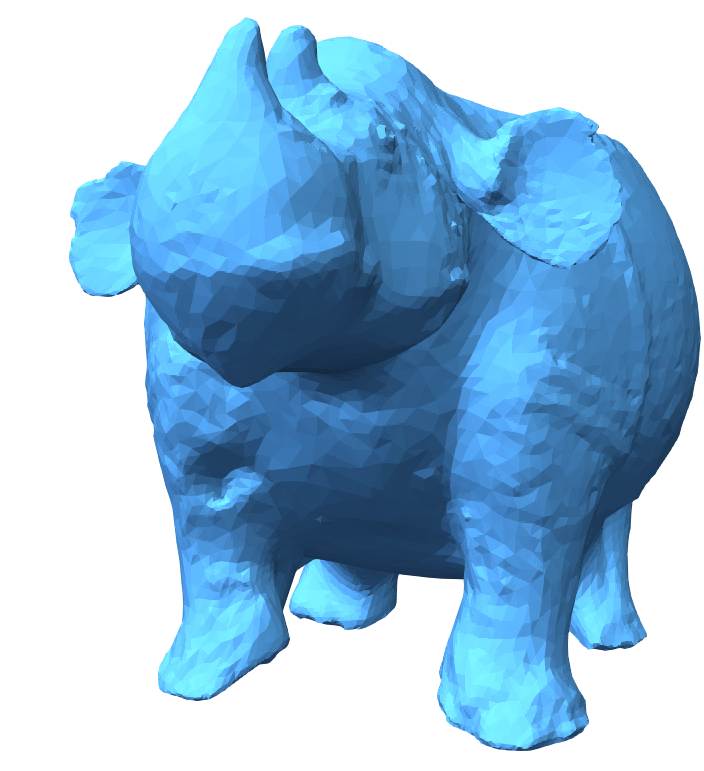} \\
    \multicolumn{2}{c}{The rhino grows taller and with lower ears.} \\
    \includegraphics[width=\linewidth]{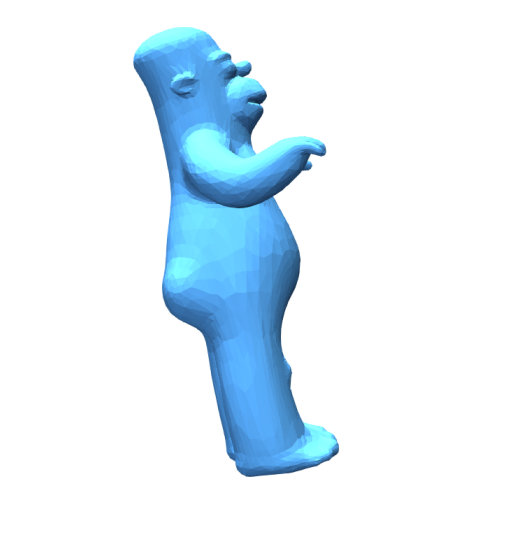} &
    \includegraphics[width=\linewidth]{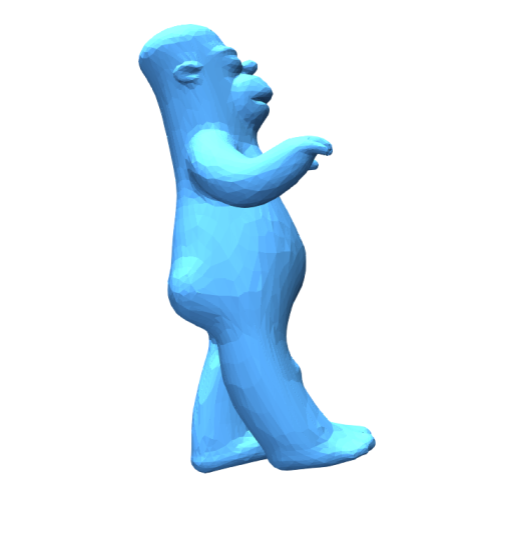} \\
    \multicolumn{2}{c}{Homer is walking.} \\
\end{tabular}%
}
\caption{\label{fig:deformation_results_1}Example visualization.}
\end{figure}

\begin{figure}[H]
\centering
\scalebox{.9}{%
\begin{tabular}{p{0.45\linewidth}p{0.45\linewidth}}
    \includegraphics[width=\linewidth]{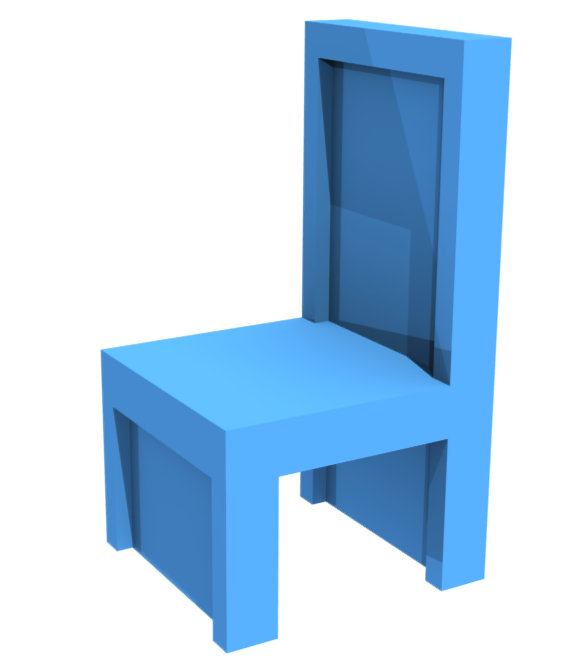} &
    \includegraphics[width=\linewidth]{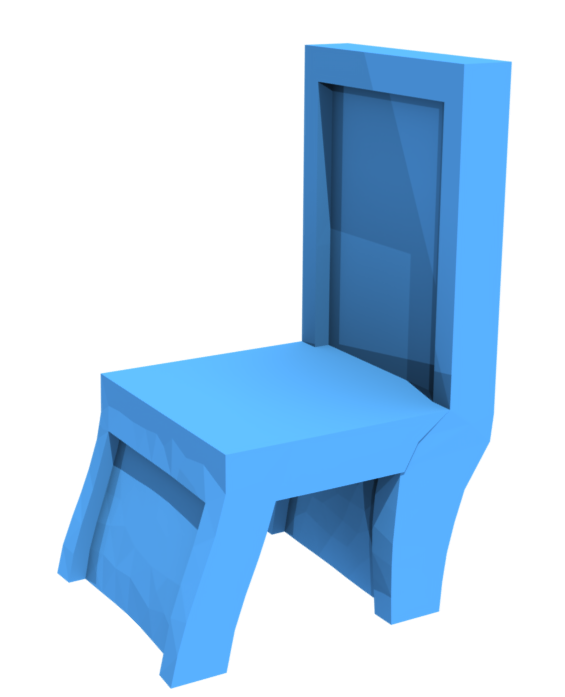} \\
    \multicolumn{2}{c}{Stretching all chair legs to the front of the chair.} \\
    \includegraphics[width=\linewidth]{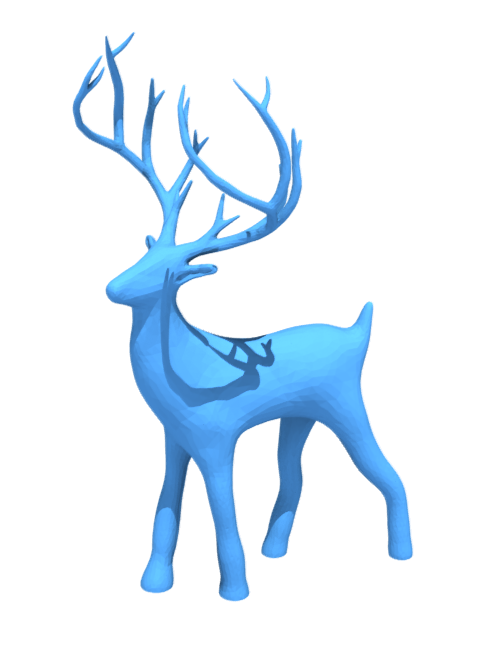} &
    \includegraphics[width=\linewidth]{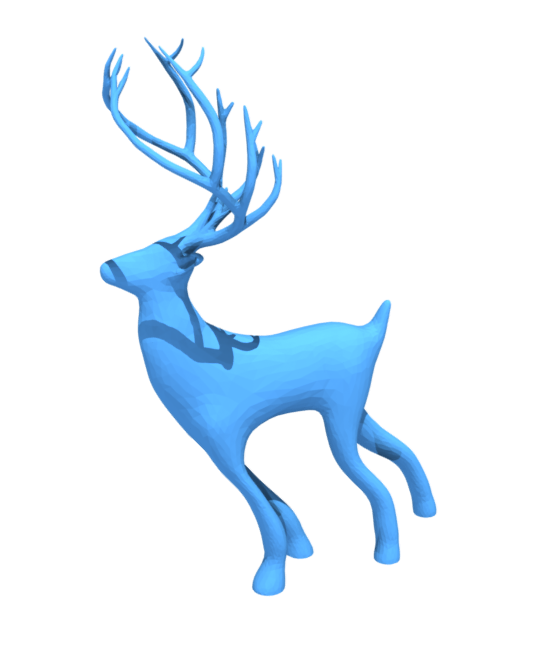} \\
    \multicolumn{2}{c}{A deer leaning forward.} \\
    \includegraphics[width=\linewidth]{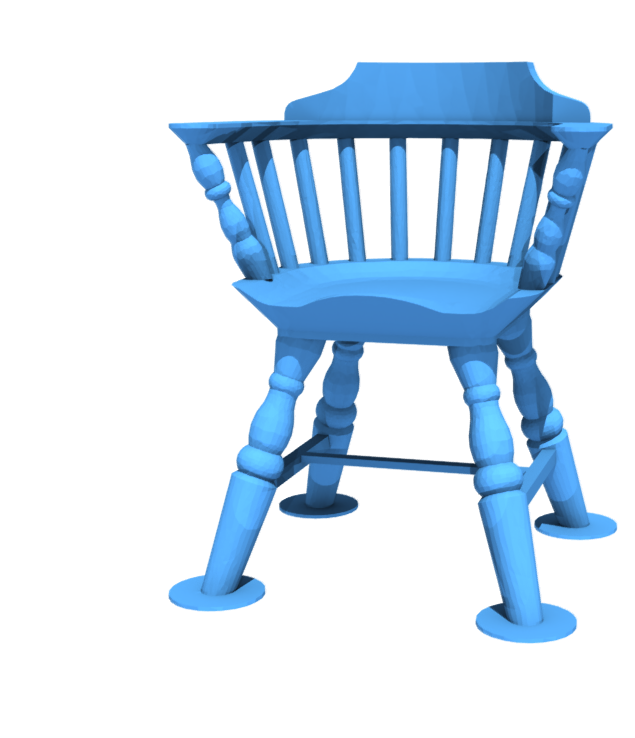} &
    \includegraphics[width=\linewidth]{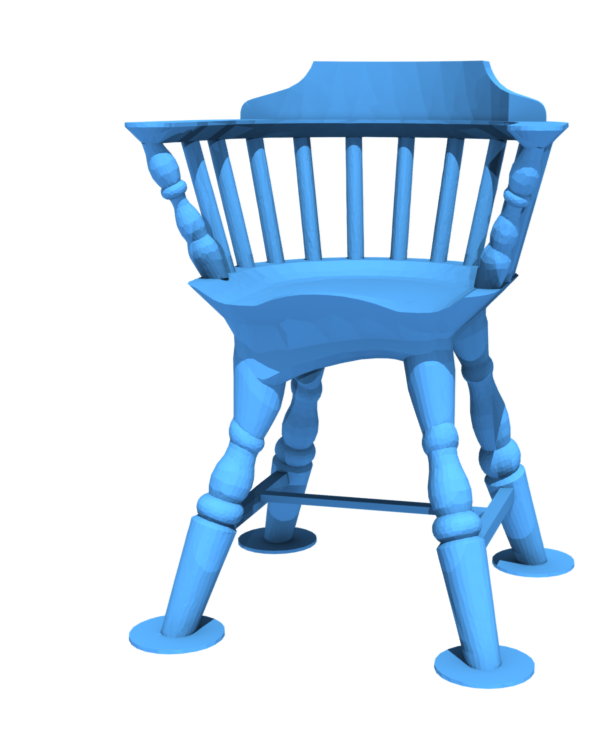} \\
    \multicolumn{2}{c}{A tall barchair.} \\
    \includegraphics[width=\linewidth]{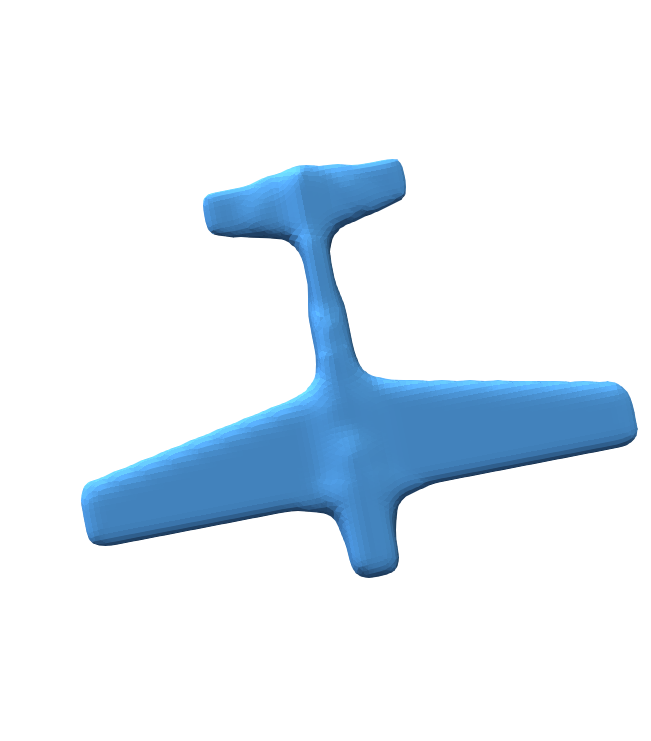} &
    \includegraphics[width=\linewidth]{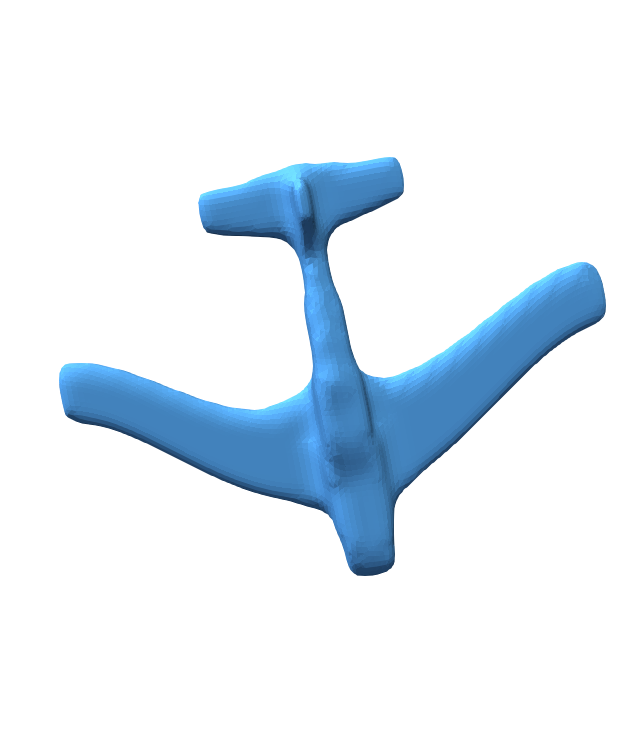} \\
    \multicolumn{2}{c}{Pull the airplane wings backwards.} \\
\end{tabular}%
}
\caption{\label{fig:deformation_results_3}Example visualization.}
\end{figure}

\begin{figure}[H]
\centering
\scalebox{.9}{%
\begin{tabular}{p{0.45\linewidth}p{0.45\linewidth}}
    \includegraphics[width=\linewidth]{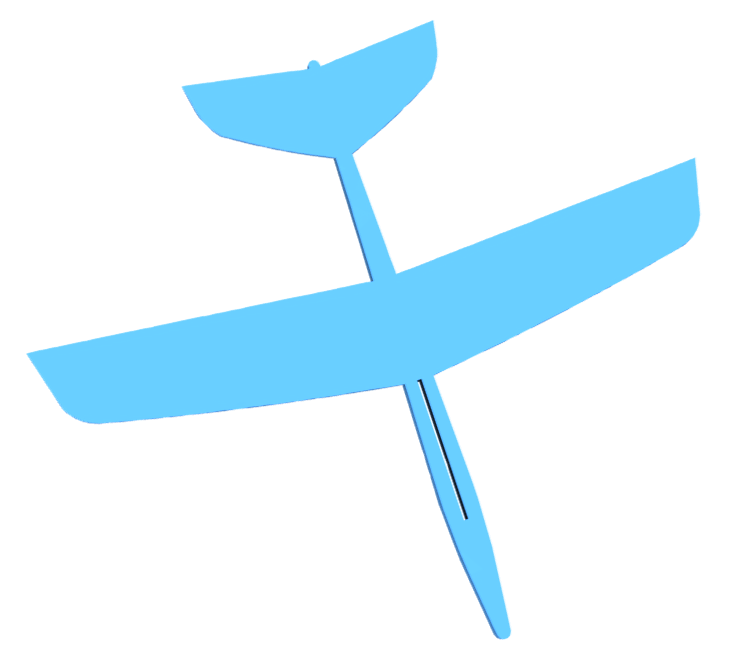} &
    \includegraphics[width=\linewidth]{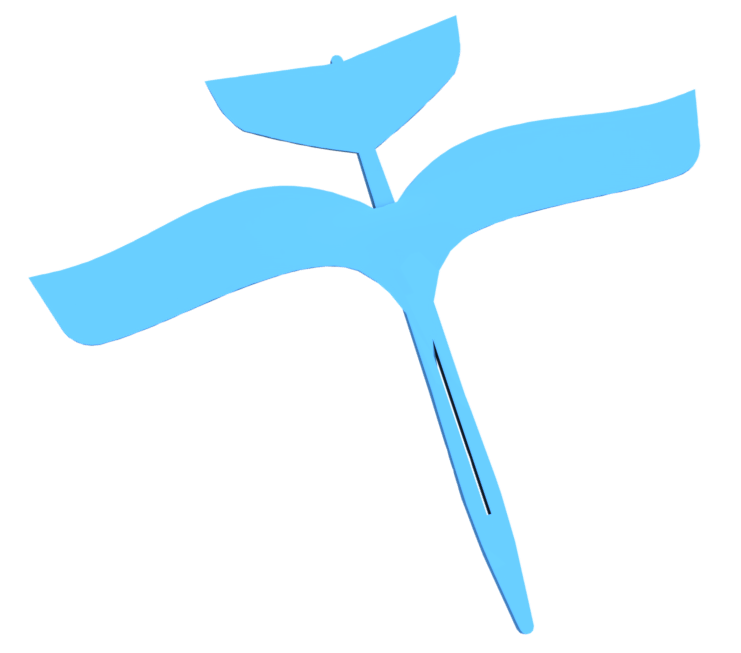} \\
    \multicolumn{2}{c}{Airplane with wings backward closer to the end of the airplane.} \\
    \includegraphics[width=\linewidth]{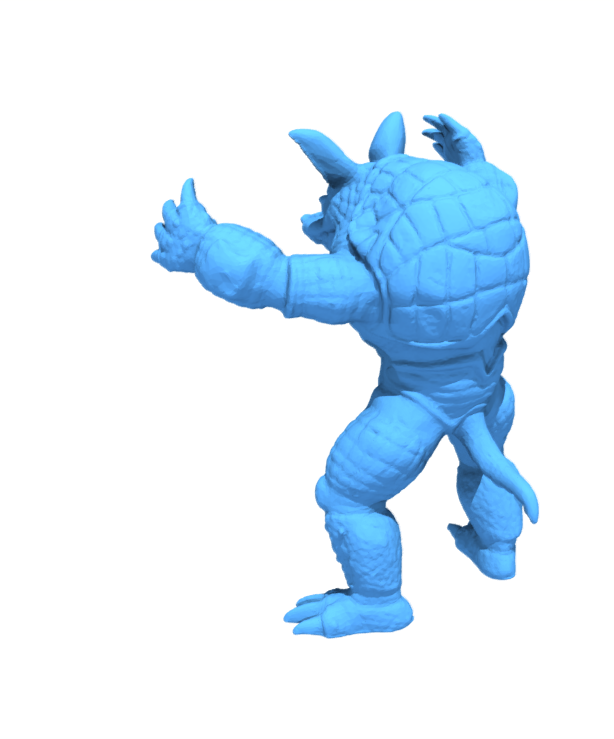} &
    \includegraphics[width=\linewidth]{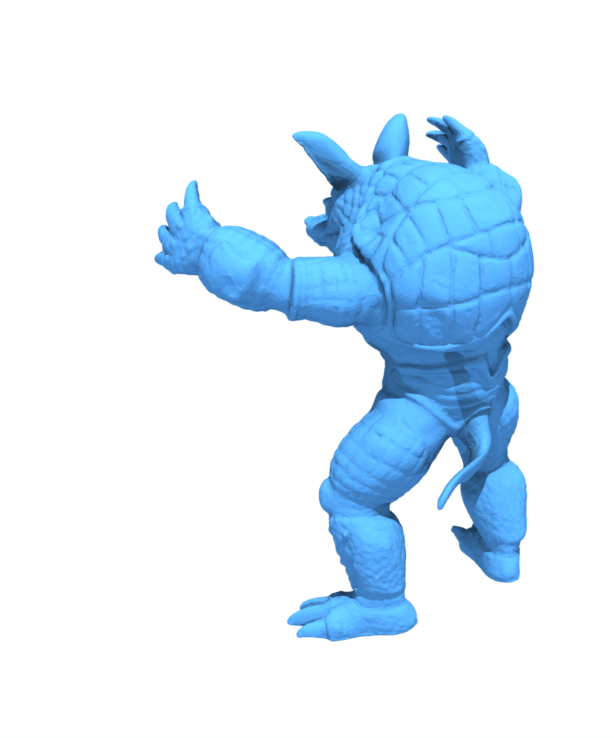} \\
    \multicolumn{2}{c}{Swift the tail to its left.} \\
    \includegraphics[width=\linewidth]{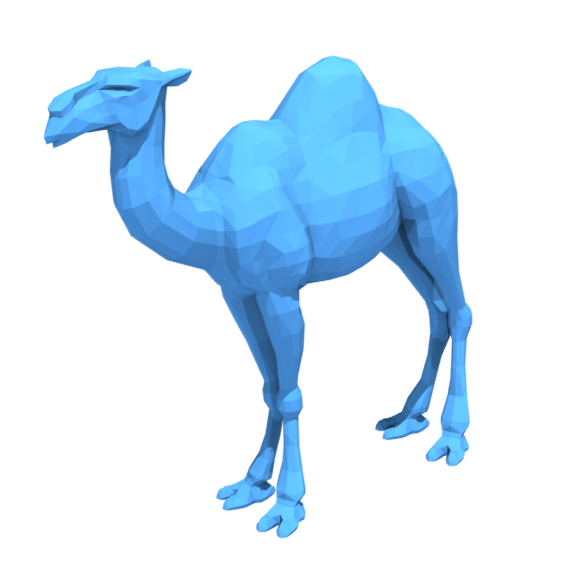} &
    \includegraphics[width=\linewidth]{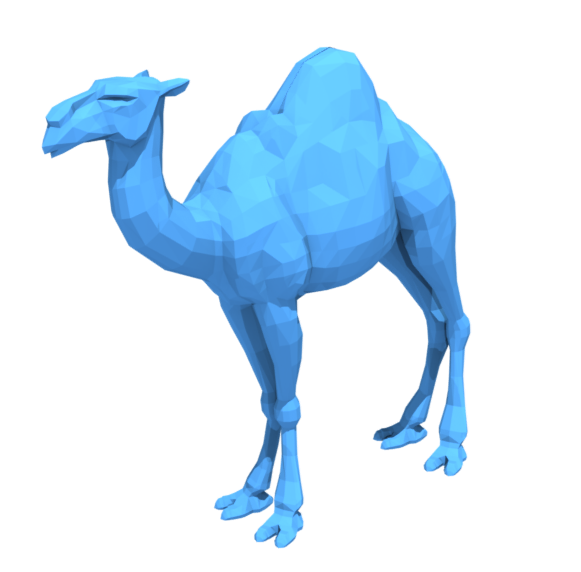} \\
    \multicolumn{2}{c}{Camel with more fat stored in its hump.} \\
    \includegraphics[width=\linewidth]{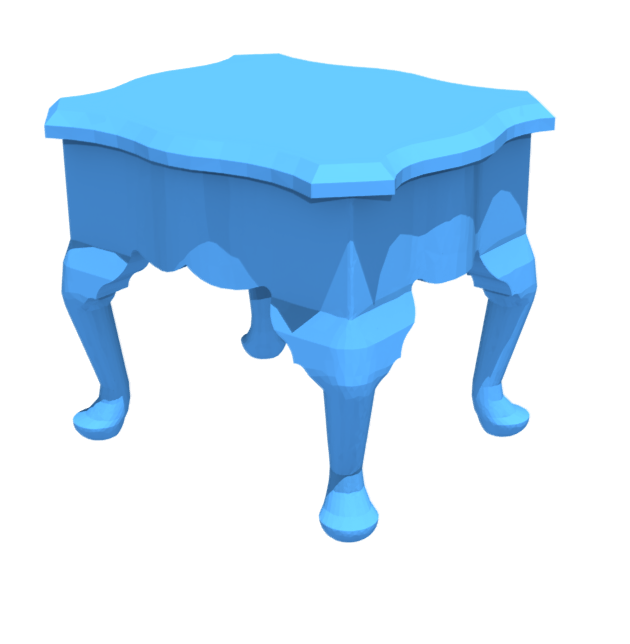} &
    \includegraphics[width=\linewidth]{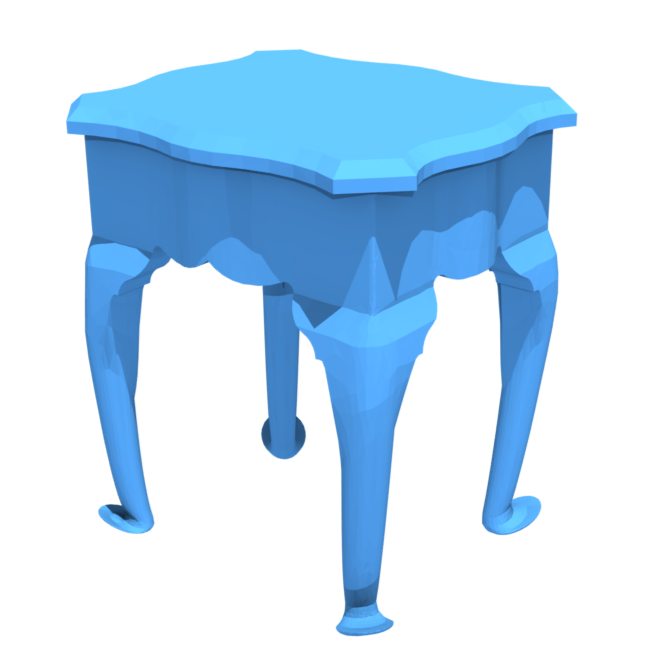} \\
    \multicolumn{2}{c}{Suitable for a tall person to seat on.} \\
    \includegraphics[width=\linewidth,
    trim=0 10cm 0 7cm,clip]{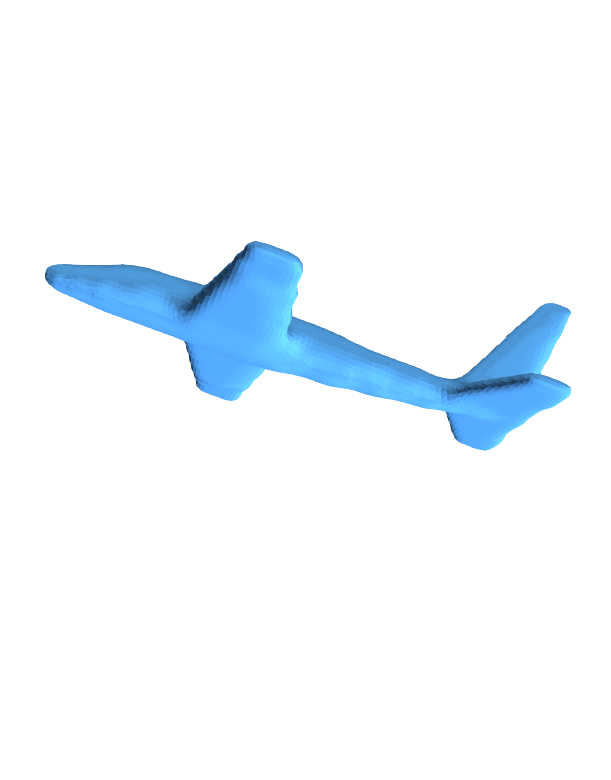} &
    \includegraphics[width=\linewidth,
    trim=0 10cm 0 7cm,clip]{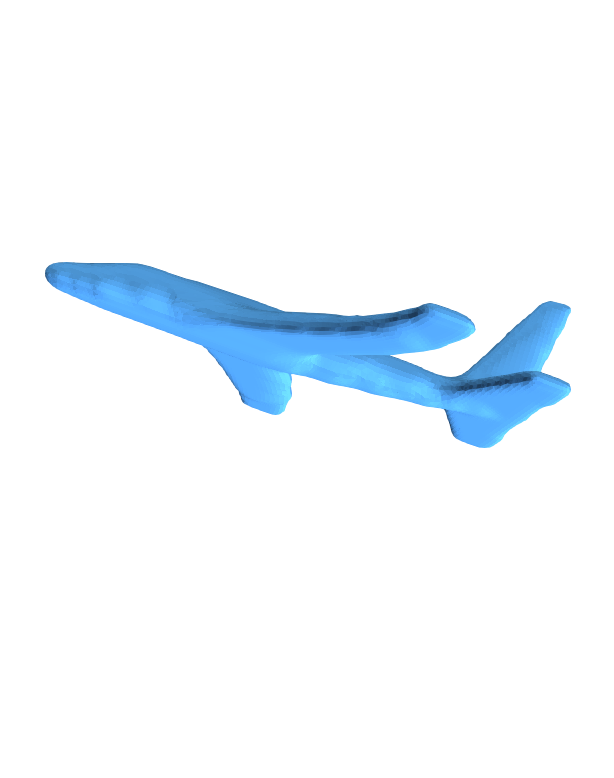} \\
    \multicolumn{2}{c}{Airplane with wings backward closer to the end of the airplane.} \\
\end{tabular}%
}
\caption{\label{fig:deformation_results_2}Example visualization.}
\end{figure}

In~\prettyref{fig:deformation_results_1}, \ref{fig:deformation_results_2}, \ref{fig:deformation_results_3}, and \ref{fig:deformation_results_4}, the left column displays the original mesh, while the right column showcases the deformation results produced by our method.


\subsection{Baseline Hyperparameters}
We compare our approach against TextDeformer, MagicClay, and MeshUp to evaluate its effectiveness. For all methods, we use the authors’ released source code with default hyperparameters, applying minor adjustments only when necessary for compatibility. Specifically, for TextDeformer, we adopt the ViT-B/32 CLIP model; for MagicClay, we set the batch size to 1 to reduce GPU memory usage and disable texture inputs, as our method does not utilize texture materials; and for MeshUp, we use the default diffusion model with the XL configuration.

\subsection{Failure Cases}
The scratch-like artifact on the horse’s front leg appears as a localized stretching or tearing effect, which is because when a handle is manipulated near regions with high curvature (like legs), the uneven weight distribution can cause localized distortions. Since our method contribution is to automatically generates text-conformed deformations on the mesh via VLM, we leave refining this issue for future improvement.
\begin{figure}[H]
\centering
\resizebox{\linewidth}{!}{%
\begin{tabular}{p{0.45\linewidth}p{0.45\linewidth}}
    \includegraphics[width=\linewidth,trim=0 8cm 0 7cm,clip]{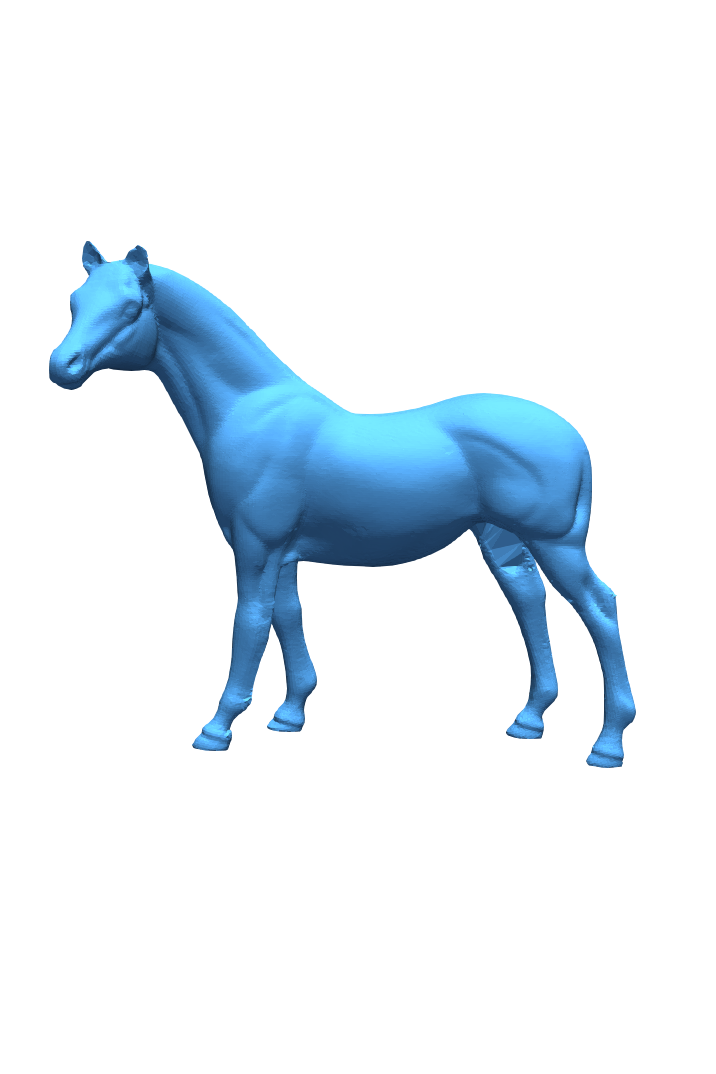} &
    \includegraphics[width=\linewidth,trim=0 8cm 0 7cm,clip]{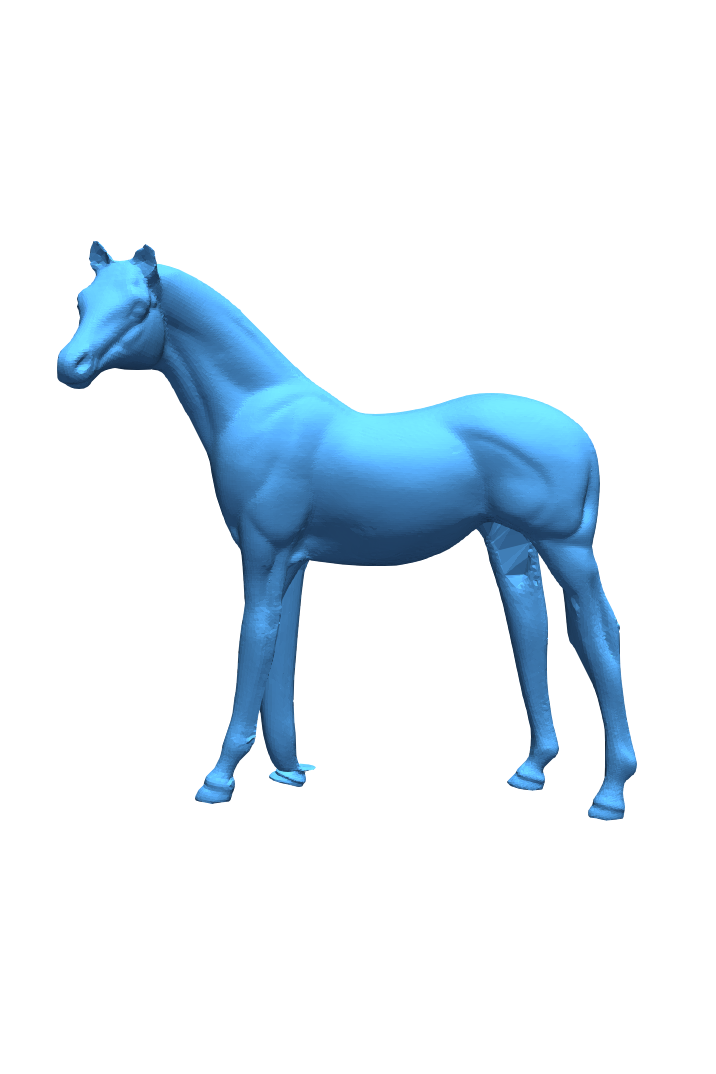} \\
    \multicolumn{2}{c}{Horse with longer legs.} \\
\end{tabular}
}
\caption{\label{fig:failture_case}The horse legs is deformed with artifacts.}
\end{figure}

\subsection{Complete Prompt for GPT-4o}
\begin{figure*}[!ht]
\caption{\label{fig:InstructionToGPT2}The instruction provided to GPT-4o for view selection.}
\begin{tcolorbox}[width=\linewidth,height=460px]
\textbf{System Instruction:}\\
You are a 3D artist and expert in 3D reasoning. I will provide you with instructions on how to deform a 3D object, along with six images of the object—one from each of its six facets. \\
Your task is to first interpret the instruction to identify which part of the object needs to be deformed. Then, you need to select images that best show the required part of the object.
You should provide detailed reasoning and output only one image that contains the required part of the 3D object. \\
Your answer must strictly follow the JSON format provided in the example below. You must strictly follow the answer format as shown in the example; do not write any additional text.\\
\textbf{Example 1:} \\
Text Input: I want to elongate the chin. \\
Image Set: From left to right are: `Camera.png', `Camera001.png', `Camera002.png', `Camera003.png', `Camera004.png', `Camera005.png' 
\begin{center}
\begin{tabular}{cccccc}
    \begin{minipage}{0.14\linewidth}
        \centering
        \includegraphics[width=\linewidth]{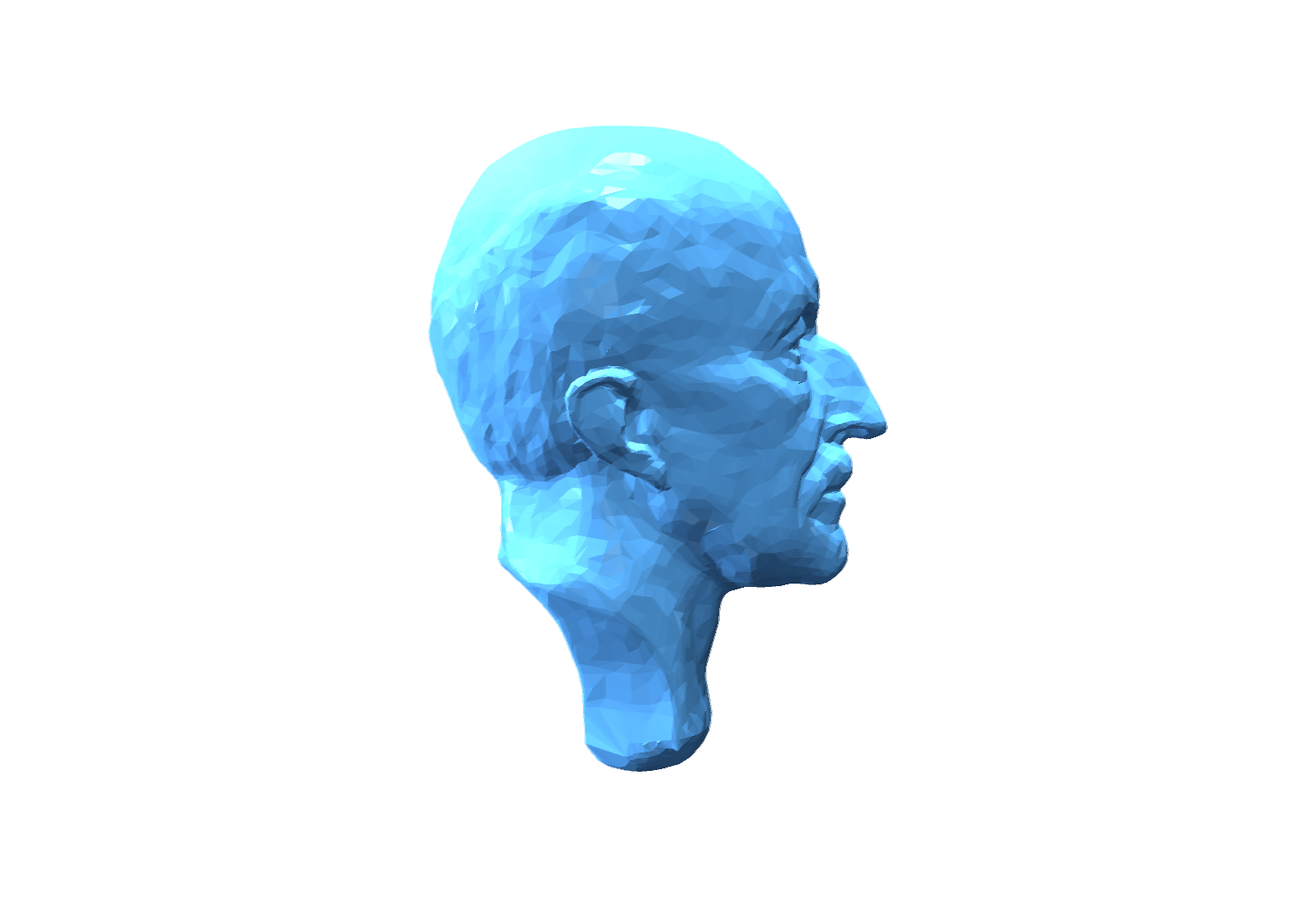}
        \label{fig:Cameraa}
    \end{minipage} &
    \begin{minipage}{0.14\linewidth}
        \centering
        \includegraphics[width=\linewidth]{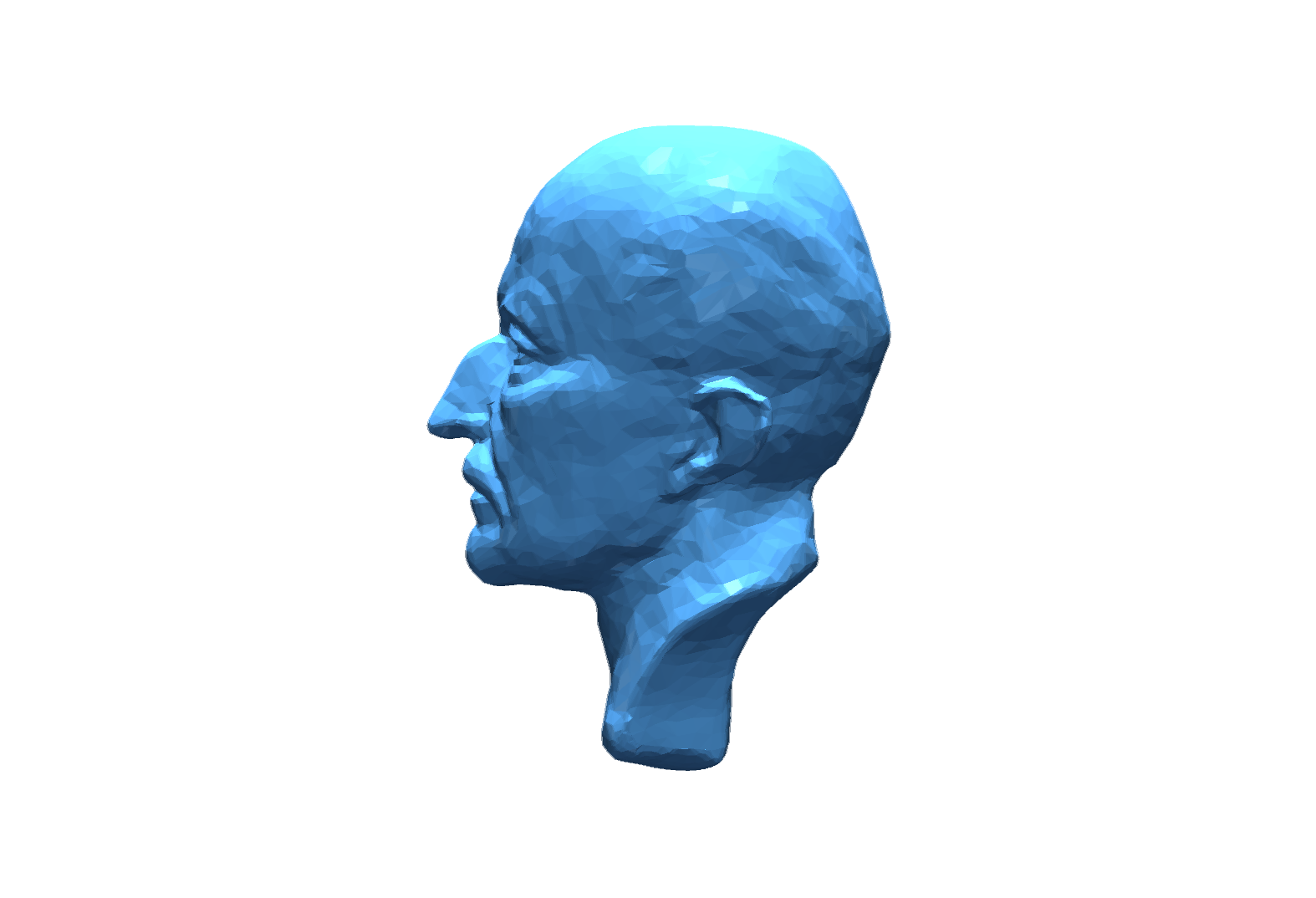}
        \label{fig:Camera001a}
    \end{minipage} &
    \begin{minipage}{0.14\linewidth}
        \centering
        \includegraphics[width=\linewidth]{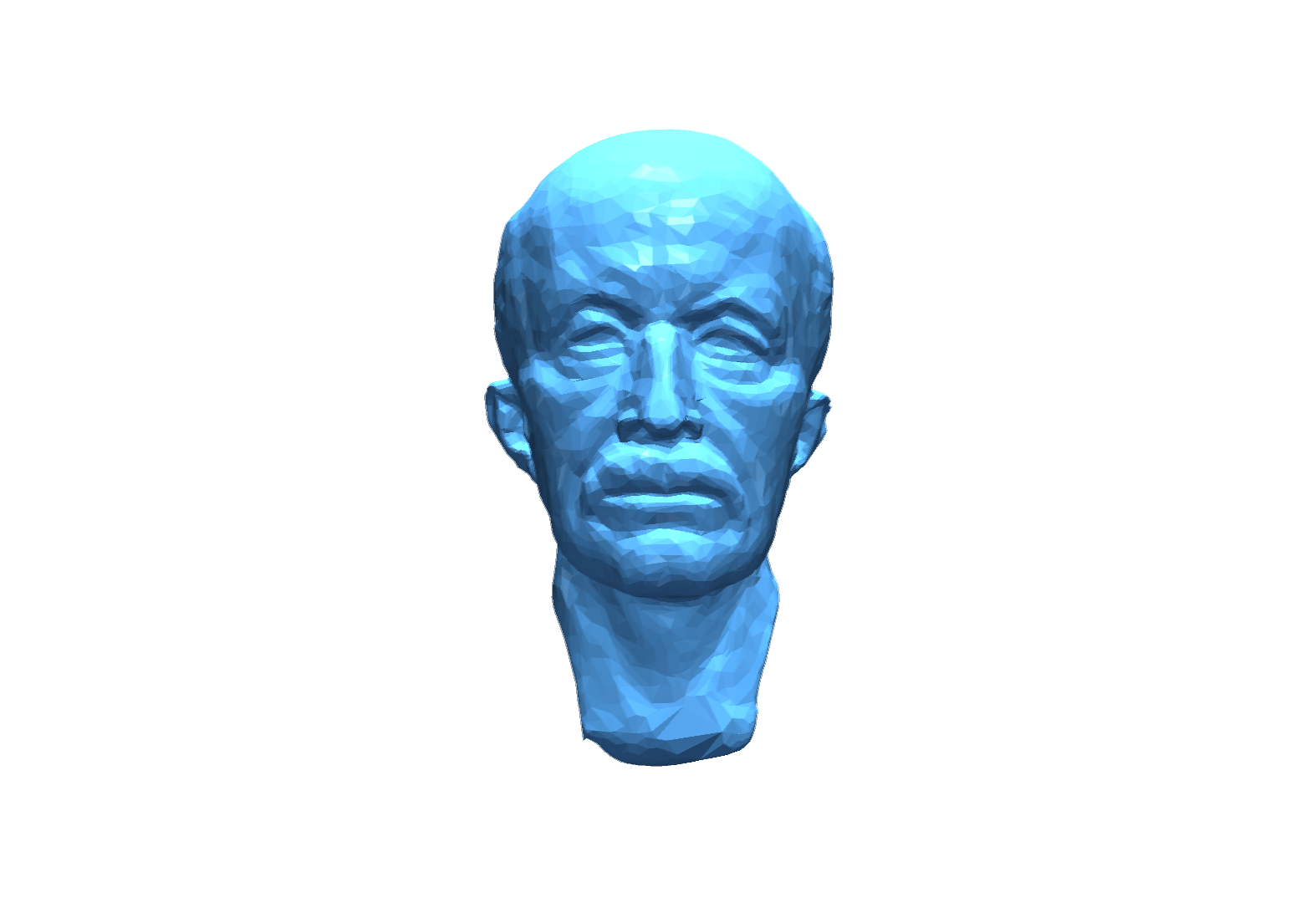}
        \label{fig:Camera002a}
    \end{minipage} &
    \begin{minipage}{0.14\linewidth}
        \centering
        \includegraphics[width=\linewidth]{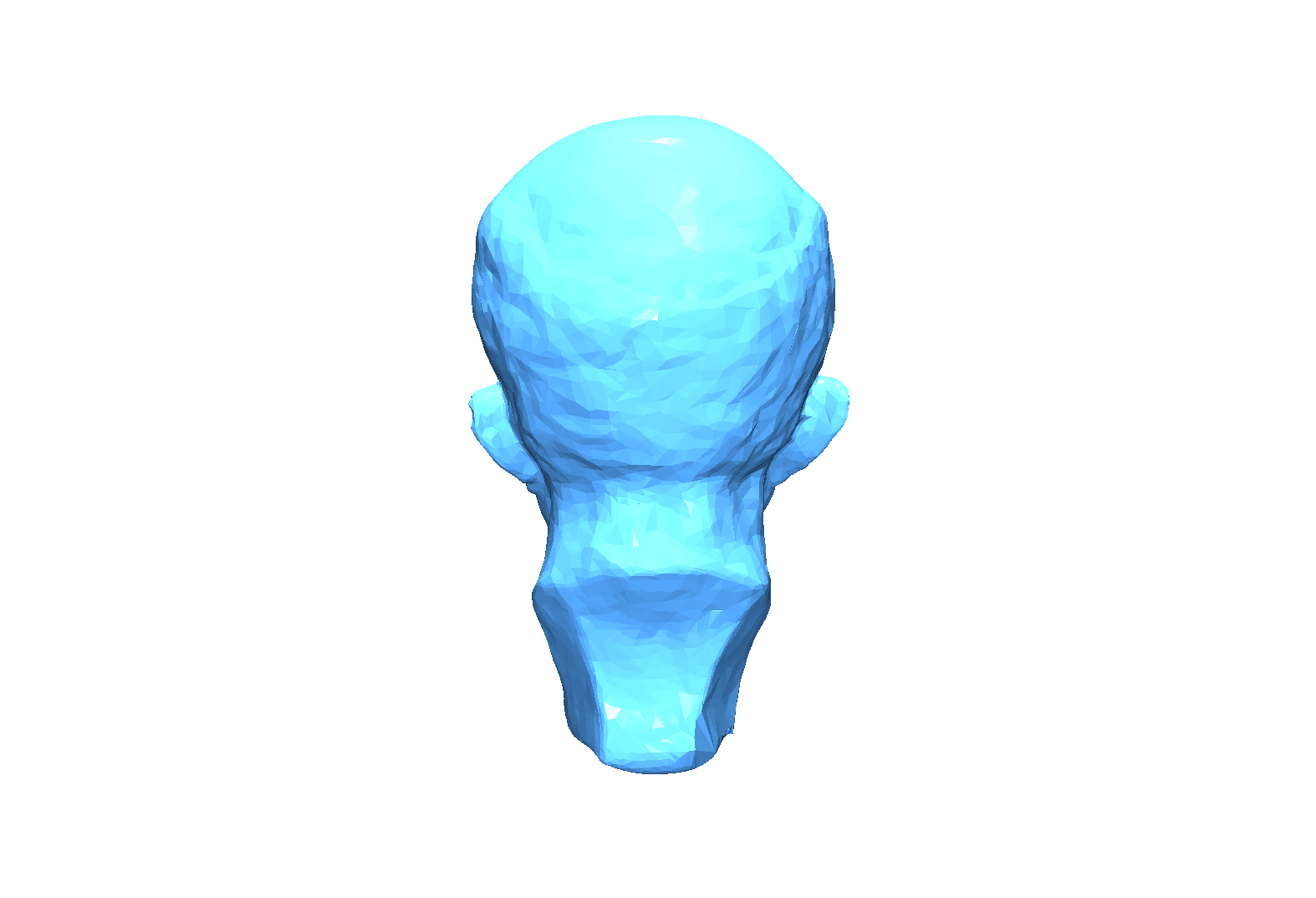}
        \label{fig:Camera003a}
    \end{minipage} &
    \begin{minipage}{0.14\linewidth}
        \centering
        \includegraphics[width=\linewidth]{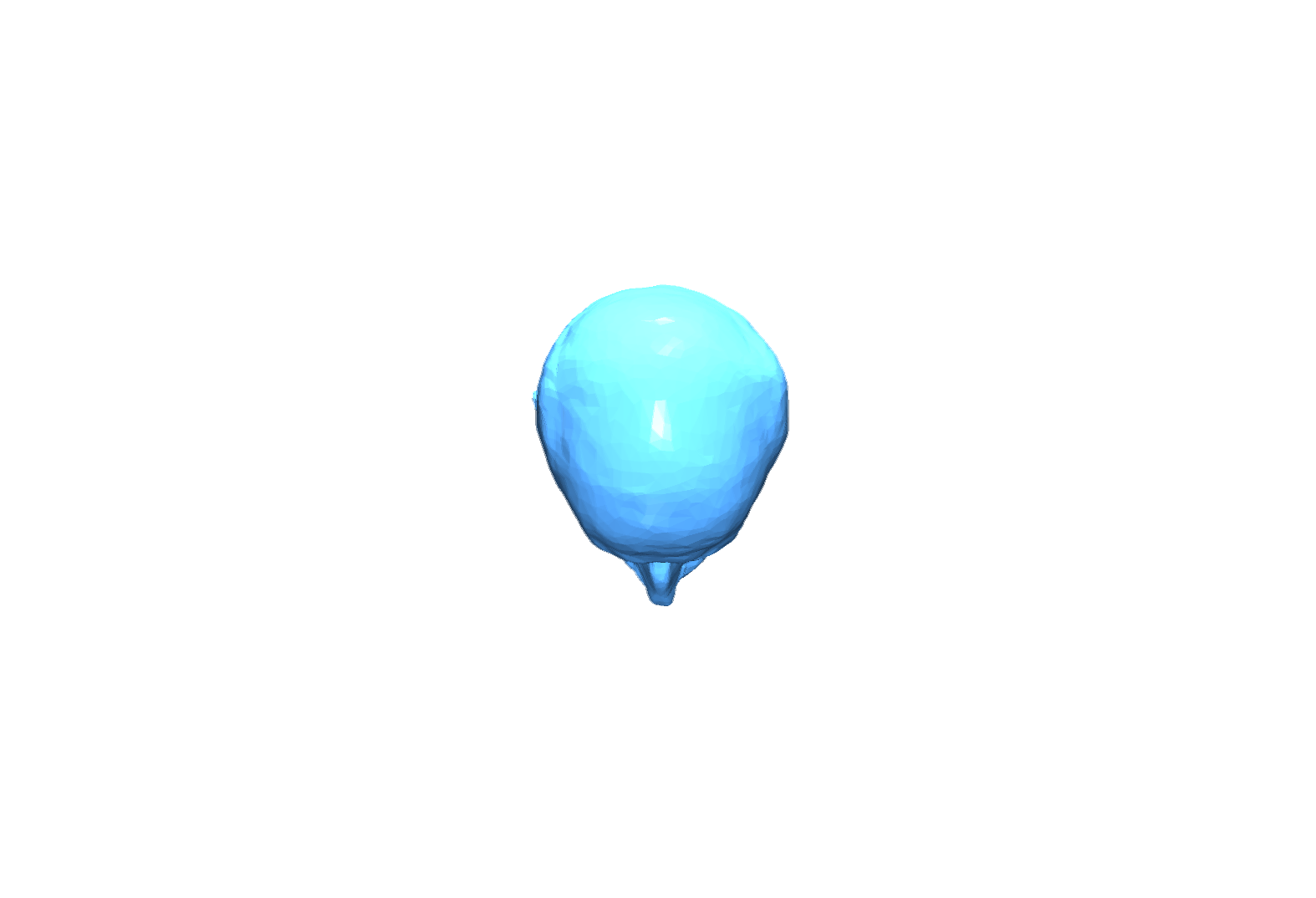}
        \label{fig:Camera004a}
    \end{minipage} &
    \begin{minipage}{0.14\linewidth}
        \centering
        \includegraphics[width=\linewidth]{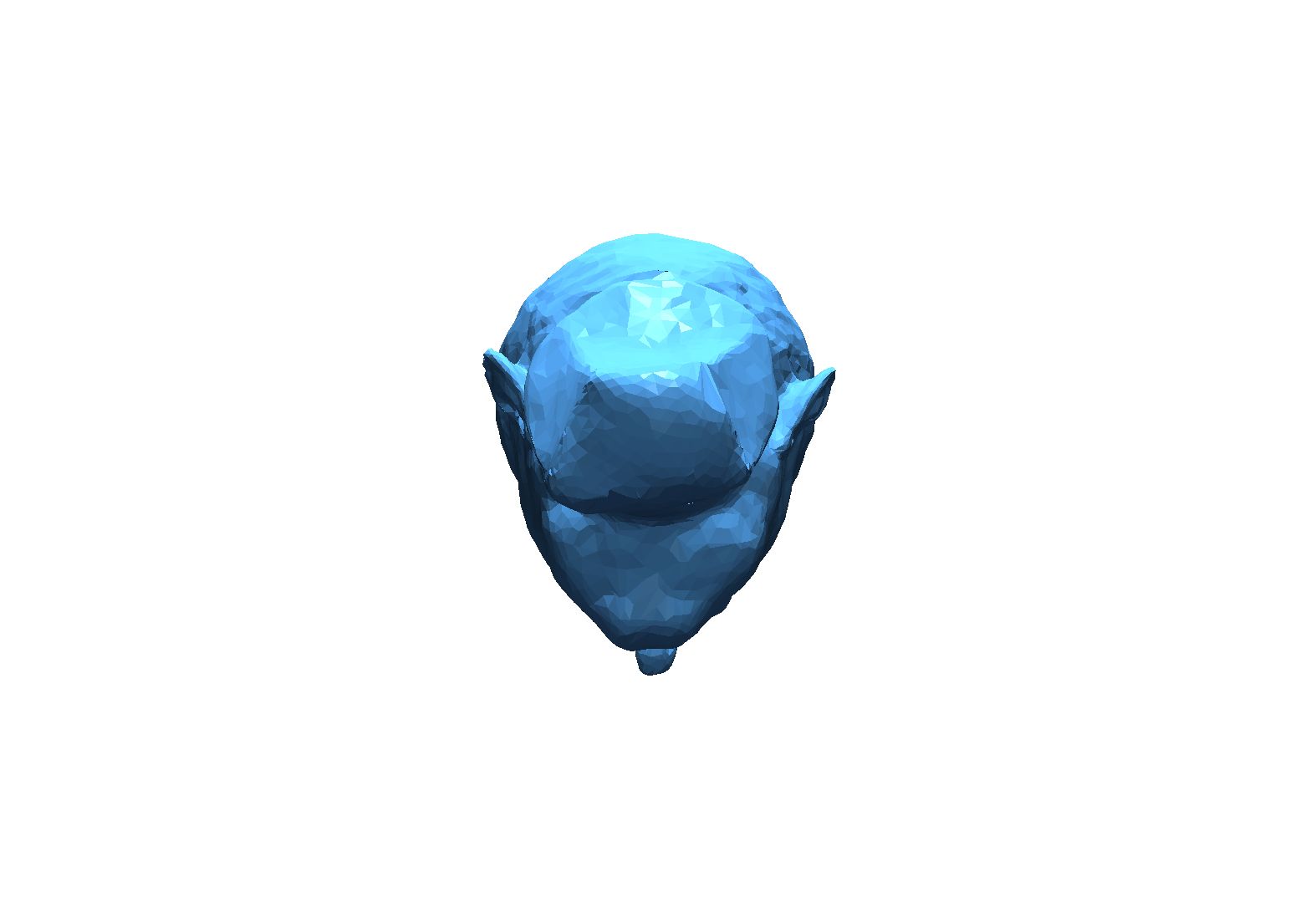}
        \label{fig:Camera005a}
    \end{minipage}
\end{tabular}
\end{center}
Reasoning: Based on the instructions, I need to find the chin. Based on the input images, the best view position for chin is `Camera.png', `Camera001.png', `Camera002.png'.\\
Part: Chin \\
Image: `Camera.png', `Camera001.png', `Camera002.png' \\
\textbf{Example 2:} \\
Text Input: I want to show the deformation result of a cow kneeling down. \\
Image Set: From left to right are: `Camera.png', `Camera001.png', `Camera002.png', `Camera003.png', `Camera004.png', `Camera005.png'
\begin{center}
\begin{tabular}{cccccc}
    \begin{minipage}{0.14\linewidth}
        \centering
        \includegraphics[width=\linewidth]{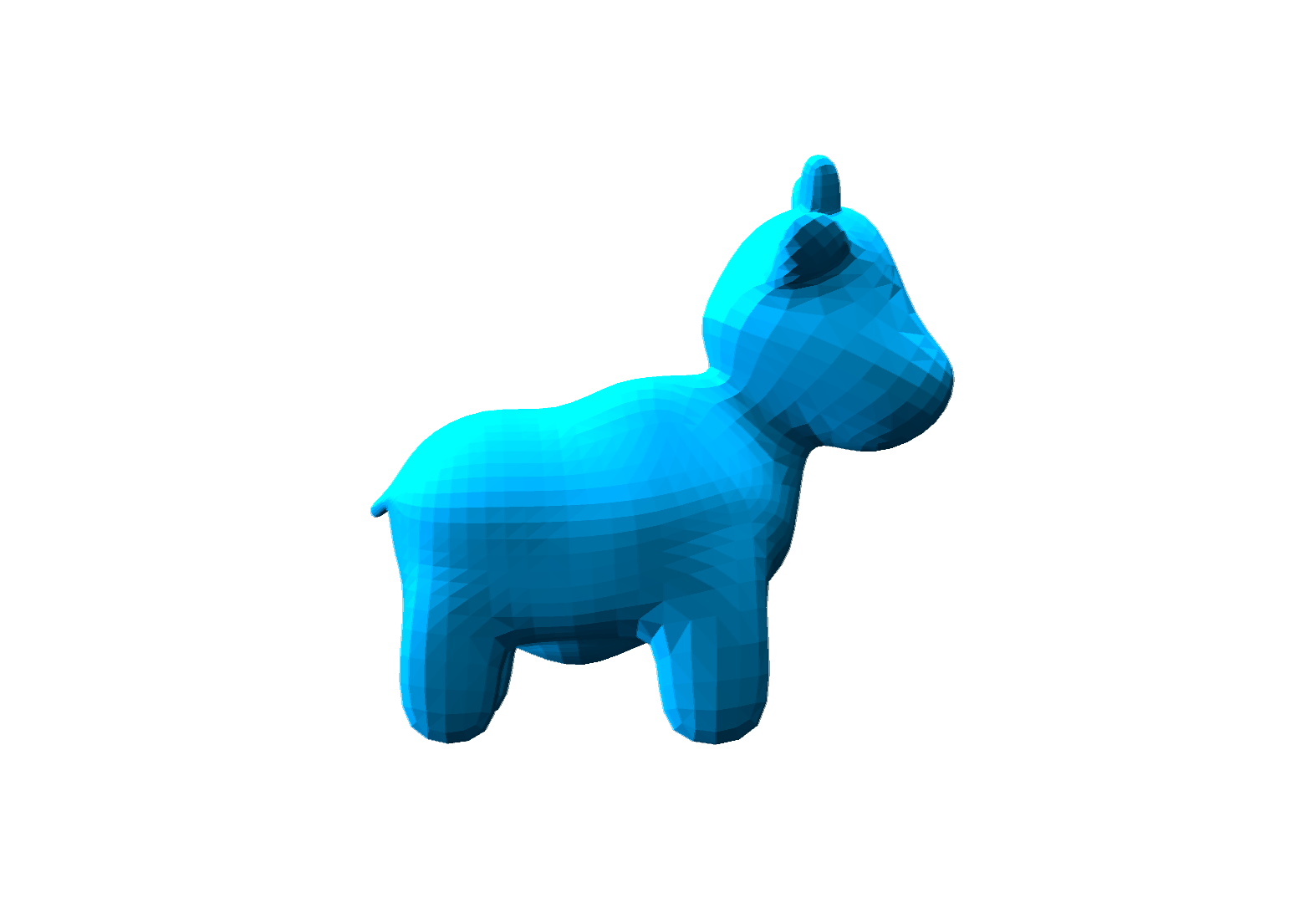}
        \label{fig:Camerab}
    \end{minipage} &
    \begin{minipage}{0.14\linewidth}
        \centering
        \includegraphics[width=\linewidth]{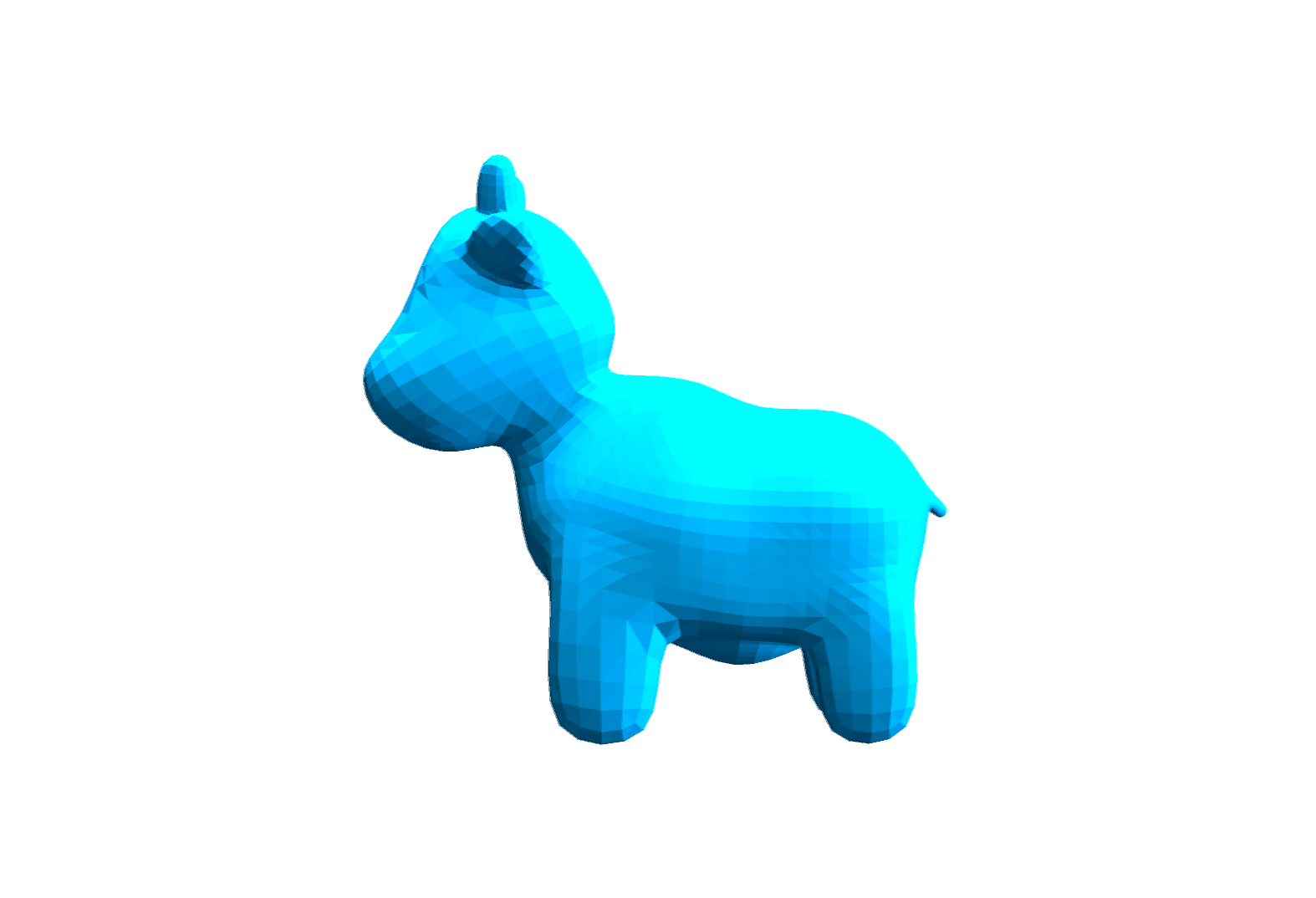}
        \label{fig:Camera001b}
    \end{minipage} &
    \begin{minipage}{0.14\linewidth}
        \centering
        \includegraphics[width=\linewidth]{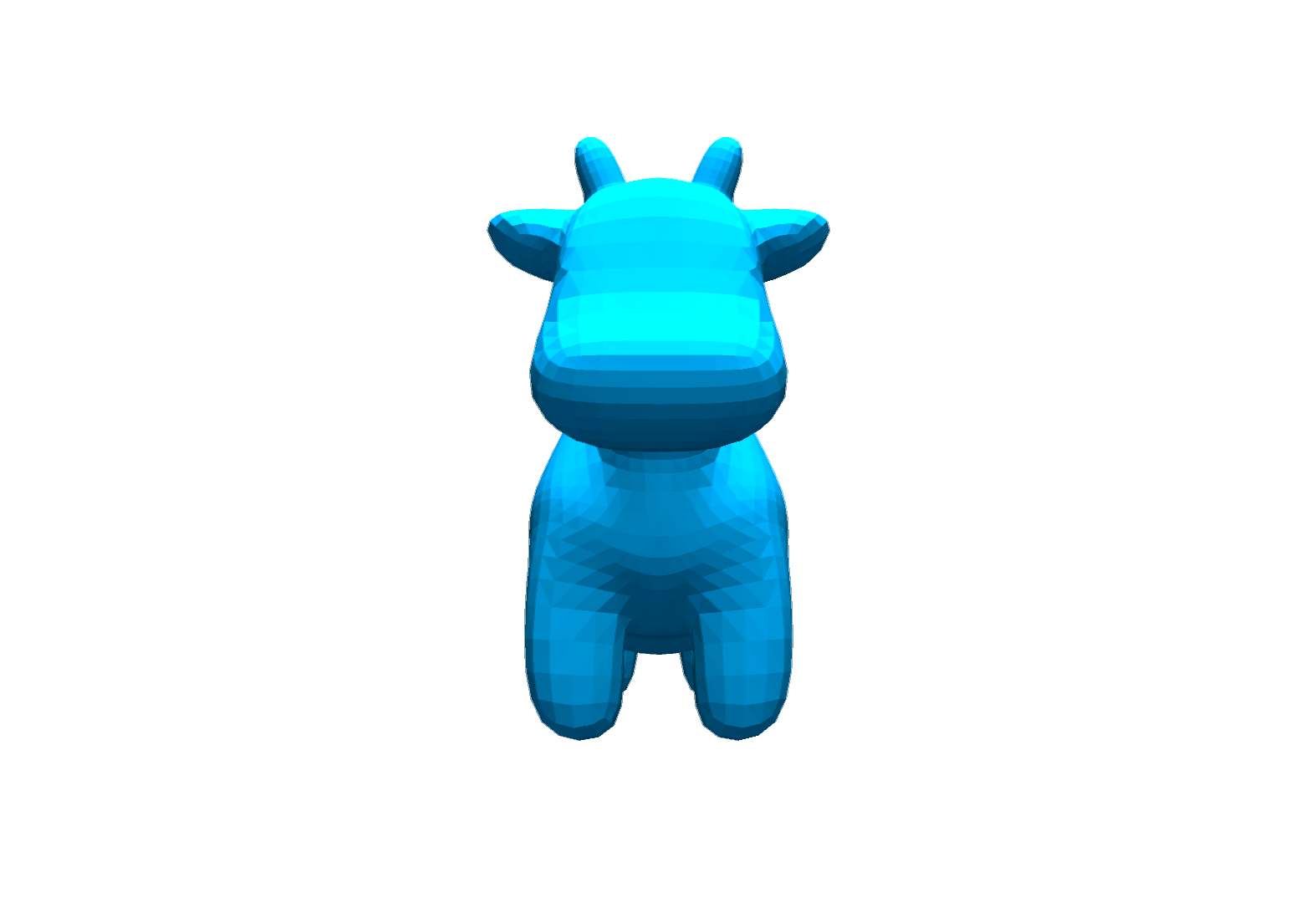}
        \label{fig:Camera002b}
    \end{minipage} &
    \begin{minipage}{0.14\linewidth}
        \centering
        \includegraphics[width=\linewidth]{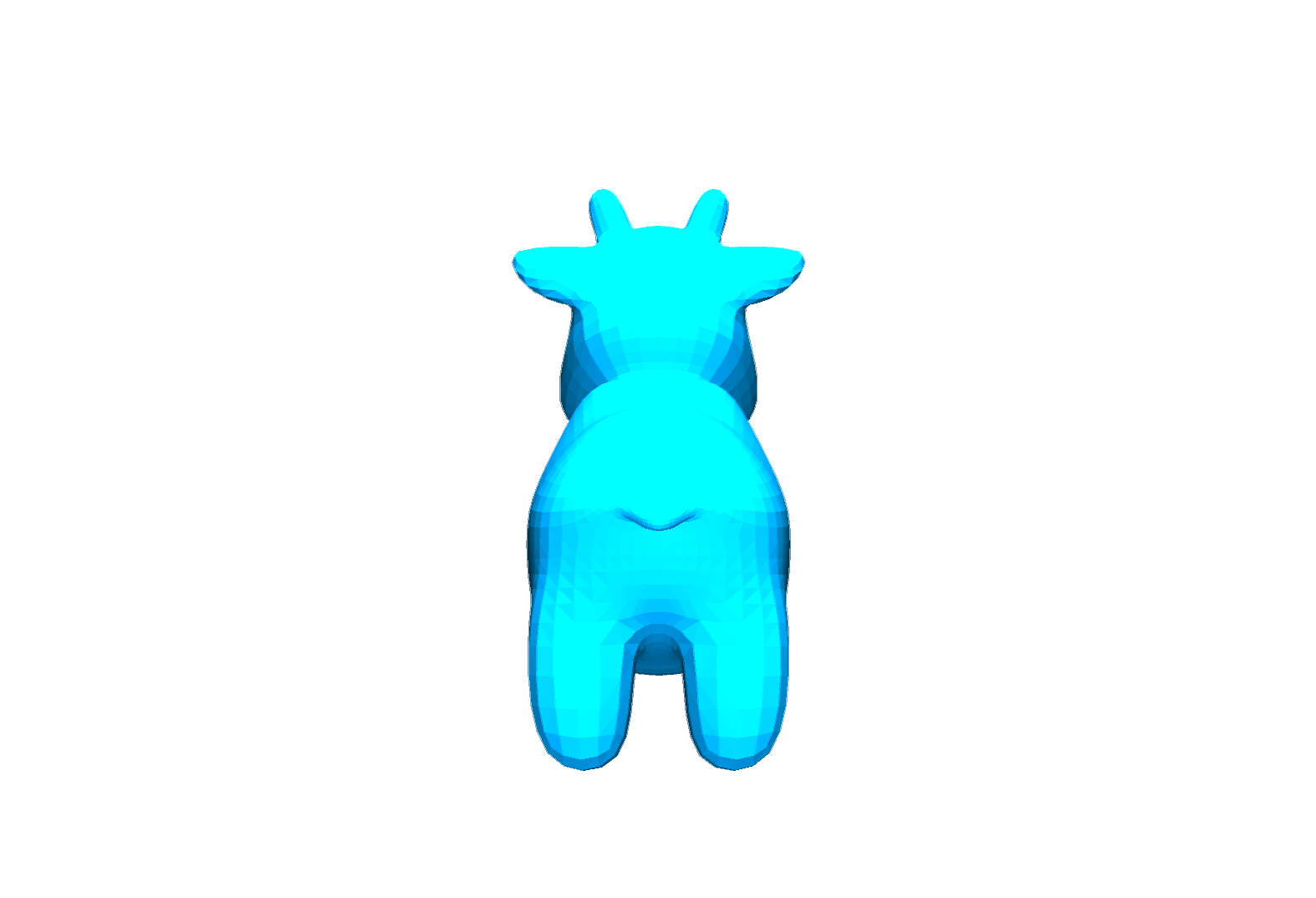}
        \label{fig:Camera003b}
    \end{minipage} &
    \begin{minipage}{0.14\linewidth}
        \centering
        \includegraphics[width=\linewidth]{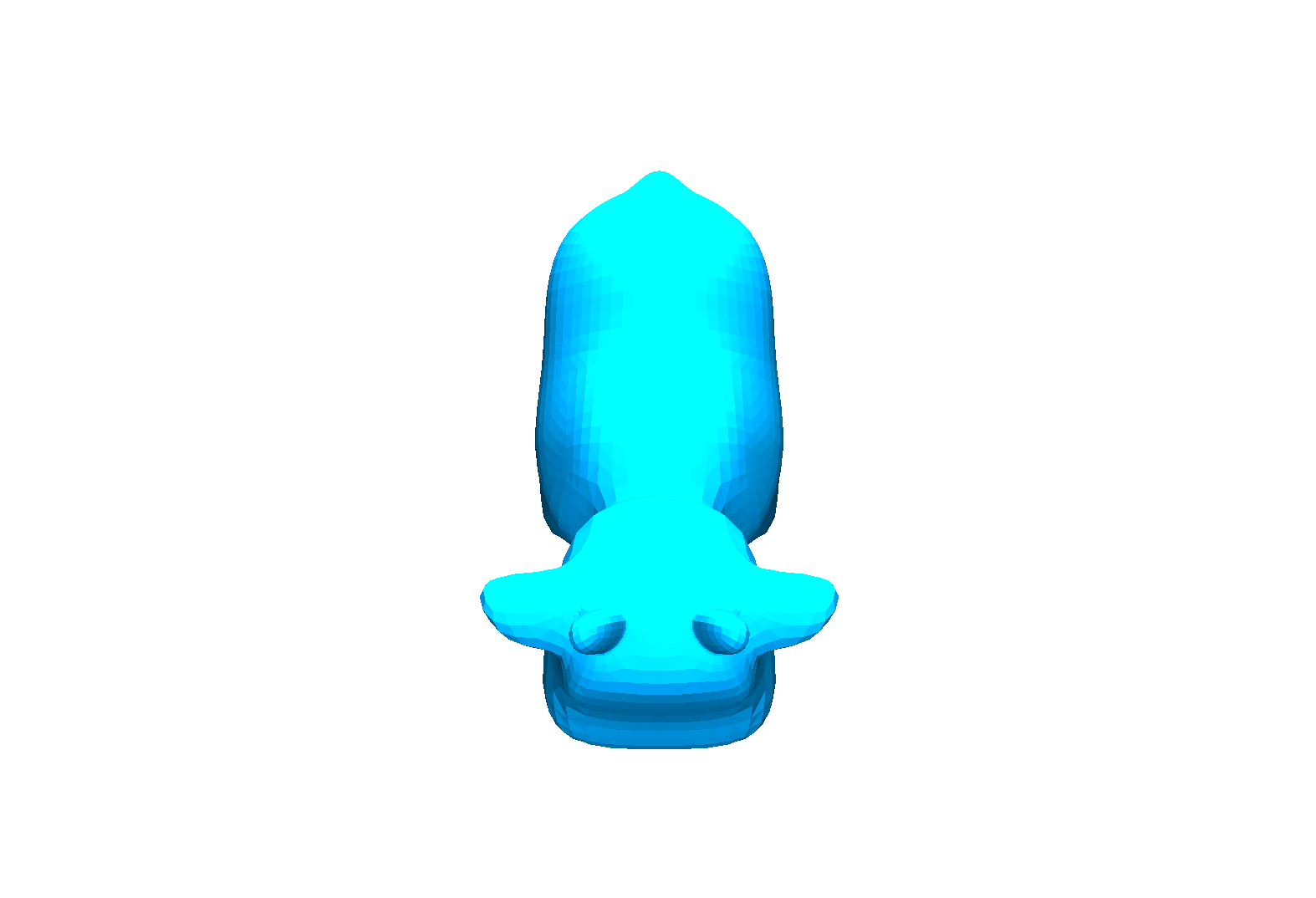}
        \label{fig:Camera004b}
    \end{minipage} &
    \begin{minipage}{0.14\linewidth}
        \centering
        \includegraphics[width=\linewidth]{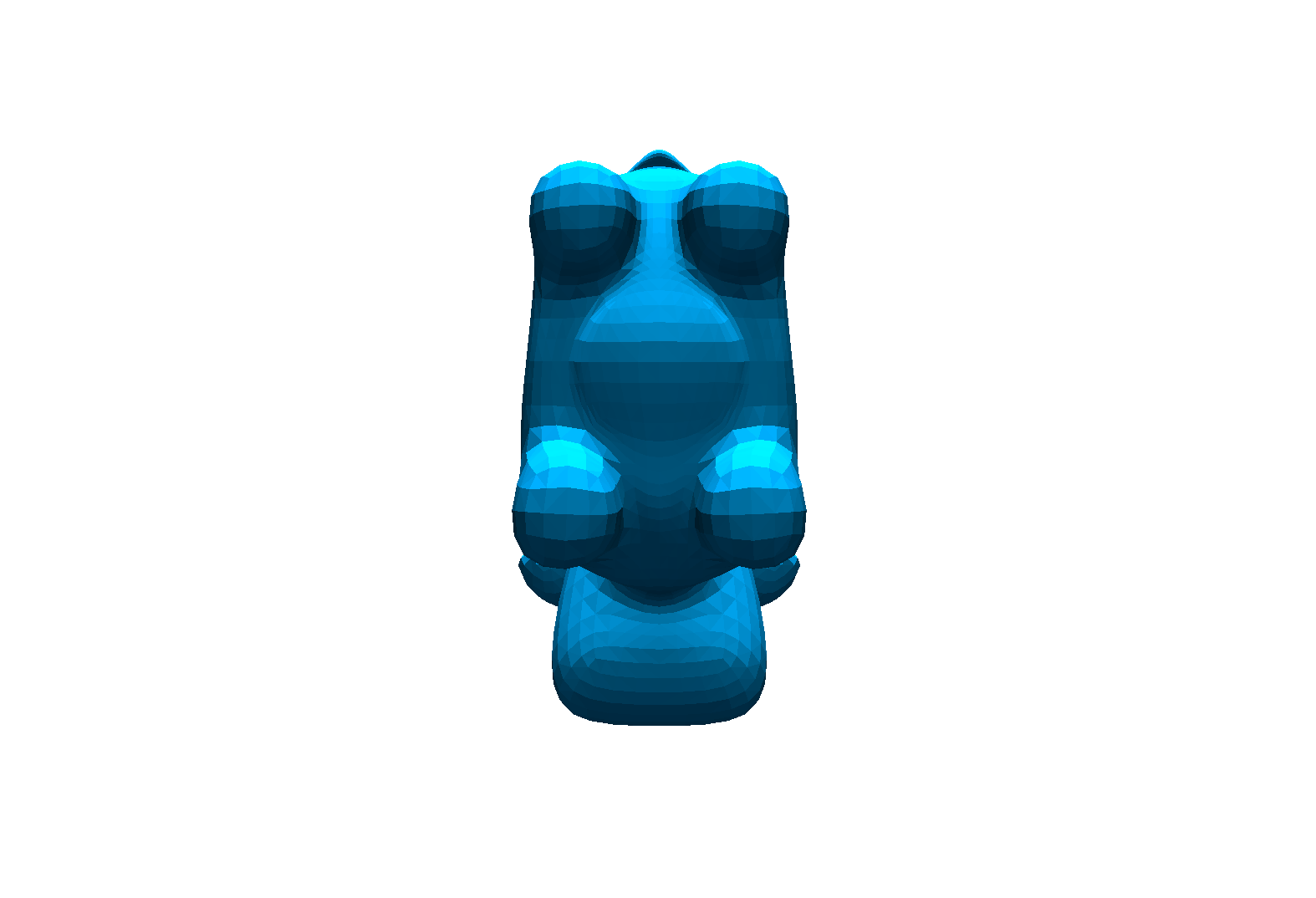}
        \label{fig:Camera005b}
    \end{minipage}
\end{tabular}
\end{center}
Reasoning: Based on the instruction, I need to find the leg of this cow and move it to the end of the body to show the effect of kneeling down. Based on the input images, the best view positions for its legs are: `Camera.png', `Camera001.png', `Camera002.png', `Camera003.png'.\\
Part: Legs \\
Image: `Camera.png', `Camera001.png', `Camera002.png', `Camera003.png' \\

\end{tcolorbox}
\end{figure*}

\begin{figure*}[ht]
\caption{\label{fig:finetune}The finetune data example provided to GPT-4o for selecting handles and directions from camera view.}
\begin{tcolorbox}[width=\linewidth,height=570px]
\textbf{System Instruction:}\\
You are an 3D artist and an expert of 3D mesh understanding. I will provide you with an instruction to deform a part of a 3D object, an image that contains the part,  and deformation handles of the 3D object. The deformation handles are the points on the object that can be moved to deform the object. The deformation handles are visualized in red spheres in the image, and also given as a list of (x,y) pixel positions in respect to the image coordinate, stored in \texttt{Handle Set}. Your task is to first interpret the instruction to identify which direction of the object needs to be deformed. Then select deformation handle and give new (x,y) pixel position in the image to move the handle towards, in order to deform the part. 
The input image dimension is 1920x1080, with the horizontal axis being x-axis, vertical axis being y-axis. so x value should be in the range of (0,1920) and y value should be in the range of (0,1080). The top-left corner of the image is (0,0).
Your answer must strictly follow the JSON format the example below provided. You must strictly follow the answer format as shown in the example, always give reasoning.

\textbf{Example 1:}\\
Instruction: Make the cow's leg longer.\\
Image: Figure \ref{fig:finetune1}\\
Handle Set: [(1023, 868), (725, 867)]\\
Reasoning: Given my camera's position, the cow faces the larger x-axis direction of the image. Since the top-left corner is (0,0) and the object faces the right side of the camera view, the legs appear at the bottom of the image. To elongate the legs, I need to select a pixel on the lower side of the leg.
From the image, I see two legs—one on the left and one on the right. To elongate both, I must increase the y-values of selected handles. The left leg has handles at (725, 867) and (738, 849), while the right leg has handles at (1023, 868) and (1019, 847). I need to pick one handle from each group and increase their y-values accordingly.\\
Direction: Down\\
Handle: [(725, 867), (1023, 868)]\\
New Position: [(725, 900), (1023, 900)]\\
\textbf{Example 2:}\\
Instruction: Make the cow's leg significantly longer.\\
Image: Figure \ref{fig:finetune2}\\
Handle Set: [(1023, 868), (725, 867)]\\
Reasoning: Given my camera's position, the cow faces the larger x-axis direction of the image. Since the top-left corner is (0,0) and the object faces the right side of the camera view, the legs appear at the bottom of the image. To elongate the legs, I need to select a pixel on the lower side of the leg.
From the image, I see two legs—one on the left and one on the right. To elongate both, I must increase the y-values of selected handles. The left leg has handles at (725, 867) and (738, 849), while the right leg has handles at (1023, 868) and (1019, 847). I need to pick one handle from each group and increase their y-values accordingly. Make sure increase them significantly.\\
Direction: Down\\
Handle: [(725, 867), (1023, 868)]\\
New Position: [(725, 1000), (1023, 1000)]\\

\begin{center}
\begin{tabular}{cc}
    \begin{minipage}{0.25\linewidth}
        \centering
        \includegraphics[width=\linewidth]{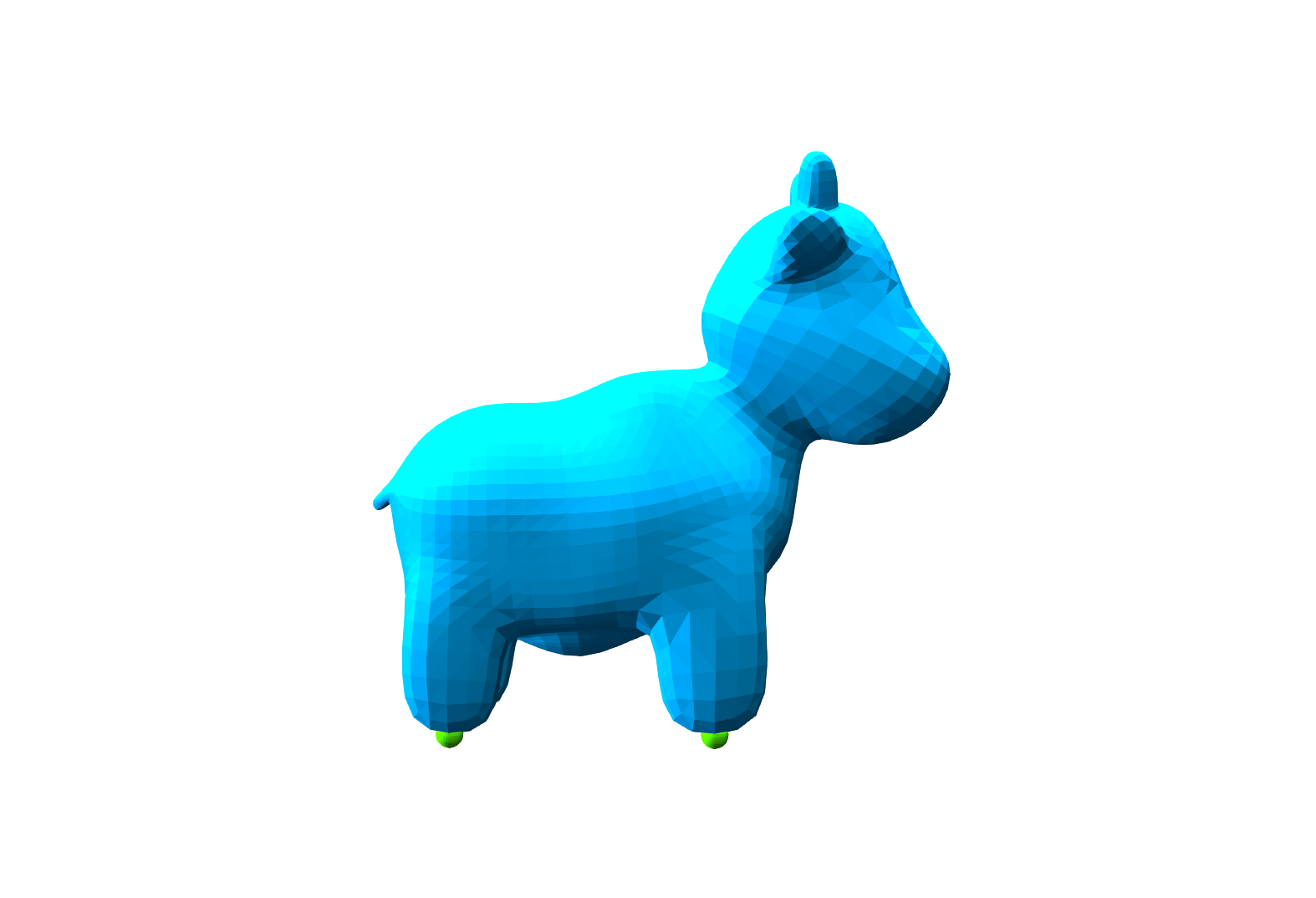}
        \caption{Example 1 image}
        \label{fig:finetune1}
    \end{minipage} &
    \begin{minipage}{0.25\linewidth}
        \centering
        \includegraphics[width=\linewidth]{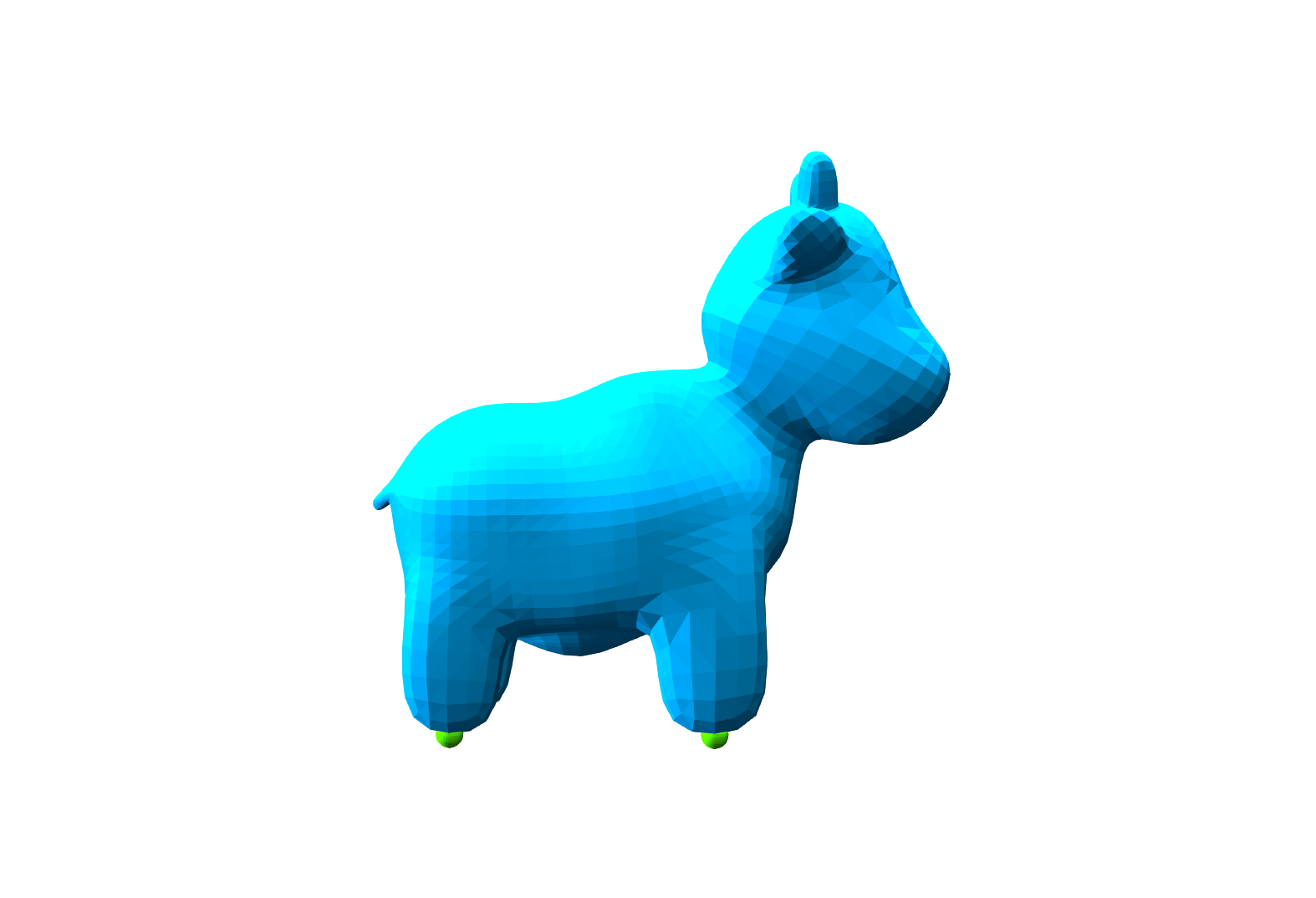}
        \caption{Example 2 image}
        \label{fig:finetune2}
    \end{minipage}
\end{tabular}
\end{center}
\end{tcolorbox}
\end{figure*}

\begin{figure*}[!ht]
\caption{\label{fig:gpteval3d}The instruction for using GPT Evaluation based on GPTEval3D.}
\begin{tcolorbox}[width=\linewidth,height=610px]
Our task here is to compare two 3D objects, both generated from the same text description.
We want to decide which one is better according to the provided criteria.\\

1. Text prompt and Asset Alignment. Focus on how well they correspond to the given text description. An ideal model should accurately reflect all objects and surroundings mentioned in the text prompt, capturing the corresponding attributes as described. Please first describe each of the two models and then evaluate how well it covers all the attributes in the original text prompt.\\
2. 3D Plausibility. Look at both the RGB and normal images and imagine a 3D model from the multi-view images. Determine which model appears more natural, solid, and plausible. Pay attention to any irregularities, such as abnormal body proportions, duplicated parts, or the presence of noisy or meaningless 3D structures. An ideal model should possess accurate proportions, shapes, and structures that closely resemble the real-world object or scene.\\
3. Low-Level Geometry Details. Focus on the local parts of the normal maps. The geometry should accurately represent the intended shape. Note that meaningless noise is not considered as high-frequency details. Determine which one has a more well-organized and efficient structure, which one exhibits intricate details, and which one is more visually pleasing and smooth.\\
4. Considering all the degrees above, which one is better overall?\\

Take a really close look at each of the multi-view images for these two 3D objects before providing your answer. When evaluating these aspects, focus on one of them at a time. Try to make independent decisions between criteria.\\
To provide an answer, please provide a short analysis for each of the abovementioned evaluation criteria. The analysis should be very concise and accurate.
For each of the criteria, you need to make a decision using these three options:\\
1. Left (object 1) is better;\\
2. Right (object 2) is better;\\
3. Cannot decide.\\
IMPORTANT: PLEASE USE THE THIRD OPTION SPARSELY.

Then, in the last row, summarize your final decision by \texttt{<option for criterion 1> <option for criterion 2> <option for criterion 3> <option for criterion 4>}.\\

An example output looks like follows:

1. Text prompt \& Asset Alignment: The left one xxxx; The right one xxxx;
The left/right one is better or cannot decide.

2. 3D Plausibility. The left one xxxx; The right one xxxx;
The left/right one is better or cannot decide.

3. Low-Level Geometry Details. The left one xxxx; The right one xxxx;
The left/right one is better or cannot decide.

4. Overall, xxxx
The left/right one is better or cannot decide.

Final answer:

x x x x  (e.g., 1 2 3 1 / 3 3 2 1 / 3 2 1 1)

\begin{center}
\begin{tabular}{ccc}
    \begin{minipage}{0.3\linewidth}
        \centering
        \includegraphics[width=\linewidth]{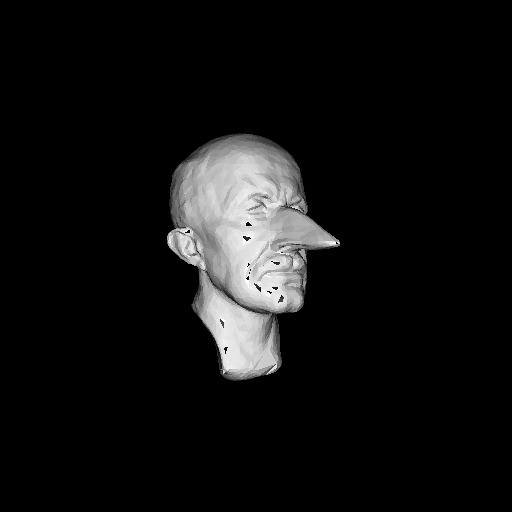}
        \caption{Render}
        \label{fig:gpteval1}
    \end{minipage} &
    \begin{minipage}{0.3\linewidth}
        \centering
        \includegraphics[width=\linewidth]{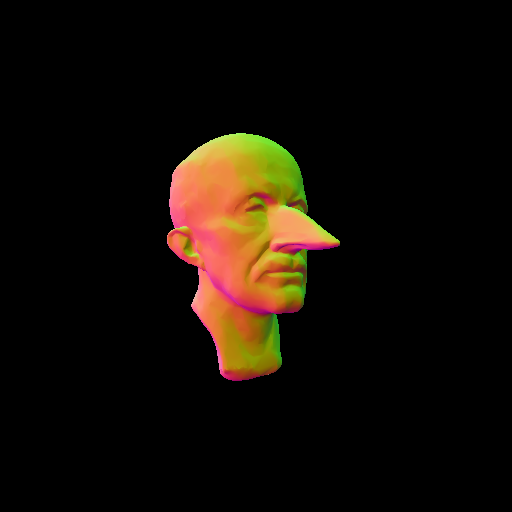}
        \caption{Surface normal}
        \label{fig:gpteval2}
    \end{minipage} &
    \begin{minipage}{0.3\linewidth}
        \centering
        \includegraphics[width=\linewidth]{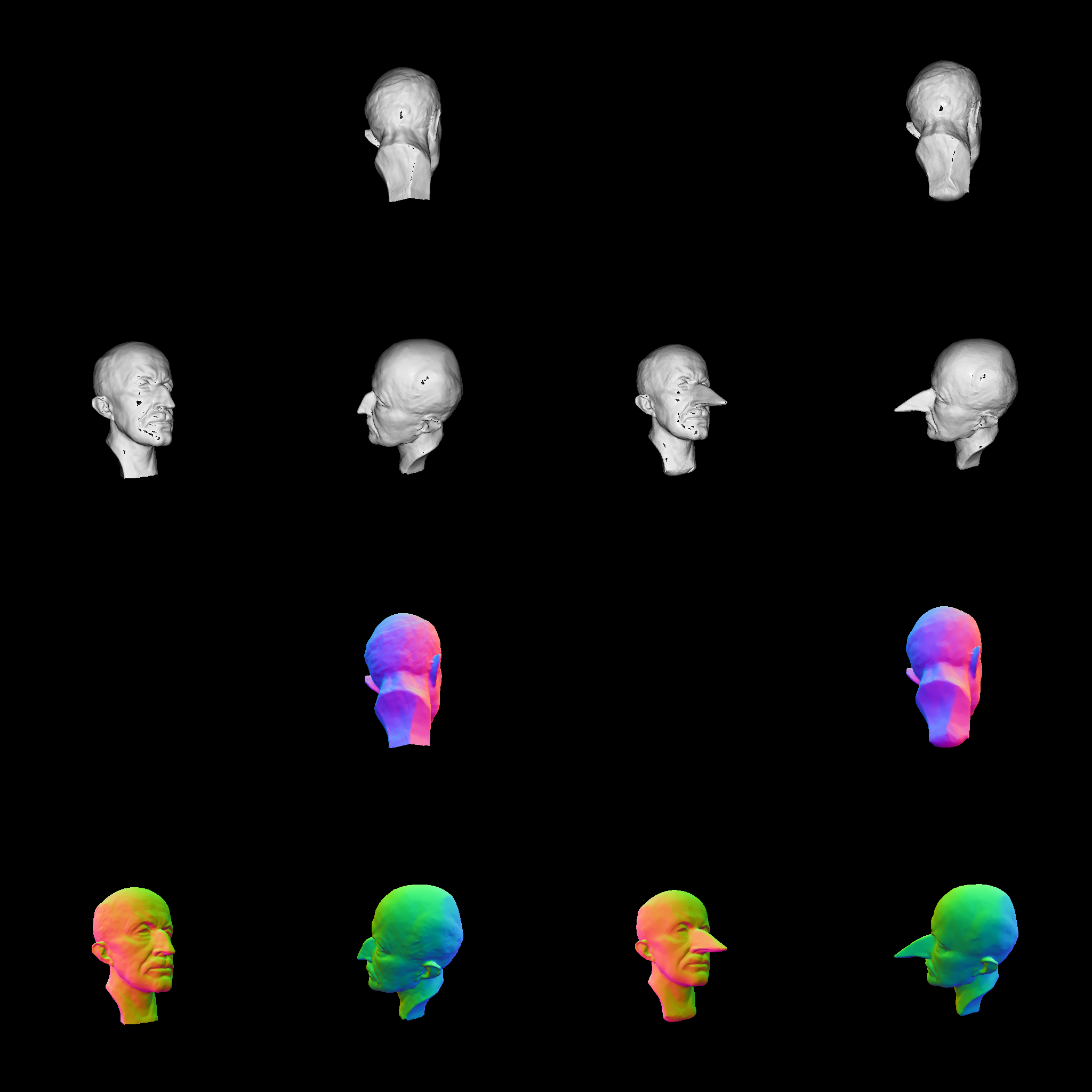}
        \caption{Image feed into GPT}
        \label{fig:gpteval3}
    \end{minipage}
\end{tabular}
\end{center}

\end{tcolorbox}
\end{figure*}

We summarize the three stages of prompt engineering. In~\prettyref{fig:InstructionToGPT}, we show the prompt for sub-instruction extraction. In~\prettyref{fig:InstructionToGPT2}, we show the prompt for detecting relevant mesh parts and selecting camera views for part segmentation. In~\prettyref{fig:InstructionToGPT3}, we show the prompt for handle selection and dragging direction. In~\prettyref{fig:finetune}, we show the data example for fine-tuning GPT4o for handle selection and movement direction prediction. We collected 102 data pairs to fine-tune the model. Finally, in~\prettyref{fig:gpteval3d}, we show the prompt for computing GPTEval3D scores.

\label{sec:prompt}
\begin{figure*}[!ht]
\caption{\label{fig:InstructionToGPT}The instruction provided to GPT-4o for parsing user input.}
\begin{tcolorbox}[width=\linewidth,height=125px]
\textbf{System Instruction:}\\
You are a skilled 3D artist with advanced expertise in 3D object manipulation and geometric reasoning. Your task is to interpret the instructions given for deforming 3D objects, analyze them thoroughly, and deconstruct them into a series of clear, actionable deformation sub-instructions. Your answer must strictly follow the JSON format provided in the example below. Do not write any additional text. \\
\textbf{Example 1:}\\
Input: I want to shorten the chin of this human. Output: ``Shorten the chin" \\
\textbf{Example 2:}\\
Input: Adjust the chair to have lower armrests and higher backrest.
Output: ``Lower armrests”, ``Higher backrest”
\end{tcolorbox}
\vspace{-10px}
\end{figure*}

\begin{figure*}[ht]
\caption{\label{fig:InstructionToGPT3}The instruction provided to GPT-4o for selecting handles and directions from camera view.}
\begin{tcolorbox}[width=\linewidth,height=595px]
\textbf{System Instruction:}\\
You are a 3D artist and an expert in 3D understanding. I will provide you with instructions on how to deform a part of a 3D object, an image that contains the part, 
and the deformation handles of the 3D object. The deformation handles are visualized in green spheres in the image, and also given as a list of (x,y) pixel positions in the image.\\
Your task is to first interpret the instruction to identify which direction of the object needs to be deformed. Then select only one deformation handle and give a new (x,y) pixel position in the image to move the handle towards, in order to deform the part. \\
The input image dimension is 1920x1080, with the horizontal axis being x-axis, vertical axis being y-axis. so x value should be in the range of (0, 1920) and y value should be in the range of (0, 1080). The top-left corner of the image is (0, 0).\\
Your answer must strictly follow the JSON format provided in the example below. You must strictly follow the answer format as shown in the example, always give reasoning.\\

\textbf{Example 1:}\\
Text Input: The deformation instruction: Deform the cow head lower to the feet. Given the handles on cow head in (x,y) format: [(806, 344),(888, 270),(897, 321),(1017, 481),(1113, 344),(1012, 500)]; what is deform direction? Which handle should I use? Please output the handle's new xy pixel position in scale of the given image.\\
Image:  Figure \ref{fig:cone1}\\

Reasoning: Based on the information provided, the cow head is in the middle-top of the image. Given the top-left corner of the image is (0,0), object is facing the camera view,
I need to pull the head downward along the vertical axis to deform it lower, and thus to the down side of the image. Thus, I should deform the head by increasing the handle's y-axis value.
From the image, I can see several handles on cow's head, including ears, horn, and mouth. To deform the whole head lower, I should pick a handle that's on the bottom of the head with largest y value, which is (1012, 500) on the mouth.
So, to deform the cow head lower, I need to increase the (1012, 500)'s y value, and thus move it to (1012, 650).\\
Direction: Down\\
Handle: [(1012, 500)]\\
New Position: [(1012, 650)]\\

\textbf{Example 2:}\\
Text Input: The deformation instruction: Turn the cow head to look right. Given the handles on cow head in (x,y) format: [(1040, 459), (1128, 231), (1117, 297), (1135, 219), (1121, 280), (1052, 473), (1097, 327), (1265, 497)]; what is deform direction? Which handle should I use? Please output the handle's new xy pixel position in scale of the given image.\\
Image:  Figure \ref{fig:cone2}\\

Reasoning: Based on the information provided, the cow head is in the right side of the image, facing right. Given the top-left corner of the image is (0,0), object is facing the right side of the camera view, I need to select a pixel on left side of the head to turn it facing its left. Thus, I should deform the head by decreasing the handle's x-axis value.
From the image, I can see several handles on cow's head, including ears, horn, and mouth. To deform the whole head turn left, I should pick a handle that's on the right most of the head with largest x value, which is (1265, 497) on the mouth.
Turning the head to right means not only displacement but also rotation, so the handle movement should be large enough to make the head turn right.
So, to deform the cow head turn left, I need to decrease the (1265, 497)'s x value, and thus move it to (660,490).
\\
Direction: Left\\
Handle: [(1265, 497)]\\
New Position: [(660, 490)]

\end{tcolorbox}
\end{figure*}

\begin{figure*}[ht]
\caption{\label{fig:InstructionToGPT3-c}The instruction provided to GPT-4o for selecting handles and directions from camera view (continued).}
\begin{tcolorbox}[width=\linewidth,height=570px]
\textbf{Example 3:}\\
Text Input: The deformation instruction: Turn the cow head to look right. Given the handles on cow head in (x,y) format: [(926, 584), (1004, 643), (855, 625), (911, 641), (917, 632), (998, 593),  (1064, 625), (992, 715)]; what is deform direction? Which handle should I use? Please output the handle's new xy pixel position in scale of the given image.\\
Image: Figure \ref{fig:cone3} \\

Reasoning: Based on the information provided, my camera is on top of the cow and cow is facing to the larger y-axis direction of the image. Given the top-left corner of the image is (0,0), object is facing the higher y-axis side of the camera view,
in order to turn the cow facing its own right, which is actually the left side of the image given my camera's orientation, I need to select a pixel on left side of the head to turn it facing its right. Thus, I should deform the head by decreasing the handle's x-axis value.
From the image, I can see several handles on cow's head, including ears, horn, and mouth. To deform the whole head facing to its right, I should pick a handle that's on the right most of the head with largest x value, which is (1064, 625) on its left ear. 
Turning the head to right means not only displacement but also rotation, so the handle movement should be large enough to make the head turn right.
So, to deform the cow head turn to its right, I need to decrease the (1064, 625)'s x value, and thus move it to (520,600).\\
Direction: Right\\
Handle: [(1064,600)]\\
New Position: [(520,625)]\\

\textbf{Example 4:}\\
Text Input: The deformation instruction: Shorten the cow's leg. Given the handles on cow legs in (x,y) format: [(1023, 868), (725, 867)] ; what is deform direction? Which handle should I use? Please output the handle's new xy pixel position in scale of the given image.\\
Image: Figure \ref{fig:cone4}\\

Reasoning:  Based on the information provided, my camera is on side of the cow and cow is facing to the larger x-axis direction of the image. 
Given the top-left corner of the image is (0,0), object is facing the larger x-axis side of the camera view, in order to shorten the legs, which is actually the bottom side of the image given my camera's orientation, I need to select a pixel on bottom side of the leg. Thus, I should deform the head by decreasing the handle's y-axis value.
From the image, I can see 2 legs, one on the left side, another on the right side, so I need to shorten both of them in order to fullfill the instruction. We have 2 handle, (725, 867) and (738, 849), on left side leg and 2 handles, (1023, 868) and (1019, 847), on right side leg.
So, I need to pick a handle from left group, and another handle from right group, and shorten them by decreasing their y value.\\
Direction: Up\\
Handle: [(725, 867), (1023, 868)]\\
New Position: [(725, 800), (1023, 800)]\\

\begin{center}
\begin{tabular}{cccc}
    \begin{minipage}{0.22\linewidth}
        \centering
        \includegraphics[width=\linewidth]{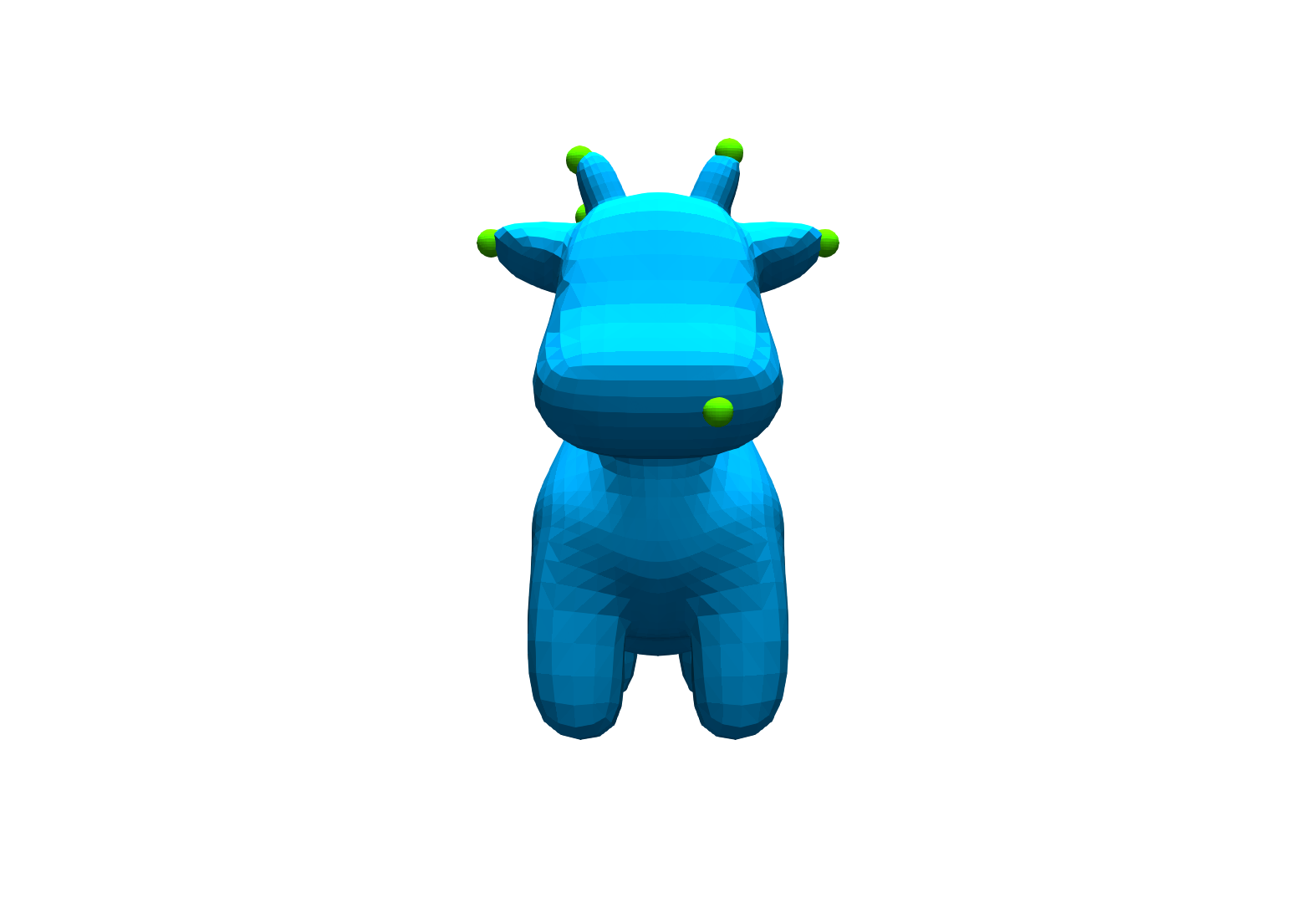}
        \caption{Example 1 image}
        \label{fig:cone1}
    \end{minipage} &
    \begin{minipage}{0.22\linewidth}
        \centering
        \includegraphics[width=\linewidth]{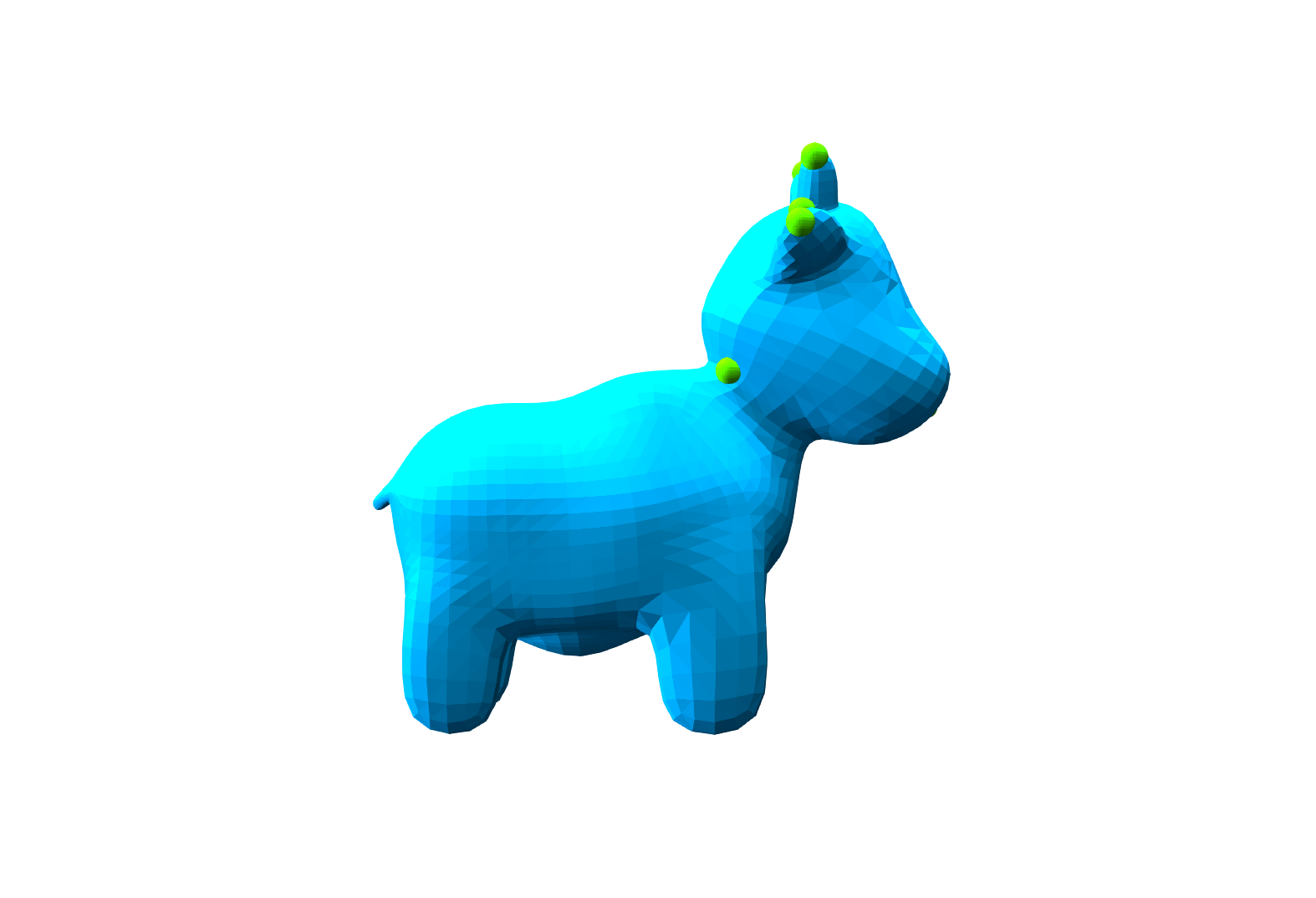}
        \caption{Example 2 image}
        \label{fig:cone2}
    \end{minipage} &
    \begin{minipage}{0.22\linewidth}
        \centering
        \includegraphics[width=\linewidth]{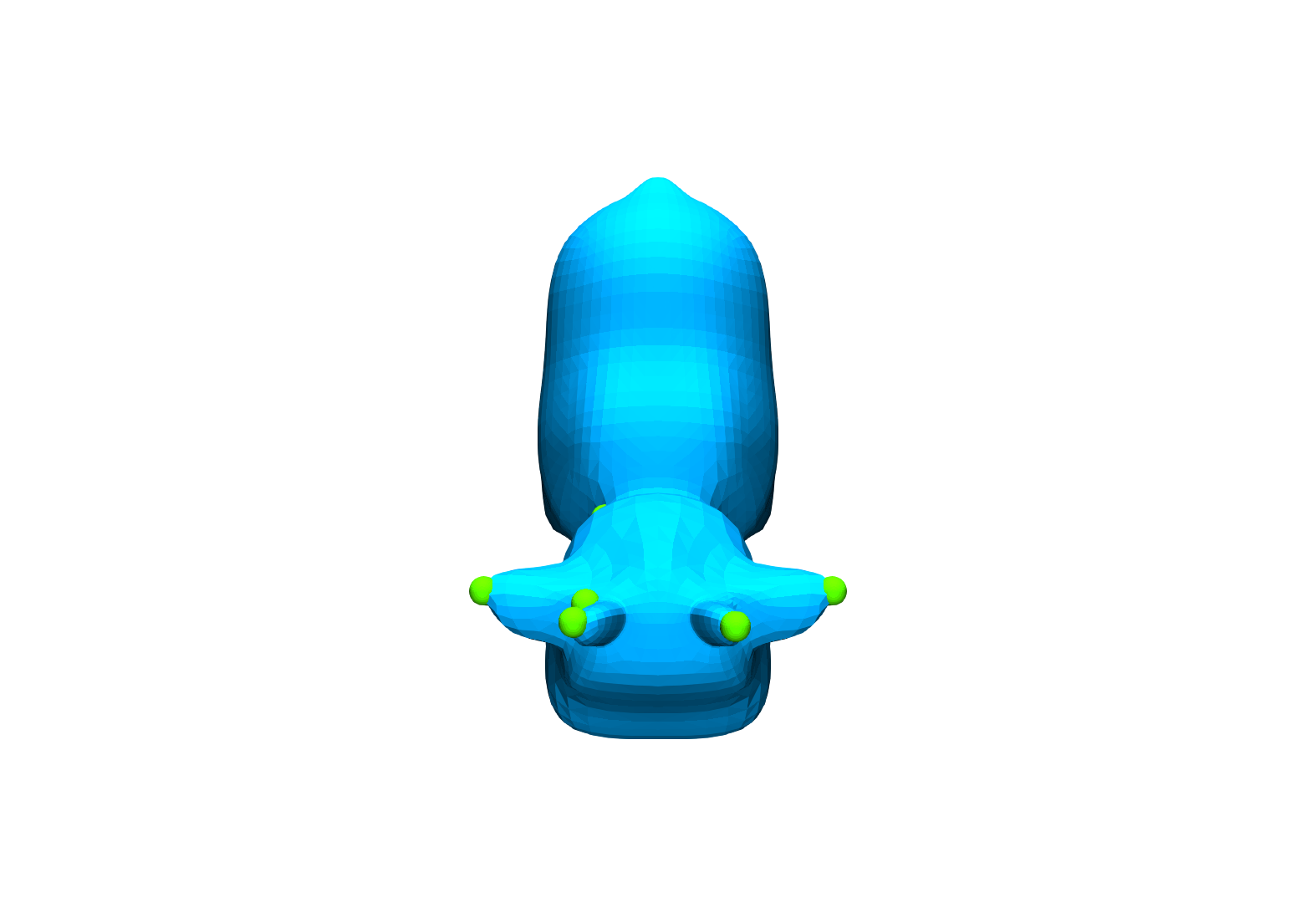}
        \caption{Example 3 image}
        \label{fig:cone3}
    \end{minipage} &
    \begin{minipage}{0.22\linewidth}
        \centering
        \includegraphics[width=\linewidth]{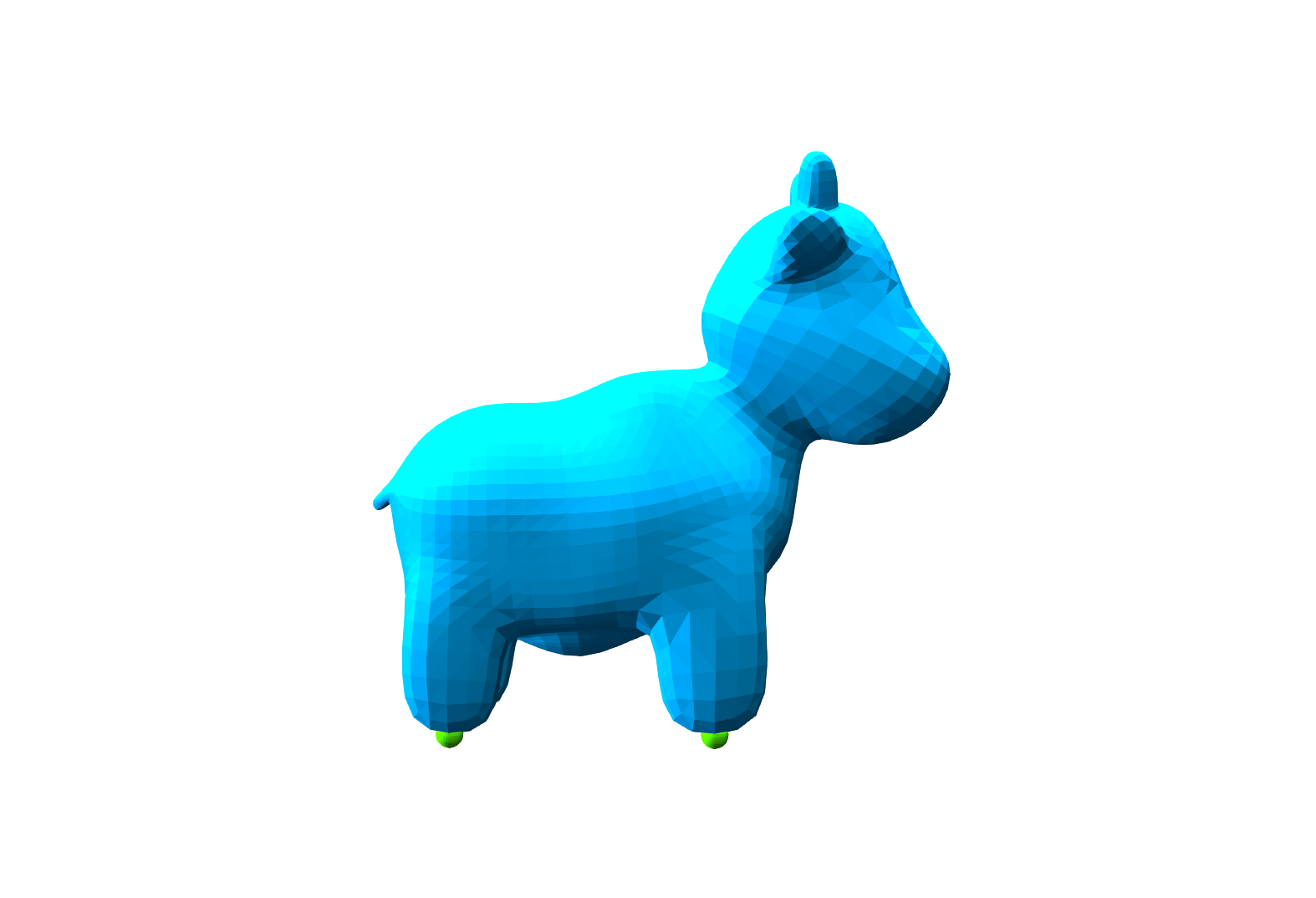}
        \caption{Example 4 image}
        \label{fig:cone4}
    \end{minipage}
\end{tabular}
\end{center}
\end{tcolorbox}
\end{figure*}

